\documentclass[longauth, usenatbib]{aa}  
\usepackage{hyperref}
\usepackage{newtxtext,newtxmath}
\usepackage{textgreek}
\usepackage{fontawesome}
\usepackage{xcolor}

\usepackage[T1]{fontenc}
\usepackage{ae,aecompl}

\usepackage{graphicx}   
\usepackage{amsmath}    
\usepackage{amssymb}    
\usepackage{caption}
\usepackage{subfig}

\usepackage[para, online, flushleft]{threeparttable}

\newcommand{\Hunit}{km s$^{-1}{\rm Mpc}^{-1}$}
\newcommand{\Htdcosmo}{74.5$^{+5.6}_{-6.1}$}
\newcommand{\HtdcosmoFlatOmega}{75.5$^{+7.0}_{-6.9}$}
\newcommand{\Htdcosmoifu}{73.3$^{+5.8}_{-5.8}$}
\newcommand{\HtdcosmoSLACS}{67.4$^{+4.3}_{-4.7}$}
\newcommand{\Hjoint}{67.4$^{+4.1}_{-3.2}$}

\begin{document}
%
%
%
\title{TDCOSMO IV: Hierarchical time-delay cosmography - joint inference of the Hubble constant and galaxy density profiles}
\titlerunning{Hierarchical time-delay cosmography}
\author{
S.~Birrer\inst{1}\fnmsep\thanks{E-mail: sibirrer@stanford.edu}
\and
A.~J.~Shajib\inst{2}
\and
A.~Galan\inst{3}
\and
M.~Millon\inst{3}
\and
T.~Treu\inst{2}
\and
A.~Agnello\inst{4}
\and
M.~Auger\inst{5, 6}
\and
G.~C.-F.~Chen\inst{7}
\and
L.~Christensen\inst{4}
\and
T.~Collett\inst{8}
\and
F.~Courbin\inst{3}
\and
C.~D.~Fassnacht\inst{7, 9}
\and
L.~V.~E.~Koopmans\inst{10}
\and
P.~J.~Marshall\inst{1}
\and
J.-W.~Park\inst{1}
\and
C.~E.~Rusu\inst{11}
\and
D.~Sluse\inst{12}
\and
C.~Spiniello\inst{13, 14}
\and
S.~H.~Suyu\inst{15, 16, 17}
\and
S.~Wagner-Carena\inst{1}
\and
K.~C.~Wong\inst{18}
\and
M.~Barnab\`e
\and
A.~S.~Bolton\inst{19}
\and
O.~Czoske\inst{20}
\and
X.~Ding\inst{2}
\and
J.~A.~Frieman\inst{21, 22}
\and
L.~Van de Vyvere\inst{12}
}
\institute{
Kavli Institute for Particle Astrophysics and Cosmology and Department of Physics, Stanford University, Stanford, CA 94305, USA
\and 
Physics and Astronomy Department, University of California, Los Angeles CA 90095, USA
\and 
Institute of Physics, Laboratory of Astrophysics, Ecole Polytechnique F\'ed\'erale de Lausanne (EPFL), Observatoire de Sauverny, 1290 Versoix, Switzerland
\and 
DARK, Niels-Bohr Institute, Lyngbyvej 2, 2100 Copenhagen, DEN
\and 
Institute of Astrononmy, University of Cambridge, Madingley Road, Cambridge CB30HA, UK
\and 
Kavli Institute for Cosmology, University of Cambridge, Madingley Road, Cambridge CB30HA, UK
\and 
Physics Dept., University of California, Davis, 1 Shields Ave., Davis, CA 95616
\and 
Institute of Cosmology and Gravitation, University of Portsmouth, Burnaby Rd, Portsmouth, PO1 3FX, UK
\and 
Carnegie Visiting Scientist
\and 
Kapteyn Astronomical Institute, University of Groningen, P.O.Box 800, 9700AV Groningen, The Netherlands
\and 
National Astronomical Observatory of Japan, 2-21-1 Osawa, Mitaka, Tokyo 181-0015, Japan
\and 
STAR Institute, Quartier Agora - All\'ee du six Ao\^ut, 19c B-4000 Li\`ege, Belgium
\and 
Department of Physics, University of Oxford, Denys Wilkinson Building, Keble Road, Oxford OX1 3RH, UK
\and 
INAF, Osservatorio Astronomico di Capodimonte, Via Moiariello  16, 80131, Naples, IT
\and 
Max-Planck-Institut f{\"u}r Astrophysik, Karl-Schwarzschild-Str.~1, 85748 Garching, Germany
\and 
Physik-Department, Technische Universit\"at M\"unchen, James-Franck-Stra\ss{}e~1, 85748 Garching, Germany
\and 
Academia Sinica Institute of Astronomy and Astrophysics (ASIAA), 11F of ASMAB, No.1, Section 4, Roosevelt Road, Taipei 10617, Taiwan
\and 
Kavli IPMU (WPI), UTIAS, The University of Tokyo, Kashiwa, Chiba 277-8583, Japan
\and 
NSF's National Optical-Infrared Astronomy Research Laboratory, 950 N.~Cherry Ave., Tucson, AZ 85719, USA
\and 
University of Vienna, Department of Astrophysics, T\"urkenschanzstr.~17, 1180 Wien, Austria
\and 
Fermi National Accelerator Laboratory, P.O. Box 500, Batavia, IL 60510, USA
\and 
Kavli Institute for Cosmological Physics, Department of Astronomy \& Astrophysics, The University of Chicago, Chicago, IL 60637, USA
}
%
\date{Accepted XXX. Received YYY; in original form ZZZ}
\abstract{
The H0LiCOW collaboration inferred via strong gravitational lensing time delays a Hubble constant value of $H_0 = 73.3^{+1.7}_{-1.8}$  \Hunit, describing deflector mass density profiles by either a power-law or stars (constant mass-to-light ratio) plus standard dark matter halos.
The mass-sheet transform (MST) that leaves the lensing observables unchanged is considered the dominant source of residual uncertainty in $H_0$.
We quantify any potential effect of the MST with a flexible family of mass models, which directly encodes it, and they are hence maximally degenerate with $H_0$.
Our calculation is based on a new hierarchical Bayesian approach in which the MST is only constrained by stellar kinematics. The approach is validated on mock lenses, which are generated from hydrodynamic simulations.
We first applied the inference to the TDCOSMO sample of seven lenses, six of which are from H0LiCOW, and measured $H_0 = $ \Htdcosmo~\Hunit. 

Secondly, in order to further constrain the deflector mass density profiles, we added imaging and spectroscopy for a set of 33 strong gravitational lenses from the Sloan Lens ACS (SLACS) sample. For nine of the 33 SLAC lenses, we used resolved kinematics to constrain the stellar anisotropy. From the joint hierarchical analysis of the TDCOSMO+SLACS sample, we measured $H_0 = $ \Hjoint~\Hunit. This measurement assumes that the TDCOSMO and SLACS galaxies are drawn from the same parent population.
The {\bf blind} H0LiCOW, TDCOSMO-only and TDCOSMO+SLACS analyses are in mutual statistical agreement. The TDCOSMO+SLACS analysis prefers marginally shallower mass profiles than H0LiCOW or TDCOSMO-only. Without relying on the form of the mass density profile used by H0LiCOW, we achieve a $\sim$5\% measurement of $H_0$.
While our new hierarchical analysis does not statistically invalidate the mass profile assumptions by H0LiCOW -- and thus the $H_0$ measurement relying on them -- it demonstrates the importance of understanding the mass density profile of elliptical galaxies. 
The uncertainties on $H_0$ derived in this paper can be reduced by physical or observational priors on the form of the mass profile, or by additional data. The full analysis is available  \faGithub\href{https://github.com/TDCOSMO/hierarchy_analysis_2020_public/}{~here}.
}
\maketitle
\section{Introduction} \label{sec:introduction}
There is a discrepancy in the reported measurements of the Hubble constant from early universe and late universe distance anchors. If confirmed, this discrepancy would have profound consequences and would require new or unaccounted physics to be added to the standard cosmological model.
Early universe measurements in this context are primarily calibrated with sound horizon physics. This includes the cosmic microwave background (CMB) observations from \textit{Planck} with $H_0 = 67.4\pm 0.5$ \Hunit \citep{Planck:2018param}, galaxy clustering and weak lensing measurements of the Dark Energy Survey (DES) data in combination with baryon acoustic oscillations (BAO) and Big Bang nucleosynthesis (BBN) measurements, giving $H_0 = 67.4\pm1.2$ \Hunit \citep{DES:2018_H0}, and using the full-shape BAO analysis in the BOSS survey in combination with BBN, giving $H_0 = 68.4\pm1.1$ \Hunit \citep{Philcox:2020}.
All of these measurements provide a self-consistent picture of the growth and scales of structure in the Universe within the standard cosmological model with a cosmological constant, $\Lambda$, and cold dark matter ($\Lambda$CDM).

Late universe distance anchors consist of multiple different methods and underlying physical calibrators. The most well established one is the local distance ladder, effectively based on radar observations on the Solar system scale, the parallax method, and a luminous calibrator to reach the Hubble flow scale.
The SH$0$ES team, using the distance ladder method with supernovae (SNe) of type Ia and Cepheids, reports a measurement of $H_0 = 74.0\pm1.4$ \Hunit \citep{Riess:2019}.
The Carnegie--Chicago Hubble Project (CCHP) using the distance ladder method with SNe Ia and the tip of the red giant branch measures $H_0 = 69.6\pm1.9$~\Hunit \citep{Freedman:2019, Freedman:2020}.
\cite{Huang:2020} used the distance ladder method with SNe Ia and Mira variable stars and measured $H_0 = 73.3\pm4.0$ \Hunit.

Among the measurements that are independent of the distance ladder are the Megamaser Cosmology Project (MCP), which uses water megamasers to measure $H_0 = 73.9\pm3.0$ \Hunit \citep{Pesce:2020}, gravitational wave standard sirens with $H_0 = 70.0^{+12.0}_{-8.0}$ \Hunit \citep{ligo_h0_gw} and the TDCOSMO collaboration\footnote{\url{http://tdcosmo.org}} (formed by members of H0LiCOW, STRIDES, COSMOGRAIL and SHARP), using time-delay cosmography with lensed quasars \citep{Wong:2020, Shajib0408, Millon:2020}.
Time-delay cosmography \citep{Refsdal:1964} provides a one-step inference of absolute distances on cosmological scales -- and thus the Hubble constant. Over the past two decades, extensive and dedicated efforts have transformed time-delay cosmography from a theoretical idea to a contender for precision cosmology
 \citep{Vanderriest:1989, Keeton:1997, Schechter:1997, Kochanek:2003, Koopmans:2003, Saha:2006, Read:2007, Oguri:2007, Coles:2008, Vuissoz:2008, Suyu:2010, Fedely:2010, Suyu:2013, Suyu:2014, SerenoParaficz:2014, RathnaKumar:2015, Birrer:2016, Wong:2017, Birrer:2019, Rusu:2020, Chen:2019, Shajib0408}.

The keys to precision time-delay cosmography are: Firstly, precise and accurate measurements of relative arrival time delays of multiple images; Secondly, understanding of the large-scale distortion of the angular diameter distances along the line of sight; and thirdly, accurate model of the mass distribution within the main deflector galaxy.
The first problem has been solved by high cadence and high precision photometric monitoring, often with dedicated telescopes \citep[e.g.,][]{Fassnacht:2002,Tewes:2013,Courbin:2018}. The time delay measurement procedure has been validated via simulations by the Time Delay Challenge (TDC1) \citep{Dobler:2015, Liao:2015}.
The second issue has been addressed by statistically correcting the effect of the line of sights to strong gravitational lenses by comparison with cosmological numerical simulations \citep[e.g.,][]{Fassnacht:2011,Suyu:2013,Greene:2013,Collett:2013}. \citet{Millon:2020} recently showed that residuals from the line of sight correction based on this methodology are smaller than the current overall errors.
Progress on the third issue has been achieved by analyzing high quality images of the host galaxy of the lensed quasars with provided spatially resolved information that can be used to constrain lens models \citep[e.g.,][]{Suyu:2009}. By modeling extended sources with complex and flexible source surface brightness instead of just the quasar images positions and fluxes, modelers have been able to move from extremely simplified models like singular isothermal ellipsoids \citep{Kormann:1994, Schechter:1997} to more flexible ones like power laws or stars plus standard dark matter halos \citep[][hereafter NFW]{NFW}.
The choice of elliptical power-law and stars plus NFW profiles was motivated by their generally good description of stellar kinematics and X-ray data in the local Universe. It was validated post-facto by the small residual corrections found via pixellated models \citep{Suyu:2009}, and by the overall goodness of fit they provided to the data.

Building on the advances in the past two decades, the H0LiCOW and SHARP collaborations analyzed six individual lenses \citep{Suyu:2010, Suyu:2014, Wong:2017, Birrer:2019, Rusu:2020, Chen:2019} and measured $H_0$ for each lens to a precision in the range 4.3-9.1\%.
The STRIDES collaboration measured $H_0$ to 3.9\% from one single quadruply lensed quasar \citep{Shajib0408}. The seven measurements follow an approximately standard (although evolving over time) procedure \citep[see e.g.][]{Suyu:2017} and incorporate single-aperture stellar kinematics measurements for each lens.
The H0LiCOW collaboration combined their six quasar lenses, of which five had their analysis blinded, assuming uncorrelated individual distance posteriors and arrived at $H_0 = 73.3_{-1.8}^{+1.7}$ \Hunit, a $2.4\%$ measurement of $H_0$ \citep{Wong:2020}. Adding the blind measurement by \cite{Shajib0408} further increases the precision to $\sim$ 2\% \citep{Millon:2020}.

Given the importance of the Hubble tension, it is crucial, however, to continue to investigate potential causes of systematic errors in time-delay cosmography. After all, extraordinary claims, like physics beyond $\Lambda$CDM, require extraordinary evidence.

The first and main source of residual modeling error in time-delay cosmography is due to the mass-sheet transform (MST) \citep{Falco:1985}. MST is a mathematical degeneracy that leaves the lensing observables unchanged, while rescaling the absolute time delay, and thus the inferred $H_0$. This degeneracy is well known and frequently discussed in the literature \citep[e.g.,][]{ Gorenstein:1988, Kochanek:2002, SahaWilliams:2006, Kochanek:2006saasfee, Read:2007, SchneiderSluse:2013, SchneiderSluse:2014, Coles:2014, Xu:2016, Birrer:2016, Unruh:2017, Sonnenfeld:2018, Wertz:2018, Kochanek:2020, Blum:2020}. Lensing-independent tracers of the gravitational potential of the deflector galaxy, such as stellar kinematics, can break this inherent degeneracy \citep[e.g.,][]{GorinNarayan:1996, RomanowskyKochanek:1999, TreuKoopmans:2002}.
Another way to break the degeneracy is to make assumptions on the mass density profile, which is primarily the strategy adopted by the H0LiCOW/STRIDES collaboration \citep{Millon:2020}. \citet{Millon:2020} showed that the two classes of radial mass profiles considered by the collaboration, power-law and stars and a Navarro Frenk \& White \citep[NFW, ][]{NFW} dark matter halo, yield consistent results\footnote{For the NFW profile parameters, priors on the mass-concentration relation were imposed on the individual analyses.}. \cite{Sonnenfeld:2018, Kochanek:2020, Kochanek:2020b} argued that the error budget of individual lenses obtained under the assumptions of power-law or stars + NFW are underestimated and that, given the MST, the typical uncertainty of the kinematic data does not allow one to constrain the mass profiles to a few percent precision\footnote{See \cite{Birrer:2016} for an analysis explicitly constraining the MST with kinematic data that satisfies the error budget of \cite{Kochanek:2020}.}.

A second potential source of uncertainty in the combined TDCOSMO analysis is the assumption of no correlation between the errors of each individual lens system. The TDCOSMO analysis shows that the scatter between systems is consistent with the estimated errors, and the random measurement errors of the observables are indeed uncorrelated \citep{Wong:2020,Millon:2020}.
However, correlations could be introduced by the modeling procedure and assumptions made, such as the form and prior on the mass profile and the distribution of stellar anisotropies in elliptical galaxies.

In this paper we address these two dominant sources of potential residual uncertainties by introducing a Bayesian hierarchical framework to analyze and interpret the data. Addressing these uncertainties is a major step forward in the field, however it should be noted that the scope of this framework is broader than just these two issues. Its longer term goal is to take advantage of the expanding quality and quantity of data to trade theoretical assumptions for empirical constraints. Specifically, this framework is designed to meet the following requirements: (1) Theoretical assumptions should be explicit and, whenever possible, verified by data or replaced by empirical constraints; (2) Kinematic assumptions and priors must be justified by the data or the laws of physics; (3) The methodology must be validated with realistic simulations. By using this framework we present an updated measurement of the Hubble constant from time-delay cosmography and we lay out a roadmap for further improvements of the methodology to enable a measurement of the Hubble constant from strong lensing time-delay measurements with 1\% precision and accuracy.

In practice, we adopt a parameterization that allows us to quantify the full extent MST in our analysis, addressing point (1) listed above. We discuss the assumptions on the kinematic modeling and the impact of the priors chosen. We deliberately choose an uninformative prior, addressing point (2). We make use of a blind submission to the time-delay lens modeling challenge (TDLMC) \citep{Ding_tdlmc2018, Ding:2020} and validate our approach end to end, including imaging analysis, kinematics analysis and MST mitigation, addressing point (3)
\footnote{Noting however the caveats on the realism of the TDLMC simulations discussed by \citet{Ding:2020}.}.

In our new analysis scheme, the MST is exclusively constrained by the kinematic information of the deflector galaxies, and thus fully accounted for in the error budget. Under these minimal assumptions, we expect that the data currently available for the individual lenses in our TDCOSMO sample will not constrain $H_0$ to the 2\% level.
In addition, we take into account covariances between the sample galaxies, by formulating the priors on the stellar anisotropy distribution and the MST at the population level and globally sampling and marginalizing over their uncertainties.

To further improve the constraints on the mass profile and the MST on the population level, we incorporate a sample of 33 lenses from the Sloan Lens ACS (SLACS) survey \citep{Bolton:2006} into our analysis. We make use of the lens model inference results presented by \cite{Shajib_slacs:2020}, which follow the standards of the TDCOSMO collaboration.
We assess the assumptions in the kinematics modeling and incorporate integral field unit (IFU) spectroscopy from VIMOS 2D data of a subset of the SLACS lenses from \cite{Czoske:2012} in our analysis. This dataset allows us to improve constraints on the stellar anisotropy distribution in massive elliptical galaxies at the population level and thus reduces uncertainties in the interpretation of the kinematic measurements, hence improving the constraints on the MST and $H_0$.
Our joint hierarchical analysis is based on the assumption that the massive elliptical galaxies acting as lenses in the SLACS and the TDCOSMO sample represent the same underlying parent population in regard of their mass profiles and kinematic properties.
The final $H_0$ value derived in this work is inferred from the joint hierarchical analysis of the SLACS and TDCOSMO samples.

The paper is structured as follows: Section \ref{sec:individual_lens} revisits the analysis performed on individual lenses and assesses potential systematics due to MST and mass profile assumptions. Section \ref{sec:hierarchy} describes the hierarchical Bayesian analysis framework to mitigate assumptions and priors associated to the MST to a sample of lenses.
We first validate this approach in Section \ref{sec:tdlmc} on the TDLMC data set \citep{Ding_tdlmc2018} and then move to perform this very same analysis on the TDCOSMO data set in Section \ref{sec:tdcosmo_analysis}. Next, we perform our hierarchical analysis on the SLACS sample with imaging and kinematics data to further constrain uncertainties in the mass profiles and the kinematic behavior of the stellar anisotropy in Section \ref{sec:slacs_analysis}. We present the joint analysis and final inference on the Hubble constant in Section \ref{sec:joint_analysis}. We discuss the limitations of the current work and lay out the path forward in Section \ref{sec:discussion} and finally conclude in Section \ref{sec:conclusion}.

All the software used in this analysis is open source and we share the analysis scripts and pipeline with the community  \faGithub\href{https://github.com/TDCOSMO/hierarchy_analysis_2020_public}{~here}\footnote{\url{https://github.com/TDCOSMO/hierarchy_analysis_2020_public/}}. Numerical tests on the impact of the MST are performed with \textsc{lenstronomy}\footnote{ \, \url{https://github.com/sibirrer/lenstronomy}} \citep{Birrer_lenstronomy, Birrer:2015}. The kinematics is modeled with the \textsc{lenstronomy.Galkin} module. The reanalysis of the SLACS lenses imaging data is performed with \textsc{dolphin}\footnote{  \, \url{https://github.com/ajshajib/dolphin}}, a wrapper around \textsc{lenstronomy} for automated lens modeling \citep{Shajib_slacs:2020} and we introduce \textsc{hierArc}\footnote{ \, \url{https://github.com/sibirrer/hierarc}} (this work) for the hierarchical sampling in conjunction with \textsc{lenstronomy}. All components of the analysis - including analysis scripts and software - were reviewed internally by people not previously involved in the analysis of the sample before the joint inference was performed. All uncertainties stated are given in 16th, 50th and 84th percentiles. Error contours in plots represent 68th and 95th credible regions.

As in previous work by our team - in order to avoid experimenter bias - we keep our analysis blind by using previously blinded analysis products, and all additional choices made in this analysis, such as considering model parameterization and including or excluding of data, are assessed blindly in regard to $H_0$ or parameters directly related to it. All sections, except Section \ref{sec:tension_status}, of this paper have been written and frozen before the unblinding of the results.

\section{Cosmography from individual lenses and the mass-sheet degeneracy}\label{sec:individual_lens}

In this section we review the principles of time-delay cosmography and the underlying observables (Section \ref{sec:td_cosmo} for lensing and time delays and Section \ref{sec:vel_disp} for the kinematic observables). We emphasize how an MST affects the observables and thus the inference of cosmographic quantities (Section \ref{sec:mst}).
We separate the physical origin of the MST into the line-of-sight (external MST, Section \ref{sec:los}) and mass-profile contributions (internal MST, Section \ref{sec:external_vs_internal}) and then provide the limits on the internal mass profile constraints from imaging data and plausibility arguments in Section \ref{sec:approx_mst}. We provide concluding remarks on the constraining power of individual lenses for time-delay cosmography in Section \ref{sec:conclusion_mst}.

\subsection{Cosmography with strong lenses} \label{sec:td_cosmo}
In this section we state the relevant governing physical principles and observables in terms of imaging, time delays, and stellar kinematics.
The phenomena of gravitational lensing can be described by the lens equation, which maps the source plane $\boldsymbol{\beta}$ to the image plane $\boldsymbol{\theta}$ (2D vectors on the plane of the sky)
\begin{equation} \label{eqn:lens_equation}
  \boldsymbol{\beta} = \boldsymbol{\theta} - \boldsymbol{\alpha}(\boldsymbol{\theta}),
\end{equation}
where $\boldsymbol{\alpha}$ is the angular shift on the sky between the original unlensed and the lensed observed position of an object.

For a single lensing plane, the lens equation can be expressed in terms of the physical deflection angle $\hat{\boldsymbol{\alpha}}$ as
\begin{equation} \label{eqn:lens_equation_single_plane}
  \boldsymbol{\beta} = \boldsymbol{\theta} - \frac{D_{\rm s}}{D_{\rm ds}}\hat{\boldsymbol{\alpha}}(\boldsymbol{\theta}),
\end{equation}
with $D_{\rm s}$, $D_{\rm ds}$ is the angular diameter distance from the observer to the source and from the deflector to the source, respectively.
In the single lens plane regime we can introduce the lensing potential $\psi$ such that
\begin{equation}
    \boldsymbol{\alpha}(\boldsymbol{\theta}) = \nabla \psi(\boldsymbol{\theta})
\end{equation}
and the lensing convergence as
\begin{equation}
    \kappa(\boldsymbol{\theta}) =  \frac{1}{2}\nabla^2 \psi(\boldsymbol{\theta}).
\end{equation}
The relative arrival time between two images $\boldsymbol{\theta}_{\rm A}$ and $\boldsymbol{\theta}_{\rm B}$, $\Delta t_{\rm AB}$, originated from the same source is
\begin{equation}\label{eqn:time_delay}
    \Delta t_{\rm AB} = \frac{D_{\Delta t}}{c} \left(\phi(\boldsymbol{\theta}_{\rm A}, \boldsymbol{\beta}) - \phi(\boldsymbol{\theta}_{\rm B}, \boldsymbol{\beta}) \right),
\end{equation}
where $c$ is the speed of light,
\begin{equation}\label{eqn:fermat_potential}
    \phi(\boldsymbol{\theta}, \boldsymbol{\beta}) = \left[ \frac{\left(\boldsymbol{\theta} - \boldsymbol{\beta} \right)^2}{2} - \psi(\boldsymbol{\theta})\right]
\end{equation}
is the Fermat potential \citep{Schneider:1985, Blanford:1986}, and
\begin{equation} \label{eqn:ddt_definition}
    D_{\Delta t} \equiv \left(1 + z_{\rm d}\right) \frac{D_{\rm d}D_{\rm s}}{D_{\rm ds}},
\end{equation}
is the time-delay distance \citep{Refsdal:1964, Schneider:1992, Suyu:2010}; $D_{\rm d}$, $D_{\rm s}$, and $D_{\rm ds}$ are the angular diameter distances from the observer to the deflector, the observer to the source, and from the deflector to the source, respectively.

Provided constraints on the lensing potential, a measured time delay allows us to constrain the time-delay distance $D_{\Delta t}$ from Equation \ref{eqn:time_delay}:
\begin{equation} \label{eqn:ang_dist_delay}
    D_{\Delta t} = \frac{c\Delta t_{\rm AB}}{\Delta\phi_{\rm AB}}.
\end{equation}
The Hubble constant is inversely proportional to the absolute scales of the Universe and thus scales with $D_{\Delta t}$ as
\begin{equation} \label{eqn:H0_ddt}
	H_0 \propto D_{\Delta t}^{-1}.
\end{equation}

\subsection{Deflector velocity dispersion}\label{sec:vel_disp}
The line-of-sight projected stellar velocity dispersion of the deflector galaxy, $\sigma^{\rm P}$, can provide a dynamical mass estimate of the deflector independent of the lensing observables and joint lensing and dynamical mass estimates have been used to constrain galaxy mass profiles \citep{GorinNarayan:1996, RomanowskyKochanek:1999, TreuKoopmans:2002}.

The modeling of the kinematic observables in lensing galaxies range in complexity from spherical Jeans modeling to Schwarzschild \citep{Schwarzschild:1979} methods. For example, \cite{Barnabe:2007, Barnabe:2009} use axisymmetric modeling of the phase-space distribution function with a two-integral Schwarzschild method by \cite{Cretton:1999, VerolmeZeeuw:2002}.
In this work, the kinematics and their interpretation are a key component of the inference scheme and thus we provide the reader with a detailed background and the specific assumptions in the modeling we apply.

The dynamics of stars with the density distribution $\rho_*(r)$ in a gravitational potential $\Phi(r)$ follows the Jeans equation.
In this work, we assume spherical symmetry and no rotation in the Jeans modeling.
In the limit of a relaxed (vanishing time derivatives) and spherically symmetric system, with the only distinction between radial, $\sigma_{\rm r}^2$, and tangential, $\sigma_{\rm t}^2$, dispersions, the Jeans equation results in \citep[e.g.,][]{BinneyTremain:2008}

\begin{equation} \label{eqn:jeans}
  \frac{\partial ( \rho_*\sigma_{\rm r}^2(r))}{\partial r} + \frac{2 \beta_{\text{ani}}(r) \rho_*(r)\sigma_{\rm r}^2(r)}{r} = - \rho_*(r)\frac{\partial \Phi(r)}{\partial r},
\end{equation}
with the stellar anisotropy parameterized as
\begin{equation} \label{eqn:anisotropy_definition}
  \beta_{\rm ani}(r) \equiv 1 -  \frac{\sigma_{\rm t}^2(r)}{\sigma_{\rm r}^2(r)}.
\end{equation}
The solution of Equation \ref{eqn:jeans} can be formally expressed as \citep[e.g.,][]{vanderMarel:1994}
\begin{equation}
 \sigma_{\rm r}^2 = \frac{G}{\rho_*(r)} \int_{r}^{\infty} \frac{M(s)\rho_*(s)}{s^2} J_{\beta}(r, s) ds
\end{equation}
where $M(r)$ is the mass enclosed in a three-dimensional sphere with radius $r$ and
\begin{equation}
  J_{\beta}(r, s) = {\rm exp} \left[ \int_r^s 2\beta(r')dr'/r' \right]
\end{equation}
is the integration factor of the Jeans Equation (Eqn. \ref{eqn:jeans}).
The modeled luminosity-weighted projected velocity dispersion $\sigma_{\rm s}$ is given by \citep{BinneyMamon:1982}
\begin{equation} \label{eqn:I_R_sigma2}
  \Sigma_*(R) \sigma_{\rm s}^2 = 2\int_R^{\infty} \left(1-\beta_{\text{ani}}(r)\frac{R^2}{r^2}\right) \frac{\rho_* \sigma_{\rm r}^2 r \mathrm{d}r}{\sqrt{r^2-R^2}},
\end{equation}
where $R$ is the projected radius and $\Sigma_*(R)$ is the projected stellar density
\begin{equation}
  \Sigma_*(R) = 2\int_R^{\infty} \frac{\rho_*(r) r dr}{\sqrt{r^2-R^2}}.
\end{equation}

The observational conditions have to be taken into account when comparing a model prediction with a data set. In particular, the aperture $\mathcal{A}$ and the PSF convolution of the seeing, $\mathcal{P}$, need to be folded in the modeling. The luminosity-weighted line of sight velocity dispersion within an aperture, $\mathcal{A}$, is then \citep[e.g.,][]{TreuKoopmas:2004, Suyu:2010}
\begin{equation} \label{eqn:sigma_convolved}
  (\sigma^\text{P})^2 = \frac{\int_{\mathcal{A}}\left[\Sigma_*(R) \sigma_{\rm s}^2 * \mathcal{P} \right]\mathrm{d}A }{\int_{\mathcal{A}}\left[\Sigma_*(R) * \mathcal{P} \right]\mathrm{d}A},
\end{equation}
where $\Sigma_*(R) \sigma_{\rm s}^2$ is taken from Equation~\ref{eqn:I_R_sigma2}.

The prediction of the stellar kinematics requires a three-dimensional stellar density $\rho_*(r)$ and mass $M(r)$ profile.
In terms of imaging data, we can extract information about the parameters of the lens mass surface density with parameters $\boldsymbol{\xi}_{\rm mass}$ and the surface brightness of the deflector with parameters $\boldsymbol{\xi}_{\rm light}$.
When assuming a constant mass-to-light ratio across the galaxy, the integrals in the Jeans equation can be performed on the light distribution and $\Sigma_*(R)$ can be taken to be the surface brightness $I(R)$.
To evaluate the three-dimensional distributions, we rely on assumptions on the de-projection to the three-dimensional mass and light components. In this work, we use spherically symmetric models with analytical projections/de-projections to solve the Jeans equation.

An additional ingredient in the calculation of the velocity dispersion is the anisotropy distribution of the stellar orbits, $\beta_{\rm ani}(r)$.
It is impossible to disentangle the anisotropy in the velocity distribution and the gravitational potential from velocity dispersion and rotation measurements alone. This is known as the mass-anisotropy degeneracy \citep{BinneyMamon:1982}.

Finally, the predicted velocity dispersion requires angular diameter distances from a background cosmology. Specifically, the prediction of any $\sigma^\text{P}$ from any model can be decomposed into a cosmological-dependent and cosmology-independent part, as \citep{Birrer:2016, Birrer:2019}

\begin{equation} \label{eqn:kinematics_cosmography}
    (\sigma^{\rm P})^2 =
    \frac{D_{\rm s}}{D_{\rm ds}} c^2 J(\boldsymbol{\xi}_{\rm mass}, \boldsymbol{\xi}_{\rm light}, \beta_{\rm ani}),
\end{equation}
where $J(\boldsymbol{\xi}_{\rm mass}, \boldsymbol{\xi}_{\rm light}, \beta_{\rm ani})$ is the dimensionless and cosmology-independent term of the Jeans equation only relying on the angular units in the light, mass and anisotropy model.
The term $\xi_{\rm light}$ in Equation (\ref{eqn:kinematics_cosmography}) includes the deflector light contribution. The deflector light is required for the Jeans modeling ($\Sigma_*$ and deconvolved $\rho_*$ terms in the equations above). In practice, the inference of the deflector light profile is jointly fit with other light components, such as source light and quasar flux.

Inverting Equation \ref{eqn:kinematics_cosmography} illustrates that a measured velocity dispersion, $\sigma^{\rm P}$, allows us to constrain the distance ratio $D_{\rm s}/D_{\rm ds}$, independent of the cosmological model and time delays but while relying on the same lens model, $\boldsymbol{\xi}_{\text{lens}}$,
\begin{equation} \label{eqn:ang_dist_kin}
  \frac{D_{\text{s}}}{D_{\text{ds}}} = \frac{(\sigma^\text{P})^2}{c^2 J(\boldsymbol{\xi}_{\text{lens}}, \boldsymbol{\xi}_{\text{light}}, \beta_{\text{ani}})}.
\end{equation}
We note that the distance ratio $D_{\rm s}/D_{\rm ds}$ can be constrained without time delays being available. If one has kinematic and time-delay data, instead of expressing constraints on $D_{\rm s}/D_{\rm ds}$, one can also express the cosmologically independent constraints in terms of $D_{\rm d}$ \citep[e.g.,][]{Paraficz:2009, Jee:2015, Birrer:2019} as
\begin{equation} \label{eqn:D_d}
  D_{\text{d}} = \frac{1}{(1 + z_{\text{d}})}\frac{c\Delta t_{\rm AB}}{\Delta\phi_{\rm AB}(\boldsymbol{\xi}_{\text{lens}})}
  \frac{c^2 J(\boldsymbol{\xi}_{\text{lens}}, \boldsymbol{\xi}_{\text{light}}, \beta_{\text{ani}})}{(\sigma^\text{P})^2}.
\end{equation}
In this work, we do not transform the kinematics constraints into $D_{\rm s}/D_{\rm ds}$ or $D_{\rm d}$ constraints but work directly on the likelihood level of the velocity dispersion when discriminating between different cosmological models.

In Appendix \ref{app:anisotropy} we illustrate the radial dependence on the model predicted velocity dispersion, $\sigma^\text{P}$, for different stellar anisotropy models. Observations at different projected radii can partially break the mass-anisotropy degeneracy provided that we have independent mass profile estimates from lensing observables.

\subsection{Mass-sheet transform} \label{sec:mst}
The mass-sheet transform (MST) is a multiplicative transform of the lens Equation (Eqn. \ref{eqn:lens_equation}) \citep{Falco:1985}

\begin{equation} \label{eqn:lens_equation_MST}
  \lambda \boldsymbol{\beta} = \boldsymbol{\theta} - \lambda \boldsymbol{\alpha}(\boldsymbol{\theta}) - (1 - \lambda) \boldsymbol{\theta},
\end{equation}
which preserves image positions (and any higher order relative differentials of the lens equation) under a linear source displacement $\boldsymbol{\beta} \rightarrow \lambda\boldsymbol{\beta}$.
The term $(1 - \lambda) \boldsymbol{\theta}$ in Equation \ref{eqn:lens_equation_MST} above describes an infinite sheet of convergence (or mass), and hence the name mass-sheet transform. Only observables related to the absolute source size, intrinsic magnification or to the lensing potential are able to break this degeneracy.

The convergence field transforms according to
\begin{equation}\label{eqn:mst}
    \kappa_{\lambda}(\theta) = \lambda \times \kappa(\theta) + \left( 1 - \lambda\right).
\end{equation}
The same relative lensing observables can result if the mass profile is scaled by the factor $\lambda$ with the addition of a sheet of convergence (or mass) of $\kappa(\boldsymbol{\theta}) = (1-\lambda)$.

The different observables described in Section \ref{sec:td_cosmo} \& \ref{sec:vel_disp} transform by an MST term $\lambda$ as follow:
The image positions remain invariant
\begin{equation}
    \boldsymbol{\theta}_{\lambda} = \boldsymbol{\theta}.
\end{equation}
The source position scales with $\lambda$
\begin{equation}
    \boldsymbol{\beta}_{\lambda} = \lambda \boldsymbol{\beta}.
\end{equation}
The time delay scales with $\lambda$
\begin{equation}\label{eqn:time_delay_mst}
    \Delta t_{\rm AB \, \lambda} =  \lambda \Delta t_{\rm AB}
\end{equation}
and the velocity dispersion scales with $\lambda$ as
\begin{equation} \label{eqn:kinematics_mst}
    \sigma_{v \, \lambda}^{\text{P}} = \sqrt{\lambda} \sigma_{v}^{\text{P}}.
\end{equation}

Until now we have only stated how the MST impacts observables directly. However, it is also useful to describe how cosmographic constraints derived from a set of observables and assumptions on the mass profile are transformed when transforming the lens model with an MST (Eqn. \ref{eqn:ang_dist_delay}, \ref{eqn:ang_dist_kin}, \ref{eqn:D_d}).
The time-delay distance (Eqn. \ref{eqn:ddt_definition}) is dependent on the time delay $\Delta t$ (Eqn. \ref{eqn:time_delay})
\begin{equation} \label{eqn:ddt_mst}
    D_{\Delta t \, \lambda} = \lambda^{-1}D_{\Delta t}.
\end{equation}
The distance ratio constrained by the kinematics and the lens model scales as
\begin{equation} \label{eqn:ds_dds_mst}
    \left(D_{\rm s}/D_{\rm ds}\right)_{\lambda} = \lambda^{-1} D_{\rm s}/D_{\rm ds}.
\end{equation}
Given time-delay and kinematics data the inference on the angular diameter distance to the lens is invariant under the MST
\begin{equation} \label{eqn:dd_mst}
    D_{\rm d \, \lambda} = D_{\rm d}.
\end{equation}
The Hubble constant, when inferred from the time-delay distance, $D_{\Delta t}$, transforms as (from Eqn. \ref{eqn:H0_ddt})
\begin{equation} \label{eqn:h0_mst}
H_{0 \, \lambda} =  \lambda H_0.
\end{equation}
Mathematically, all the MSTs can be equivalently stated as a change in the angular diameter distance to the source
\begin{equation} \label{eqn:mst_ds}
    D_{\rm s} \rightarrow \lambda D_{\rm s}.
\end{equation}
In other words, if one knows the dependence of any lensing variable upon $D_{\rm s}$ one can transform it under the MST and scale all other quantities in the same way.

\subsection{Line-of-sight contribution} \label{sec:los}
Structure along the line of sight of lenses induce distortions and focusing (or de-focusing) of the light rays. The first-order shear distortions do have an observable imprint on the shape of Einstein rings and can thus be constrained as part of the modeling procedure of strong lensing imaging data. The first order convergence effect alters the angular diameter distances along the specific line of sight of the strong lens.
We define $D^{\rm lens}$ as the specific angular diameter distance along the line of sight of the lens and $D^{\rm bkg}$ as the angular diameter distance from the homogeneous background metric without any perturbative contributions. $D^{\rm lens}$ and $D^{\rm bkg}$ are related through the convergence terms as

\begin{multline}
D^{\rm lens}_{\rm d} = (1 - \kappa_{\rm d})D_{\rm d}^{\rm bkg}\\
D^{\rm lens}_{\rm s} = (1 - \kappa_{\rm s})D_{\rm s}^{\rm bkg}\\
D^{\rm lens}_{\rm ds} = (1 - \kappa_{\rm ds})D_{\rm ds}^{\rm bkg}.\\
\end{multline}
$\kappa_{\rm s}$ is the integrated convergence along the line of sight passing through the strong lens to the source plane and the term $1-\kappa_{\rm s}$ corresponds to an MST (Eqn. \ref{eqn:mst_ds})\footnote{The integral between the deflector and the source deviates from the Born approximation as the light paths are significantly perturbed \cite[see e.g.,][]{Barkana:1996, Birrer:2017}}. To predict the velocity dispersion of the deflector (Eqn. \ref{eqn:kinematics_cosmography}), the terms $\kappa_{\rm s}$ and $\kappa_{\rm ds}$ are relevant when using background metric predictions from a cosmological model ($D^{\rm bkg}$). To predict the time delays (Eqn. \ref{eqn:time_delay}) from a cosmological model, all three terms are relevant. We can define a single effective convergence, $\kappa_{\rm ext}$, that transforms the time-delay distance (Eqn. \ref{eqn:ddt_definition})

\begin{equation} \label{eqn:kappa_ext}
	D_{\Delta t}^{\rm lens} \equiv (1 - \kappa_{\rm ext}) D_{\Delta t}^{\rm bkg}
\end{equation}
with
\begin{equation}
	1 - \kappa_{\rm ext} = \frac{(1 - \kappa_{\rm d})(1 - \kappa_{\rm s})}{(1 - \kappa_{\rm ds})}.
\end{equation}

\subsection{External vs. internal mass sheet transform} \label{sec:external_vs_internal}
An MST (Eqn. \ref{eqn:mst}) is always linked to a specific choice of lens model and so is its physical interpretation.
The MST can be either associated with line-of-sight structure ($\kappa_{\rm s}$) not affiliated with the main deflector or as a transform of the mass profile of the main deflector itself \citep[e.g.,][]{Koopmans:2004, SahaWilliams:2006, SchneiderSluse:2013, Birrer:2016, Shajib0408}.

There are different observables and physical priors related to these two distinct physical causes and we use the notation $\kappa_{\rm s}$ to describe the external convergence aspect of the MST and $\lambda_{\rm int}$ to describe the internal profile aspect of the MST. The total transform which affects the time delays and kinematics (see Eqn. \ref{eqn:time_delay_mst} \& \ref{eqn:kinematics_mst}) is the product of the two transforms

\begin{equation}\label{eqn:mst_split}
    \lambda = (1 - \kappa_{\rm s}) \times \lambda_{\rm int}.
\end{equation}

The line-of-sight contribution can be estimated by tracers of the larger scale structure, either using galaxy number counts \citep[e.g.,][]{Rusu:2017} or weak lensing of distant galaxies by all the mass along the line of sight \citep[e.g.,][]{Tihhonova:2018}, and can be estimated with a few per cent precision per lens.
The internal MST requires either priors on the form of the deflector profile or exquisite kinematic tracers of the gravitational potential. The $\lambda_{\rm int}$ component is the focus of this work.

\subsection{Approximate internal mass-sheet transform} \label{sec:approx_mst}
Imposing the physical boundary condition, $\lim_{r \to \infty} \kappa(r) = 0$, violates the mathematical form of the MST\footnote{We note that the mean cosmological background density is already fully encompassed in the background metric and we effectively only require to model the enhancement matter density \cite[see e.g.,][]{Wucknitz:2008, Birrer:2017}.}. However, approximate MSTs that satisfy the boundary condition of a finite physically enclosed mass may still be possible and encompass the limitations and concerns of strong gravitational lensing in providing precise constraints on the Hubble constant. We specify an approximate MST as a profile
 without significantly impacting imaging observables around the Einstein radius and resulting in the transforms of the time delays (Eqn. \ref{eqn:time_delay_mst}) and kinematics (Eqn. \ref{eqn:kinematics_mst}).

Cored mass components, $\kappa_{\rm c}(r)$, can serve as physically motivated approximations to the MST \citep{Blum:2020}.
We can write a physically motivated approximate internal MST with a parameter $\lambda_{\rm c}$ as

\begin{equation} \label{eqn:approx_mst_core}
    \kappa_{\lambda_{\rm c}}(\boldsymbol{\theta}) = \lambda_{\rm c} \kappa_{\rm model}(\boldsymbol{\theta}) + (1-\lambda_{\rm c})\kappa_{\rm c}(\boldsymbol{\theta}),
\end{equation}
where $\kappa_{\rm model}$ corresponds to the model used in the reconstruction of the imaging data and $\lambda_{\rm c}$ describes the scaling between the cored and the other model components, in resemblance to $\lambda_{\rm int}$.
Approximating a physical cored transform with the pure MST means that:
\begin{equation}\label{eqn:core_profile_approx}
	\lambda_{\rm int} \approx \lambda_{\rm c}
\end{equation}
in deriving all the observable scalings in Section \ref{sec:mst}.

\cite{Blum:2020} showed that several well-chosen cored 3D mass profiles, $\rho(r)$, can lead to approximate MST's in projection, $\kappa_{\rm c}(r)$, with physical interpretations, such as

\begin{equation}\label{eqn:core_profile_3d}
    \rho(r) = \frac{2}{\pi} \Sigma_{\rm crit} \frac{R_{\rm c}^2} {\left(R_{\rm c}^2 + r^2 \right)^{3/2}},
\end{equation}
resulting in the projected convergence profile
\begin{equation}\label{eqn:core_profile_projected}
    \kappa_{\rm c}(\boldsymbol{\theta}) = \frac{R_{\rm c}^2}{R_{\rm c}^2  + \boldsymbol{\theta}^2},
\end{equation}
where $\Sigma_{\rm crit}$ is the critical surface density of the lens. The specific functional form of the profile listed above (\ref{eqn:core_profile_3d}) resemble the outer slope of the NFW profile with $\rho(r) \propto r^{-3}$.

\begin{figure*}
  \centering
  \includegraphics[angle=0, width=160mm]{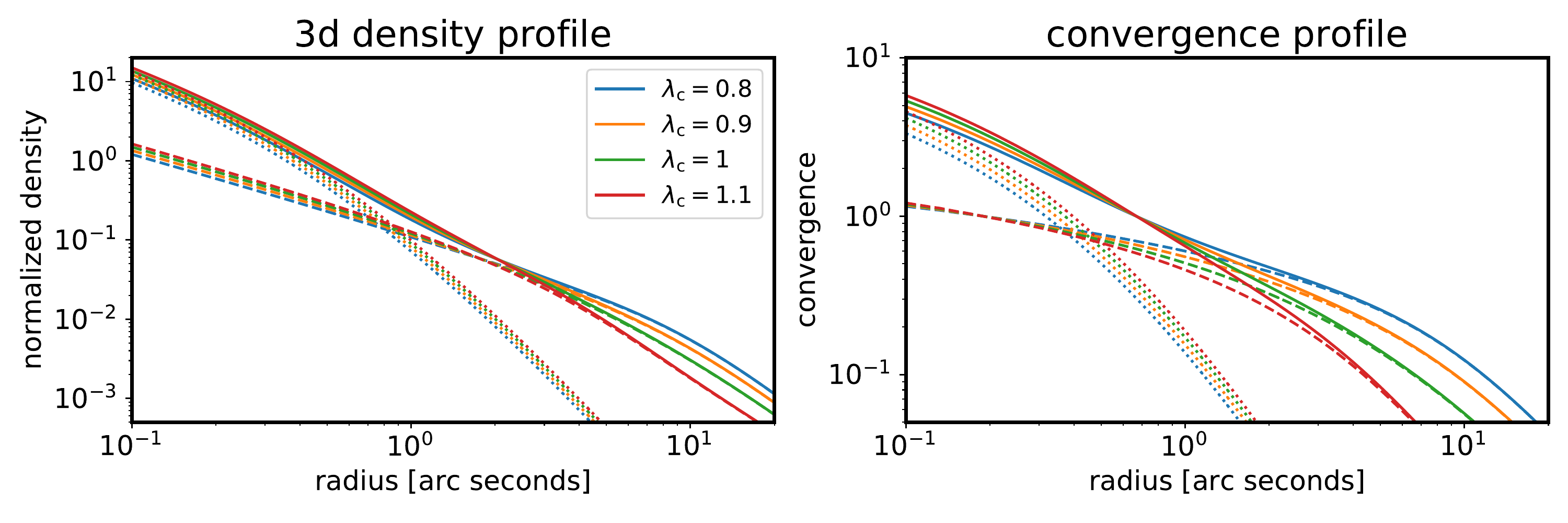}
  \caption{Illustration of a composite profile consisting of a stellar component (Hernquist profile, dotted lines) and a dark matter component (NFW + cored component (Eqn. \ref{eqn:core_profile_projected}), dashed lines) which transform according to an approximate MST (joint as solid lines). The stellar component gets rescaled by the MST while the cored component transforms the dark matter component.
  \textbf{Left:} profile components in three dimensions. \textbf{Right:} profile components in projection. The transforms presented here cannot be distinguished by imaging data alone and require i.e., stellar kinematics constraints.  \faGithub\href{https://github.com/TDCOSMO/hierarchy_analysis_2020_public/blob/6c293af582c398a5c9de60a51cb0c44432a3c598/MST_impact/MST_composite_cored.ipynb}{~source} }
\label{fig:composte_mst_profile}
\end{figure*}

Figure \ref{fig:composte_mst_profile} illustrates a composite profile consisting of a stellar component (Hernquist profile) and a dark matter component (NFW + cored component, Eqn. \ref{eqn:core_profile_3d}) which transform according to an approximate MST. The stellar component gets rescaled by the MST while the cored component is transforming only the dark matter component.

It is of greatest importance to quantify the physical plausibility of those transforms and their impact on other observables in detail.
In this section we extend the study of \cite{Blum:2020}. We perform detailed numerical experiments on mock imaging data to quantify the constraints from imaging data, time delays and kinematics, and we quantify the range of such an approximate transform with physically motivated boundary conditions. Further illustrations and details on the examples given in this section can be found in Appendix \ref{app:mst_pemd}.

\subsubsection{Imaging constraints on the internal MST} \label{sec:imaging_constraints}
In this section we investigate the extent to which imaging data is able to distinguish between different lens models with different cored mass components and their impact on the inferred time delay distance in combination with time delay information. We first generate a mock image and time delays without a cored component and then perform the inference with an additional cored component model (Eqn \ref{eqn:core_profile_projected}) parameterized with the core radius $R_{\rm c}$ and the core projected density $\Sigma_{\rm c} \equiv (1 - \lambda_{\rm c})$ (Eqn. \ref{eqn:approx_mst_core}).
In our specific example, we simulate a quadruply lensed quasar image similar to \cite{Millon:2020} (more details in Appendix \ref{app:mst_pemd} and Fig. \ref{fig:mock_lens}) with a power-law elliptical mass distribution \citep[PEMD,][]{Kormann:1994, Barkana:1998}

\begin{equation} \label{eqn:pl_profile}
  \kappa(\theta_1, \theta_2) = \frac{3-\gamma_{\rm pl}}{2} \left[\frac{\theta_{\rm E}}{\sqrt{q_{\rm m}\theta_1^2 + \theta_2^2/q_{\rm m}}} \right]^{\gamma_{\rm pl}-1}
\end{equation}
where $\gamma_{\rm pl}$ is the logarithmic slope of the profile, $q_{\rm m}$ is the axis ratio of the minor and the major axes of the elliptical profile, and $\theta_{\rm E}$ is the Einstein radius. The coordinate system is defined such that $\theta_1$ and $\theta_2$ are along the major and minor axis respectively. We also add an
external shear model component with distortion amplitude $\gamma_{\rm ext}$ and direction $\phi_{\rm ext}$. The PEMD+shear model is one of two lens models considered in the analysis of the TDCOSMO sample.
For the source and lens galaxies we use elliptical S\'ersic surface brightness profiles. We add a Gaussian Point Spread Function (PSF) with Full-Width-at-Half-Maximum (FWHM) of 0$^{\prime\prime}$.1, pixel scale of 0$^{\prime\prime}$.05 and noise properties consistent with the current TDCOSMO sample of \textit{Hubble Space Telescope} (HST) images. The time delays between the images between the first arriving image and the subsequent images are 11.7,  27.6, and  94.0 days, respectively.
We chose time-delay uncertainties of $\pm 2$ days between the three relative delays. The time-delay precision does not impact our conclusions about the MST.
The inference is performed on the pixel level of the mock image as with the real data on the TDCOSMO sample.

In the modeling and parameter inference, we add an additional cored mass component (Eqn \ref{eqn:core_profile_projected}) and perform the inference on all the lens and source parameters simultaneously, including the core radius $R_{\rm c}$ and the projected core density $\Sigma_{\rm c}$.
In the limit of a perfect MST there is a mathematical degeneracy if we only use the imaging data as constraints. We thus expect a full covariance in the parameters involved in the MST (Einstein radius of the main deflector, source position, source size etc) and the posterior inference of our problem to be inefficient in the regime where the cored profile mimics the full MST ($\kappa_{\rm c}(\boldsymbol{\theta})$ acts as $\Sigma_{\rm crit}$ for $R_{\rm c} \rightarrow \infty$). To improve the sampling, instead of modeling the cored profile $\kappa_{\rm c}(\boldsymbol{\theta})$, we model the difference between the cored component and a perfect MST, $\Delta \kappa_{\rm c} = \kappa_{\rm c}(\boldsymbol{\theta}) - \Sigma_{\rm crit}$, with $\lambda_{\rm c}$ (Eqn. \ref{eqn:approx_mst_core}) instead. $\Delta \kappa_{\rm c}$ is effectively the component of the model that does not transform under the MST and leads to a physical three-dimensional profile interpretation.

\begin{figure*}
  \centering
  \includegraphics[angle=0, width=170mm]{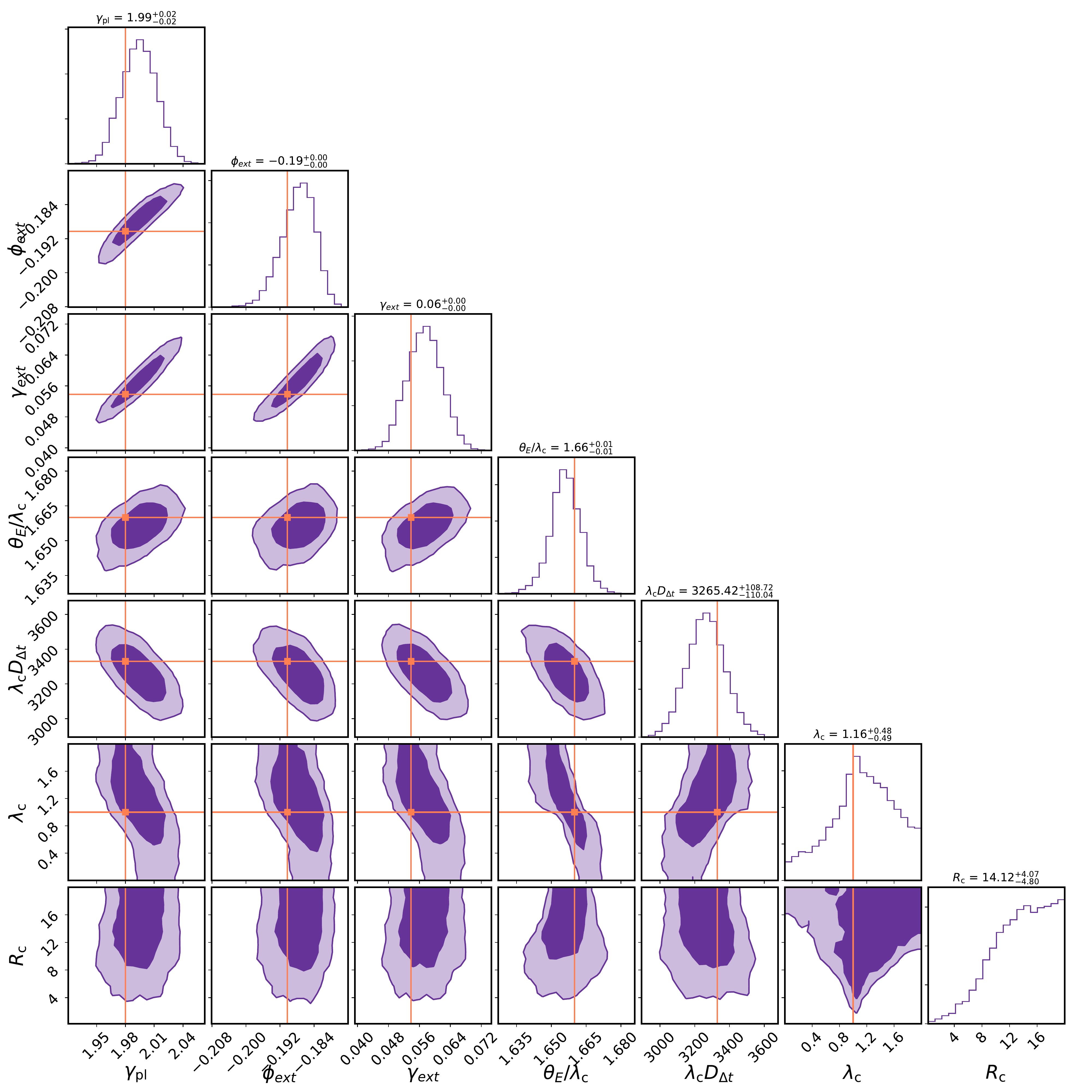}
  \caption{Illustration of the constraining power of imaging data on a cored mass component (Eqn. \ref{eqn:approx_mst_core}). Shown are the parameter inference of the power-law profile mock quadruply lensed quasar of Figure \ref{fig:mock_lens} when including a marginalization of an additional cored power law profile (Eqn. \ref{eqn:core_profile_projected}). Orange lines indicate the input truth of the model without a cored component. $\lambda_{\rm c}$ is the scaled core model parameter (Eqn. \ref{eqn:approx_mst_core}) resembling the pure MST for large core radii ($\lambda_{\rm c} \approx \lambda_{\rm int}$).  \faGithub\href{https://github.com/TDCOSMO/hierarchy_analysis_2020_public/blob/6c293af582c398a5c9de60a51cb0c44432a3c598/MST_impact/MST_pl_cored.ipynb}{~source} }
  \label{fig:posteriors_core}
\end{figure*}

Figure \ref{fig:posteriors_core} shows the inference on the relevant lens model parameters for the mock image described in Appendix \ref{app:mst_pemd}. The input parameters are marked as orange lines for the model without a cored component. We can clearly see that for small core radii, $R_{\rm c}$, the approximate MST parameter $\lambda_{\rm c}$ can be constrained. This is the limit where the additional core profile cannot mimic a pure MST at a level where the data is able to distinguish between them. For core radii $R_{\rm c} = 3 \theta_{\rm E}$, the uncertainty on the approximate MST, $\lambda_{\rm c}$, is 10\%. For core radii $R_{\rm c} > 5 \theta_{\rm E}$, the approximate MST is very close to the pure MST and the imaging information in our example is not able to constrain $\lambda_{\rm c}$ to better than $\lambda_{\rm c} \pm 0.4$. We make use of the expected constraining power on $\lambda_{\rm c}$ as a function of $R_{\rm c}$ when we discuss the plausibility of certain transforms. When looking at the inferred time-delay distance $\lambda_{\rm c} D_{\Delta t}$, we see that this quantity is constant as a function of $R_{\rm c}$ and thus the time-delay prediction is accurately being transformed by a pure MST (Eqn. \ref{eqn:time_delay_mst}).
Overall, we find that $\lambda_{\rm c} \approx \lambda_{\rm int}$ is valid for larger core radii.

Identical tests with a composite profile instead of a PEMD profile result in the same conclusions and are available  \faGithub\href{https://github.com/TDCOSMO/hierarchy_analysis_2020_public/blob/6c293af582c398a5c9de60a51cb0c44432a3c598/MST_impact/MST_composite_cored.ipynb}{~here}.

\subsubsection{Allowed cored mass components from physical boundary conditions} \label{sec:core_boundary_conditions}
In the previous section (\ref{sec:imaging_constraints}) we demonstrated that, for large core radii, there are physical profiles that approximate a pure MST ($\lambda_{\rm c} \approx \lambda_{\rm int}$). In this section we take a closer look at the physical interpretation of such large positive and negative cored component transforms with respect to a chosen mass profile. It is possible that the core model itself does not require a physical interpretation as it is overall included in the total mass distribution. The galaxy surface brightness provides constraints on the stellar mass distribution (modulo a mass-to-light conversion factor) and the focus here is a consideration of the distribution of the invisible (dark) matter component of the deflector. Our starting model is a NFW profile and we assess departures from this model by using a cored component.

We apply the following conservative boundary conditions on the distribution of the dark matter component:
Firstly, the total mass of the cored component within a three-dimensional radius shall not exceed the total mass of the NFW profile within the same volume, $M_{\rm core}(<r) \le M_{\rm NFW}(< r)$. This is not a strict bound, but violating this condition would imply changing the mass of the halo itself.
Secondly, the density profile shall never drop to negative values, $\rho_{\rm NFW+core}(r) \ge 0$.

\begin{figure}
  \centering
  \includegraphics[angle=0, width=80mm]{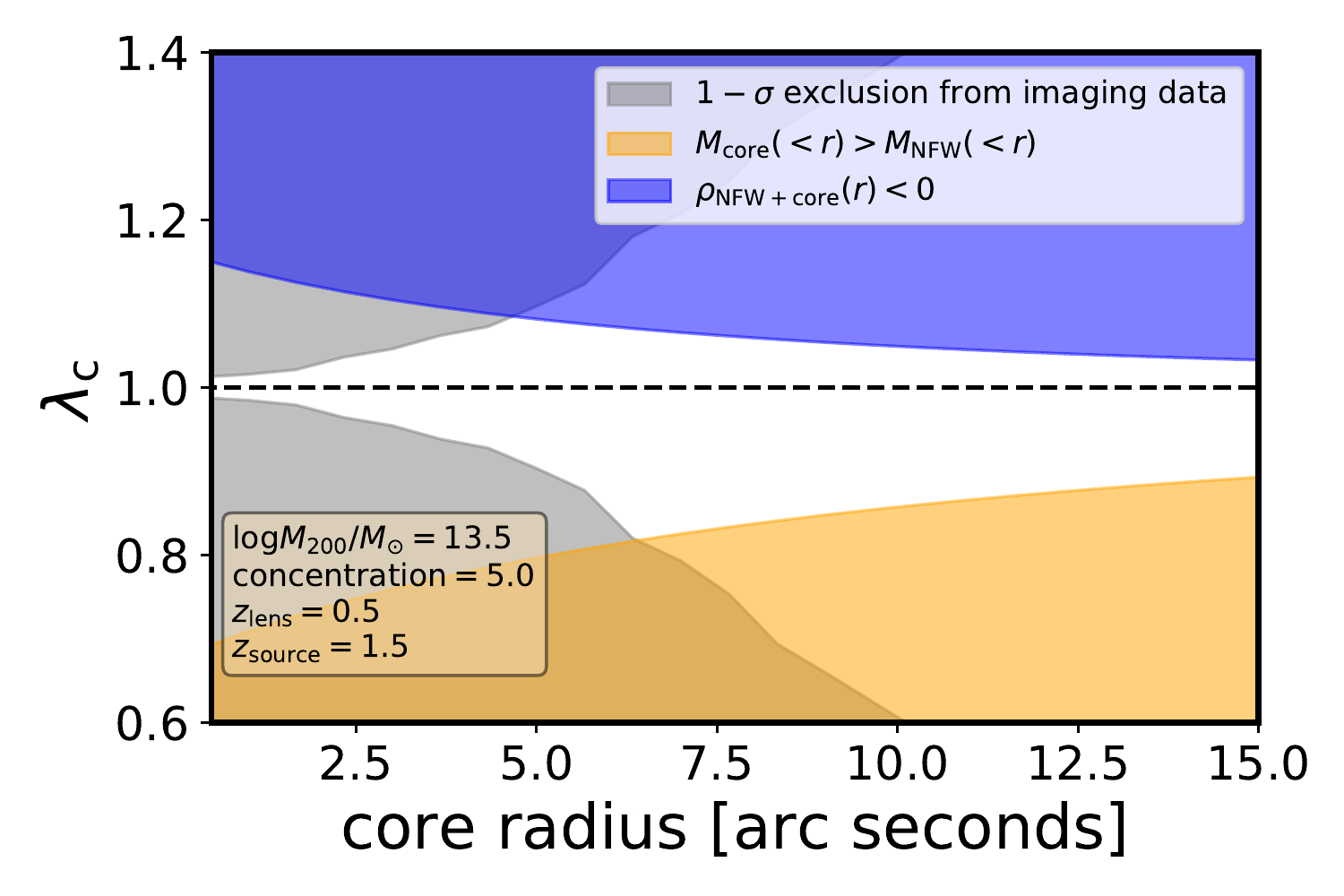}
  \caption{Constraints on an approximate internal MST transform with a cored component, $\lambda_{\rm c}$, of an NFW profile as a function of core radius. In gray are the 1-$\sigma$ exclusion limits that imaging data can provide. In orange is the region where the total mass of the core within a three-dimensional radius exceeds the mass of the NFW profile in the same sphere. In blue is the region where the transformed profile results in negative convergence at the core radius. The white region is effectively allowed by the imaging data and simple plausibility considerations and where we can use the mathematical MST as an approximation ($\lambda_{\rm c} \approx \lambda_{\rm int}$). The halo mass, concentration and the redshift configuration is displayed in the lower left box.  \faGithub\href{https://github.com/TDCOSMO/hierarchy_analysis_2020_public/blob/6c293af582c398a5c9de60a51cb0c44432a3c598/MST_impact/MST_pl_cored.ipynb}{~source} }
\label{fig:lambda_int_bounds}
\end{figure}

Those two imposed conditions define a physical interpretation of a three-dimensional mass profile as being a redistribution of matter from the dark matter component and a rescaling of the mass-to-light ratio of the luminous component. An independent estimate of the mass-to-light ratio of few per cent is below our current limits of knowledge about the stellar initial mass function, stellar evolution models and dust extinction. Moreover, the mass-to-light ratio can vary with radius.
Figure \ref{fig:lambda_int_bounds} provides the constraints from the two conditions, as well as from the imaging data constraints of Section \ref{sec:imaging_constraints}, for an expected NFW mass and concentration profile at a typical lens and source redshift configuration. The remaining white region in Figure \ref{fig:lambda_int_bounds} is effectively allowed by the imaging data and simple plausibility considerations.
We conclude that the physically allowed parameter space does encompass a pure MST with $\lambda_{\rm int} = 1^{+0.07}_{-0.15}$, with more parameter volume for $\lambda_{\rm int} < 1$, which corresponds to a positive cored component.
We emphasize that the constraining power at small core radii may be due to the angular rather than the radial imprint of the cored profile \citep[see e.g.,][]{Kochanek:2020b}. However, such a behavior would not alter our conclusions and inference method chosen in the analysis presented in subsequent sections of this work. We also performed this inference for a composite (stellar light + NFW dark matter) model and arrive at the same conclusions.

\subsubsection{Stellar kinematics of an approximate MST}
In this section we investigate the kinematics dependence on the approximate MST. To do so, we perform spherical Jeans modeling (Section \ref{sec:vel_disp}) and compute the predicted velocity dispersion in an aperture under realistic seeing conditions (Eqn. \ref{eqn:sigma_convolved}) for models with a cored mass component as an approximation of the MST.

\begin{figure}
  \centering
  \includegraphics[angle=0, width=80mm]{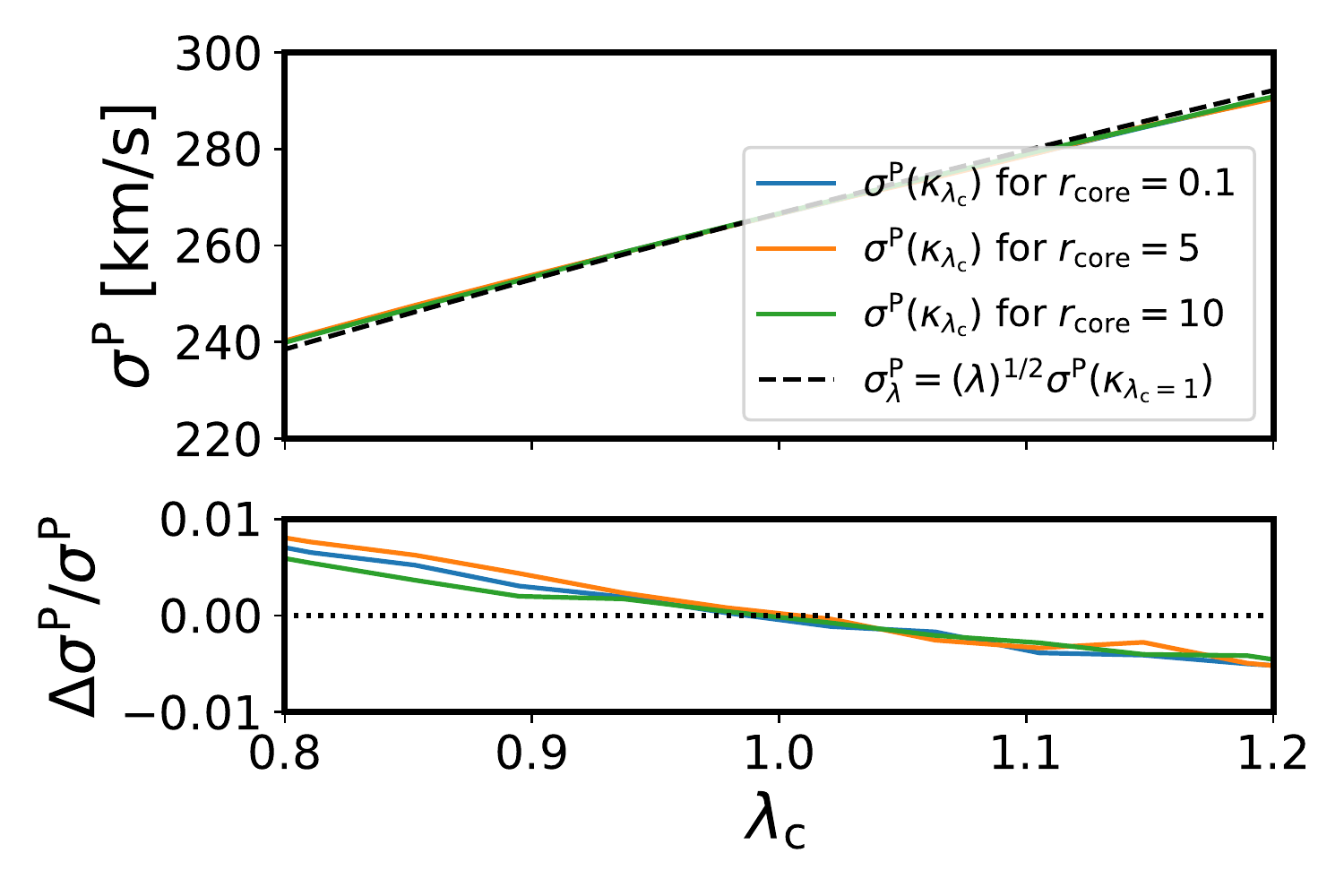}
  \caption{Comparison of the actual predicted kinematics from the modeling of the physical three-dimensional mass distribution $\kappa_{\lambda_{\rm int}}$ (Eqn. \ref{eqn:approx_mst_core}) for varying core sizes (solid) and the analytic relation of a perfect MST (Eqn. \ref{eqn:kinematics_mst}, dashed) for the mock lens presented in Figure \ref{fig:mock_lens}. Lower panel shows the fractional differences between the exact prediction and a perfect MST calculation. The MST prediction matches to <1\% in the considered range. Minor numerical noise is present at the subpercent level.  \faGithub\href{https://github.com/TDCOSMO/hierarchy_analysis_2020_public/blob/6c293af582c398a5c9de60a51cb0c44432a3c598/MST_impact/MST_pl_cored.ipynb}{~source} }
\label{fig:kinematics_mst}
\end{figure}

Figure \ref{fig:kinematics_mst} compares the actual predicted kinematics from the modeling of the physical three-dimensional mass distribution $\kappa_{\lambda_{\rm c}}$ (Eqn. \ref{eqn:approx_mst_core}) and the analytic relation of a perfect MST (Eqn. \ref{eqn:kinematics_mst}) for the mock lens presented in Appendix \ref{app:mst_pemd}. For this figure, we chose an aperture size of 1$^{\prime\prime}\times 1^{\prime\prime}$ and seeing of FWHM = 0$^{\prime\prime}$.7 and an isotropic stellar orbit distribution ($\beta_{\rm ani}(r) = 0$). For $\lambda_{\rm c}$ in the range [0.8, 1.2], the MST approximation in the predicted velocity dispersion is accurate to <1\%.
We conclude that, for the $\lambda_{\rm int}$ range considered in this work, the analytic approximation of a perfect MST is valid to reliably compute the predicted velocity dispersion. The precise dependence of the velocity dispersion only marginally depends on the specific core radius $R_{\rm c}$ and the approximation remains valid for all reasonable and non-excluded core radii and $\lambda_{\rm int}$. We tested that our conclusions also hold for different anisotropy profiles and observational conditions.

\subsection{Constraining power using individual lenses} \label{sec:conclusion_mst}
For each individual strong lens in the TDCOSMO sample, there are four data sets available: (i) imaging data of the strong lensing features and the deflector galaxy, $\mathcal{D}_{\rm img}$; (2) time-delay measurements between the multiple images, $\mathcal{D}_{\rm td}$; (3) stellar kinematics measurement of the main deflector galaxy, $\mathcal{D}_{\rm spec}$; (4) line-of-sight galaxy count and weak lensing statistics, $\mathcal{D}_{\rm los}$.

These data sets are independent and so are their likelihoods in a joint cosmographic inference. Hence, we can write the likelihood of the joint set of the data $\mathcal{D} =\{\mathcal{D}_{\rm img},  \mathcal{D}_{\rm td}, \mathcal{D}_{\rm spec}, \mathcal{D}_{\rm los}\}$ given the cosmographic parameters $\{D_{\rm d}, D_{\rm s}, D_{\rm ds} \} \equiv D_{\rm d, s, ds}$ as
\begin{eqnarray}\label{eqn:single_lens_inference}
    \mathcal{L}(\mathcal{D}| D_{\rm d, s, ds} ) =
     \int \mathcal{L}(\mathcal{D}_{\rm img} | \boldsymbol{\xi}_{\rm mass}, \boldsymbol{\xi}_{\rm light}) \\
     \times \mathcal{L}(\mathcal{D}_{\rm td} | \boldsymbol{\xi}_{\rm mass}, \boldsymbol{\xi}_{\rm light}, \lambda, D_{\Delta t})\\
     \times \mathcal{L}(\mathcal{D}_{\rm spec}| \boldsymbol{\xi}_{\rm mass}, \boldsymbol{\xi}_{\rm light}, \beta_{\rm ani}, \lambda, D_{\rm s}/D_{\rm ds})
      \mathcal{L}(\mathcal{D}_{\rm los}| \kappa_{\rm ext}) \\
      \times p(\boldsymbol{\xi}_{\rm mass}, \boldsymbol{\xi}_{\rm light}, \lambda_{\rm int}, \kappa_{\rm ext}, \beta_{\rm ani}) d\boldsymbol{\xi}_{\rm mass} d\boldsymbol{\xi}_{\rm light} d\lambda_{\rm int} d\kappa_{\rm ext} d\beta_{\rm ani}.
\end{eqnarray}
In the expression above we only included the relevant model components in the expressions of the individual likelihoods. $\boldsymbol{\xi}_{\rm light}$ formally includes the source and lens light surface brightness. For the time-delay likelihood, we only consider the time-variable source position from the set of $\boldsymbol{\xi}_{\rm light}$ parameters.
In Appendix \ref{app:likelihood} we provide details on the computation of the combined likelihood, in particular with application in the hierarchical context.

An approximate internal MST of a power law with $\lambda_{\rm int}$ of 10\% still leads to physically interpretable mass profiles with the Hubble constant changed by 10\% (see Eqn. \ref{eqn:h0_mst}). Imaging data is not sufficiently able to distinguish between models producing $H_0$ value within this 10\% range \citep[][]{Kochanek:2020}.
The kinematics are changed with good approximation by Equation \ref{eqn:kinematics_mst} through this transform. The kinematic prediction is also cosmology dependent by Equation \ref{eqn:kinematics_cosmography}. The scalings of an MST are analytical in the model-predicted time-delay distance and kinematics and thus its marginalization can be performed in post processing given posteriors for a specific lens model family that breaks the MST, such as a power-law model.

The kinematics information is the decisive factor in discriminating different profile families. The relative uncertainty in the velocity dispersion measurement directly propagates into the relative uncertainty in the MST as
\begin{equation} \label{eqn:sigma_v_lambda_error_propagation}
	\frac{\delta \lambda_{\rm int}}{\lambda_{\rm int}} = 2\frac{\delta \sigma^{\rm P}}{\sigma^{\rm P}}.
\end{equation}
The current uncertainties on the velocity dispersion measurements, on the order of 5-10\% (including the uncertainties due to stellar template mismatch and other systematic errors) limit the precise determination of the mass profile per individual lens. Uncertainties in the interpretation of the stellar anisotropy orbit distribution additionally complicates the problem.
\cite{Birrer:2016} performed such an analysis and demonstrated that an explicit treatment of the MST (in their approach parameterized as a source scale) leads to uncertainties consistent with the expectations of \cite{Kochanek:2020}.
Because the kinematic measurement of each lens is not sufficiently precise to constrain the mass profile to the desired level, in this work we marginalize over the uncertainties properly accounting for the priors.

\section{Hierarchical Bayesian cosmography}\label{sec:hierarchy}
The overarching goal of time-delay cosmography is to provide a robust inference of cosmological parameters, $\boldsymbol{\pi}$, and in particular the absolute distance scale, the Hubble constant $H_0$, and possibly other parameters describing the expansion history of the Universe (such as $\Omega_{\Lambda}$ or $\Omega_{\rm m}$), from a sample of gravitational lenses with measured time delays.
Based on the conclusions we draw from Section \ref{sec:individual_lens}, it is absolutely necessary to propagate assumptions and priors made on the analysis of an individual lens hierarchically when performing the inference on the cosmological parameters from a population of lenses.
In particular, this is relevant for parameters that we cannot sufficiently constrain on a lens-by-lens basis and parameters whose uncertainties significantly propagate to the $H_0$ inference on the population level.
In this section, we introduce three specific hierarchical sampling procedures for properties of lensing galaxies and their selection that are relevant for the cosmographic analysis.
In particular, these are: (1) an overall internal MST relative to a chosen mass profile, $\lambda_{\rm int}$, and its distribution among the sample of lenses;
(2) stellar anisotropy distribution in the sample of lenses; (3) the line-of-sight structure selection and distribution of the lens sample.

In Section \ref{sec:hierarchy_approx} we formalize the Bayesian problem and define an approximate scheme for the full hierarchical inference that allows us to keep track of key systematic uncertainties while still being able to reuse currently available inference products. In Section \ref{sec:hyper_param} we specify the hyper-parameters we sample on the population level. Section \ref{sec:approx_likelihood} details the specific approximations in the likelihood calculation.
All hierarchical computations and sampling presented in this work are implemented in the open-source software \textsc{hierArc}.

\subsection{Hierarchical inference problem} \label{sec:hierarchy_approx}
In Bayesian language, we want to calculate the probability of the cosmological parameters, $\boldsymbol{\pi}$, given the strong lensing data set, $p(\boldsymbol{\pi} | \{\mathcal{D}_{i} \}_{N})$, where $\mathcal{D}_i$ is the data set of an individual lens (including imaging data, time-delay measurements, kinematic observations and line-of-sight galaxy properties) and $N$ the total number of lenses in the sample.

In addition to $\boldsymbol{\pi}$, we introduce $\boldsymbol{\xi}$ that incorporates all the model parameters. Using Bayes rule and considering that the data of each individual lens $\mathcal{D}_{i}$ is independent, we can write:

\begin{multline} \label{eqn:full_inference}
    p(\boldsymbol{\pi} | \{\mathcal{D}_{i} \}_{N}) \propto \mathcal{L}(\{\mathcal{D}_{i} \}_{N}| \boldsymbol{\pi}) p(\boldsymbol{\pi})  = \int \mathcal{L}(\{\mathcal{D}_{i} \}_{N}| \boldsymbol{\pi}, \boldsymbol{\xi})p(\boldsymbol{\pi}, \boldsymbol{\xi}) d \boldsymbol{\xi} \\
    = \int \prod_i^N \mathcal{L}(\mathcal{D}_{i}| \boldsymbol{\pi}, \boldsymbol{\xi})p(\boldsymbol{\pi}, \boldsymbol{\xi}) d \boldsymbol{\xi}.
\end{multline}

In the following, we divide the nuisance parameter, $\boldsymbol{\xi}$, into a subset of parameters that we constrain independently per lens, $\boldsymbol{\xi}_i$, and a set of parameters that require to be sampled across the lens sample population globally, $\boldsymbol{\xi}_{\rm pop}$. The parameters of each individual lens, $\boldsymbol{\xi}_i$, include the lens model, source and lens light surface brightness and any other relevant parameter of the model to predict the data.
Hence, we can express the hierarchical inference (Eqn. \ref{eqn:full_inference}) as
\begin{multline} \label{eqn:full_inference_extended}
    p(\boldsymbol{\pi} | \{\mathcal{D}_{i} \}_{N})
     \propto \int \prod_i \left[ \mathcal{L}(\mathcal{D}_i | D_{\rm d, s, ds}(\boldsymbol{\pi}), \boldsymbol{\xi}_i,  \boldsymbol{\xi}_{\rm pop})
     p(\boldsymbol{\xi}_i) \right]
     \\ \times
     \frac{p(\boldsymbol{\pi}, \{\boldsymbol{\xi}_i \}_{N}, \boldsymbol{\xi}_{\rm pop})}{\prod_i p(\boldsymbol{\xi}_i)} d \boldsymbol{\xi}_{\{i\}} d \boldsymbol{\xi}_{\rm pop}
\end{multline}

where $\{\boldsymbol{\xi}_i \}_{N} = \{\boldsymbol{\xi}_1, \boldsymbol{\xi}_2, ..., \boldsymbol{\xi}_N \}$ is the set of the parameters applied to the individual lenses and $p( \boldsymbol{\xi}_i)$ are the interim priors on the model parameters in the inference of an individual lens. The cosmological parameters $\boldsymbol{\pi}$ are fully encompassed in the set of angular diameter distances, $\{D_{\rm d}, D_{\rm s}, D_{\rm ds} \} \equiv D_{\rm d, s, ds}$, and thus, instead of stating $\boldsymbol{\pi}$ in Equation \ref{eqn:full_inference_extended}, we now state $D_{\rm d, s, ds}(\boldsymbol{\pi})$.
Up to this point, no approximation was applied to the full hierarchical expression (Eqn. \ref{eqn:full_inference}).

From now on, we assume
\begin{equation} \label{eqn:approx_separability}
    \frac{p(\boldsymbol{\pi}, \boldsymbol{\xi}_{\{i\}}, \boldsymbol{\xi}_{\rm pop})}{\prod_i p(\boldsymbol{\xi}_i)} \approx p(\boldsymbol{\pi}, \boldsymbol{\xi}_{\rm pop}),
\end{equation}
which states that, for the parameters classified as $\boldsymbol{\xi}_{\{i\}}$, the interim priors do not propagate into the cosmographic inference and the population prior on those parameters is formally known exactly. The population parameters, $\boldsymbol{\xi}_{\rm pop}$, describe a distribution function such that the values of individual lenses, $\boldsymbol{\xi'}_{\rm pop, i}$, follow the distribution likelihood $p(\boldsymbol{\xi'}_{\rm pop, i} | \boldsymbol{\xi}_{\rm pop})$.

With this approximation and the notation of the sample distribution likelihood, we can simplify expression \ref{eqn:full_inference_extended} to

\begin{equation} \label{eqn:population_marginalization}
    p(\boldsymbol{\pi} | \{\mathcal{D}_{i} \}_{N}) \propto
     \int \prod_i \mathcal{L}(\mathcal{D}_i | D_{\rm d, s, ds}, \boldsymbol{\xi}_{\rm pop})  p(\boldsymbol{\pi}, \boldsymbol{\xi}_{\rm pop}) d \boldsymbol{\xi}_{\rm pop}
\end{equation}

where
\begin{multline}\label{eqn:single_lens_posterior}
    \mathcal{L}(\mathcal{D}_i | D_{\rm d, s, ds}, \boldsymbol{\xi}_{\rm pop}) =  \\
    \int \mathcal{L}(\mathcal{D}_i | D_{\rm d, s, ds}, \boldsymbol{\xi'}_{\rm pop, i}) p(\boldsymbol{\xi'}_{\rm pop, i} | \boldsymbol{\xi}_{\rm pop}) d\boldsymbol{\xi'}_{\rm pop, i}
\end{multline}
are the individual likelihoods from an independent sampling of each lens with access to global population parameters, $\boldsymbol{\xi}_{\rm pop}$, and marginalized over the population distribution. The integral in Equation \ref{eqn:single_lens_posterior} goes over all individual parameters where a population distribution $p(\boldsymbol{\xi}'_{\rm pop, i}| \boldsymbol{\xi}_{\rm pop})$ is applied. Equation \ref{eqn:single_lens_inference} is effectively expression \ref{eqn:single_lens_posterior} without the marginalization over parameters assigned as $\boldsymbol{\xi}_{\rm pop}$.

For parameters in the category $\boldsymbol{\xi}_{\{i\}}$, our approximation implies that there is no population prior and that the interim priors do not impact the cosmographic inference. This approximation is valid in the regime where the posterior distribution in $\boldsymbol{\xi}_{\{i\}}$ is effectively independent of the prior. Although formally this is never true, for many parameters in the modeling of high signal-to-noise imaging data the individual lens modeling parameters are very well constrained relative to the prior imposed.

In the following we highlight some key aspects of the cosmographic analysis and in particular the inference on the Hubble constant where the approximation stated in expression \ref{eqn:approx_separability} is not valid and thus fall in the category of $\boldsymbol{\xi}_{\rm pop}$. We give explicit parameterizations of these effects and provide specific expressions to allow for an efficient and sufficiently accurate sampling and marginalization, according to Equation \ref{eqn:single_lens_posterior}, for individual lenses within an ensemble.

\subsection{Lens population hyper-parameters} \label{sec:hyper_param}
In this section we discuss the choices of population level hyper-parameters we include in our analysis.

\subsubsection{Deflector lens model}
The deflectors in the quasar lenses with measured time delays of the TDCOSMO sample are massive elliptical galaxies. These galaxies, observationally, follow a tight relation in a luminosity, size and velocity dispersion parameter space \citep[e.g.,][]{FaberJackson:1976, Auger:2010, Bernardi:2020}, exhibiting a high degree of self-similarity among the population.

In Section \ref{sec:approx_mst} we defined $\lambda_{\rm c}$ as the approximate MST relative to a chosen profile of an individual lens and established the close correspondence to a perfect MST ($\lambda_{\rm c} \approx \lambda_{\rm int}$). For the inference from a sample of lenses, the sample distribution of deflector profiles is the relevant property to quantify.
For the deflector mass profile, we do not want to artificially break the MST based on imaging data and require the kinematics to constrain the mass profile.
To do so, we chose as a base-line model a PEMD (Eqn. \ref{eqn:pl_profile}) to be constrained on the lens-by-lens case and we add a global internal MST specified on the population level, $\lambda_{\rm int}$.

The PEMD lens profile inherently breaks the MST and the parameters of the PEMD profile can be precisely constrained (within few per cent) by exquisite imaging data. In this work, we avoid describing the PEMD parameters at the population level, such as redshift, mass or galaxy environment, and make use of the individual lens inference posterior products derived on flat priors.
We note that the power-law slope, $\gamma_{\rm pl}$, of the PEMD profile inferred from imaging data is a local quantity at the Einstein radius of the deflector. The Einstein radius is a geometrical quantity that depends on the mass of the deflector and lens and source redshift. Thus, the physical location of the measured $\gamma_{\rm pl}$ from imaging data depends on the redshift configuration of the lens system. In a scenario where the mass profiles of massive elliptical galaxies deviate from an MST transformed PEMD resulting in a gradient in the measured slope $\gamma_{\rm pl}$ as a function of physical projected distance, a global joint MST correction on top of the individually inferred PEMD profiles may lead to inaccuracies.

To allow for a radial trend in the applied MST relative to the imaging inferred local quantities, we parameterize the global MST population with a linear relation in $r_{\rm eff}/\theta_{\rm E}$ as

\begin{equation}\label{eqn:lambda_scaling}
 \lambda_{\rm int}(r_{\rm eff}/ \theta_{\rm E}) = \lambda_{\rm int, 0} + \alpha_{\lambda}\left( \frac{r_{\rm eff}}{\theta_{\rm E}}-1 \right),
\end{equation}
where $\lambda_{\rm int, 0}$ is the global MST when the Einstein radius is at the half-light radius of the deflector, $r_{\rm eff}/\theta_{\rm E} = 1$, and $\alpha_{\lambda}$ is the linear slope in the expected MST as a function of $r_{\rm eff}/\theta_{\rm E}$. In this form, we assume self-similarity in the lenses in regard to their half-light radii.
In addition to the global MST normalization and trend parameterization, we add a Gaussian distribution scatter with standard deviation $\sigma(\lambda_{\rm int})$ at fixed $r_{\rm eff}/\theta_{\rm E}$.

\cite{Wong:2020} and \cite{Millon:2020} showed that the TDCOSMO sample results in statistically consistent individual inferences when employing a PEMD lens model.
This implies that the global properties of the mass profiles of massive elliptical galaxies in the TDCOSMO sample can be considered to be homogeneous to the level to which the data allows to distinguish differences.

\subsubsection{External convergence}
The line-of-sight convergence, $\kappa_{\rm ext}$, is a component of the MST (Eqn. \ref{eqn:mst_split}) and impacts the cosmographic inference. When performing a joint analysis of a sample of lenses, the key quantity to constrain is the sample distribution of the external convergence. We require the global selection function of lenses to be accurately represented to provide a Hubble constant measurement. A bias in the distribution mean of $\kappa_{\rm ext}$ on the population level directly leads to a bias of $H_0$.

In this work, we do not explicitly constrain the global external convergence distribution hierarchically but instead constrain $p(\kappa_{\rm ext})$ for each individual lens independently. However, due to the multiplicative nature of internal and external MST (Eqn. \ref{eqn:mst_split}), the kinematics constrains foremost the total MST, which is the relevant parameter to infer $H_0$. The population distribution of $p(\kappa_{\rm ext})$ only changes the interpretation of the divide into internal vs. external MST and the scatter in each of the two parts.

\subsubsection{Stellar anisotropy}
The anisotropy distribution of stellar orbits (Eqn. \ref{eqn:anisotropy_definition}) can alter significantly the observed line-of-sight projected stellar velocity dispersion (see Section \ref{sec:vel_disp} and Appendix \ref{app:anisotropy}). The kinematics can constrain (together with a lens model) the angular diameter distance ratio $D_{\rm s}/D_{\rm ds}$ (Eqn. \ref{eqn:kinematics_cosmography}, \ref{eqn:ang_dist_kin}). Having a good quantitative handle on the anisotropy behavior of the lensing galaxies is therefor crucial in allowing for a robust inference of cosmographic quantities.
As is the case for an internal MST, the anisotropy cannot be constrained on a lens-by-lens basis with a single aperture velocity dispersion measurement, which impacts the derived cosmographic constraints. It is thus crucial to impose a population prior on the deflectors' anisotropic stellar orbit distribution and propagate the population uncertainty onto the cosmographic inference.

Observations suggest that typical massive elliptical galaxies are, in their central regions, isotropic or mildly radially anisotropic \citep[e.g.,][]{Grehard:2001, Cappellari:2007}; similarly, different theoretical models of galaxy formation predict that elliptical galaxies should have anisotropy varying with radius, from almost isotropic in the center to radially biased in the outskirts \citep[]{vanAlbada:1982, Hernquist:1993, Nipoti:2006}.
A simplified description of the transition can be made with an anisotropy radius parameterization, $r_{\text{ani}}$, defining $\beta_{\rm ani}$ as a function of radius $r$ \citep[][]{Osipkov:1979, Merritt:1985}
\begin{equation} \label{eqn:r_ani}
  \beta_{\text{ani}}(r) = \frac{r^2}{r_{\text{ani}}^2+r^2}.
\end{equation}
To describe the anisotropy distribution on the population level, we explicitly parameterize the profile relative to the measured half-light radius of the galaxy, $r_{\rm eff}$, with the scaled anisotropy parameter
\begin{equation} \label{eqn:a_ani}
  a_{\rm ani} \equiv \frac{r_{\rm ani}}{r_{\rm eff}}.
\end{equation}
To account for lens-by-lens differences in the anisotropy configuration, we also introduce a Gaussian scatter in the distribution of $a_{\rm ani}$, parameterized as $\sigma(a_{\rm ani})$, such that $\sigma(a_{\rm ani}) \langle a_{\rm ani}\rangle$ is the standard deviation of $a_{\rm ani}$ at sample mean $\langle a_{\rm ani}\rangle$.

\subsubsection{Cosmological parameters} \label{sec:cosmo_param}
All relevant cosmological parameters, $\boldsymbol{\pi}$, are part of the hierarchical Bayesian analysis.
\cite{Wong:2020} and \cite{Taubenberger:2019} showed that when adding supernovae of type Ia from the Pantheon \citep{Scolnic:2018} or JLA \citep{Betoule:2014} sample as constraints of an inverse distance ladder, the cosmological-model dependence of strong-lensing $H_0$ measurements is significantly mitigated.

In this work, we assume a flat $\Lambda$CDM cosmology with parameters $H_0$ and $\Omega_{\rm m}$.
We are using the inference from the Pantheon-only sample of a flat $\Lambda$CDM cosmology with $\Omega_{\rm m} = 0.298 \pm 0.022$ as our prior on the relative expansion history of the Universe in this work.

\subsection{Likelihood calculation} \label{sec:approx_likelihood}
In Section \ref{sec:hierarchy_approx} we presented the generic form of the likelihood $\mathcal{L}(\mathcal{D}_i | D_{\rm d, s, ds}, \boldsymbol{\xi}_{\rm pop})$ (Eqn. \ref{eqn:single_lens_posterior}) that we need to evaluate for each individual lens for a specific choice of hyper-parameters, and in Section \ref{sec:hyper_param} we provided the specific choices and parameterization of the hyper-parameters used in this work.
In this section, we specify the specific likelihood of Equation (\ref{eqn:single_lens_posterior}), $\mathcal{L}(\mathcal{D}_i | D_{\rm d, s, ds}, \boldsymbol{\xi'}_{\rm pop, i})$, that we use, since it is accessible and sufficiently fast to evaluate so that we can sample over a large number of lenses and their population priors.

Specifically, the parameters treated on the population level are $\boldsymbol{\xi'}_{\rm pop, i} = \{\lambda_{\rm int, 0}, \alpha_{\lambda}, \sigma(\lambda_{\rm int}), \langle a_{\rm ani}\rangle, \sigma(a_{\rm ani}) \}$.
Our choice of hyper-parameters allows us to reutilize many of the posterior products derived from an independent analysis of single lenses (Eqn. \ref{eqn:single_lens_inference}).
None of the lens model parameters, $\boldsymbol{\xi}_{\rm mass}$, except parameters describing $\lambda_{\rm int}$ and none of the light profile parameters, $\boldsymbol{\xi}_{\rm light}$, are treated on the population level and thus we can sample those independently for each lens directly from their imaging data

\begin{multline} \label{eqn:marginal_lens_light}
	\mathcal{L}(\mathcal{D}_i | D_{\rm d, s, ds}, \boldsymbol{\xi'}_{\rm pop, i}) = \int \mathcal{L}(\mathcal{D}_i | D_{\rm d, s, ds}, \boldsymbol{\xi'}_{\rm pop, i}, \boldsymbol{\xi}_{\rm mass}, \boldsymbol{\xi}_{\rm light}) \\
	\times p(\boldsymbol{\xi}_{\rm mass}, \boldsymbol{\xi}_{\rm light}) d\boldsymbol{\xi}_{\rm mass} d\boldsymbol{\xi}_{\rm light}.
\end{multline}

Furthermore, $\kappa_{\rm ext}$ and $\lambda_{\rm int}$ can be merged to a total MST parameter $\lambda$ according to their definitions (Eqn. \ref{eqn:mst_split}). All observables and thus the likelihood only respond to this overall MST parameter.

\section{Validation on the time-delay lens modeling challenge} \label{sec:tdlmc}
Before applying the hierarchical framework to real data, we use the time-delay lens modeling
challenge \citep[TDLMC][]{Ding_tdlmc2018, Ding:2020} data set to validate the hierarchical analysis and to explore different anisotropy models and priors. The TDLMC was structured with three independent submission rungs.
Each of the rungs contained 16 mock lenses with \textit{HST}-like imaging, time delays and kinematics information. The $H_0$ value used to create the mocks was hidden from the modeling teams.
The Rung1 and Rung2 mocks both used PEMD (Eqn. \ref{eqn:pl_profile}) with external shear lens models.
The Rung3 lenses were generated by ray-tracing through zoom-in hydrodynamic simulations and reflect a large complexity in their mass profiles and kinematic structure, as expected in the real Universe.

In the blind submissions for Rung1 and Rung2, different teams demonstrated that they could recover the unbiased Hubble constant within their uncertainties under realistic conditions of the data products, uncertainties in the Point Spread Function (PSF) and complex source morphology. In particular, two teams used \textsc{lenstronomy} in their submissions in a completely independent way and achieved precise constraints on $H_0$ while maintaining accuracy.
For Rung1 and Rung2, the most precise submissions used the same model parameterization in their inference, thus omitting the problems reviewed in Section \ref{sec:individual_lens}.

It is hard to draw precise conclusions from Rung3 as there are remaining issues in the simulations, such as numerical smoothing scale, sub-grid physics, and a truncation at the virial radius.
For more details of the challenge setup we refer to \cite{Ding_tdlmc2018} and on the results and the simulations used in Rung3 to \cite{Ding:2020}. For a recent study comparing spectroscopic observations with hydrodynamical simulations at $z=0$ we refer for instance to \cite{Sande:2019}.

Despite the limitations of the available simulations for accurate cosmology, the application of the hierarchical analysis scheme on TDLMC Rung3 is a stress  for the flexibility introduced by the internal MST and the kinematic modeling. Furthermore, the stellar kinematics from the stellar particle orbits provides a self-consistent and highly complex dynamical system. The analysis of TDLMC Rung3 can further help in validating the kinematic modeling aspects in our analysis. However, the removal of substructure in post-processing and truncation effects
do not allow, in this regard, conclusions below the 1\% level \citep[see][]{Ding:2020}.
For the effect of substructure on the time delays we refer, for instance, to \cite{Mao:1998, Keeton:2009} and for a study including the full line-of-sight halo population to \cite{Gilman:2020}.

We describe the analysis as follow: In Section \ref{sec:tdlmc_individual_lens} we discuss the modeling of the individual lenses. In Section \ref{sec:tdlmc_hierarchical_analysis} we describe the hierarchical analysis and priors, and present the inference on $H_0$.

\subsection{TDLMC individual lens modeling} \label{sec:tdlmc_individual_lens}
For the validation, we make use of the blind submissions of the EPFL team by A. Galan, M. Millon, F. Courbin and V. Bonvin.
The modeling of the EPFL team is performed with \textsc{lenstronomy}, including an adaptive PSF reconstruction technique and taking into account astrometric uncertainties explicitly \citep[e.g.,][]{BirrerTreu:2019}. Overall, the submissions of the EPFL team follow the standards of the TDCOSMO collaboration. The time that each investigator spent on each lens was substantially reduced due to the homogeneous mock data products, the absence of additional complexity of nearby perturbers and the line of sight, and improvements in the modeling procedure \citep[][]{Shajib:2019}.
The EPFL team achieved the target precision and accuracy requirement on Rung2, with and without the kinematic constraints, and thus showed reliable inference of lens model parameters within a mass profile parameterization for which the MST does not apply. We refer to the TDLMC paper \citep[][]{Ding:2020} for the details of the performance of all of the participating teams.

We use Rung2 as the reference result for which the MST does not apply, and Rung3 as a test case of the hierarchical analysis.
In particular, we make use of the EPFL team's blind Rung3 submission of the joint time-delay and imaging likelihood (Eqn. \ref{eqn:td_image_likelihood}) of their PEMD + external shear models to allow for a direct comparison with the Rung2 results without the kinematics constraints.
From the model posteriors of the EPFL team submission, we require the time-delay distance $D_{\Delta t}$, Einstein radius $\theta_{\rm E}$, power-law slope $\gamma_{\rm pl}$ and half-light radius $r_{\rm eff}$ of the deflector. The added external convergence is specified in the challenge setup to be drawn from a normal distribution with mean $\langle\kappa_{\rm ext}\rangle = 0$ and $\sigma(\kappa_{\rm ext}) = 0.025$.
The EPFL submission of Rung3, which is used in this work, consists of 13 lenses out of the total sample of 16. Three lenses were dropped in their analysis prior to submission due to unsatisfactory results and inconsistency with the submission sample. The uncertainty on the Einstein radius and half-light radius is at subpercent value for all the lenses and the power-law slope reached an absolute precision ranging from below 1\% to about 2\% for the least constraining lens in their sample from the imaging data alone.

In this work, we perform the kinematic modeling and the likelihood calculation within the hierarchical framework.
We use the anisotropy model of \cite{Osipkov:1979} and \cite{Merritt:1985} (Eqn. \ref{eqn:r_ani}) with a parameterization of the transition radius relative to the half-light radius (Eqn. \ref{eqn:a_ani}).
We assume a Hernquist light profile with $r_{\rm eff}$ in conjunction with the power-law lens model posteriors $\theta_{\rm E}$ and $\gamma_{\rm pl}$ to model the dimensionless kinematic quantity $J$ (Eqn. \ref{eqn:sigma_convolved}, \ref{eqn:kinematics_cosmography}), incorporating the slit mask and seeing conditions (slit 1$^{\prime\prime}\times1^{\prime\prime}$, seeing FWHM = 0$^{\prime\prime}$.6), as specified in the challenge setup.

\subsection{TDLMC hierarchical analysis} \label{sec:tdlmc_hierarchical_analysis}

For the setting of the TDLMC we only sample $H_0$ as a free cosmology-relevant parameter. The matter density $\Omega_{\rm m} = 0.27$ is provided in the challenge setup. We extend the EPFL submission by adding an internal MST distribution with a linear scaling of $r_{\rm eff}/\theta_{\rm E}$ described by $\lambda_{\rm int, 0}$ and $\alpha_{\lambda}$ (Eqn. \ref{eqn:lambda_scaling}) and Gaussian standard deviation $\sigma(\lambda_{\rm int})$ of the population at fixed $r_{\rm eff}/\theta_{\rm E}$. The anisotropy parameter $a_{\rm ani}$ is also treated on the population level with mean $\langle a_{\rm ani}\rangle$ and Gaussian standard deviation $\sigma(a_{\rm ani})$ for the population.
In the hierarchical sampling we ignore the covariances between $D_{\Delta t}$ and the model prediction of the kinematics $J$. This is justified because of the precise $\gamma_{\rm pl}$ constraints from the imaging data and the inference from the EPFL team.

The summary of the parameters and prior being used in this inference on the TDLMC is presented in Table \ref{table:param_summary_tdlmc}. We chose two different forms of the prior on the anisotropy parameter $\langle a_{\rm ani}\rangle$, one uniform in $\langle a_{\rm ani}\rangle$ and a second one uniform in $\log(\langle a_{\rm ani}\rangle)$, covering the same range in the parameter space, to investigate prior dependences in our inference. To account for the external convergence, we marginalize for each individual lens from the probability distribution $p(\kappa_{\rm ext})$ as specified in the challenge setup.\footnote{Alternatively, we could have also transformed the $D_{\Delta t}$ posteriors accordingly to account for the external convergence for each individual lens.}

\begin{table*}
\caption{Summary of the model parameters sampled in the hierarchical inference on TDLMC Rung3 in Section \ref{sec:tdlmc}.}
\begin{center}
\begin{threeparttable}
\begin{tabular}{l l l}
    \hline
    name & prior & description \\
    \hline \hline
    Cosmology  (Flat $\Lambda$CDM) \\
    $H_0$ [\Hunit] & $\mathcal{U}([0, 150])$  & Hubble constant \\
    $\Omega_{\rm m}$ & $=0.27$ & current normalized matter density \\
    \hline
    Mass profile \\
    $\lambda_{\rm int, 0}$ & $\mathcal{U}([0.5, 1.5])$ & internal MST population mean for $r_{\rm eff}/\theta_{\rm E}=1$\\
    $\alpha_{\lambda}$ & $\mathcal{U}([-1, 1])$  & slope of $\lambda_{\rm int}$ with $r_{\rm eff}/\theta_{\rm E}$ of the deflector (Eqn. \ref{eqn:lambda_scaling}) \\
    $\sigma(\lambda_{\rm int})$ & $\mathcal{U}([0, 0.2])$ & 1-$\sigma$ Gaussian scatter in $\lambda_{\rm int}$ at fixed $r_{\rm eff}/\theta_{\rm E}$ \\
    \hline
    Stellar kinematics \\
    $\langle a_{\rm ani}\rangle$ & $\mathcal{U}([0.1, 5])$ or $\mathcal{U}(\log([0.1, 5]))$ & scaled anisotropy radius (Eqn. \ref{eqn:r_ani}, \ref{eqn:a_ani}) \\
    $\sigma(a_{\rm ani})$ & $\mathcal{U}([0, 1])$  & $\sigma(a_{\rm ani}) \langle a_{\rm ani}\rangle$ is the 1-$\sigma$ Gaussian scatter in $a_{\rm ani}$\\
    \hline
    Line of sight \\
    $\langle\kappa_{\rm ext}\rangle$ & $=0$  & population mean in external convergence of lenses \\
    $\sigma(\kappa_{\rm ext})$ & $=0.025$  & 1-$\sigma$ Gaussian scatter in $\kappa_{\rm ext}$ \\
    \hline
\end{tabular}
\begin{tablenotes}
\end{tablenotes}
\end{threeparttable}
\end{center}
\label{table:param_summary_tdlmc}
\end{table*}

\begin{figure*}
  \centering
  \includegraphics[angle=0, width=140mm]{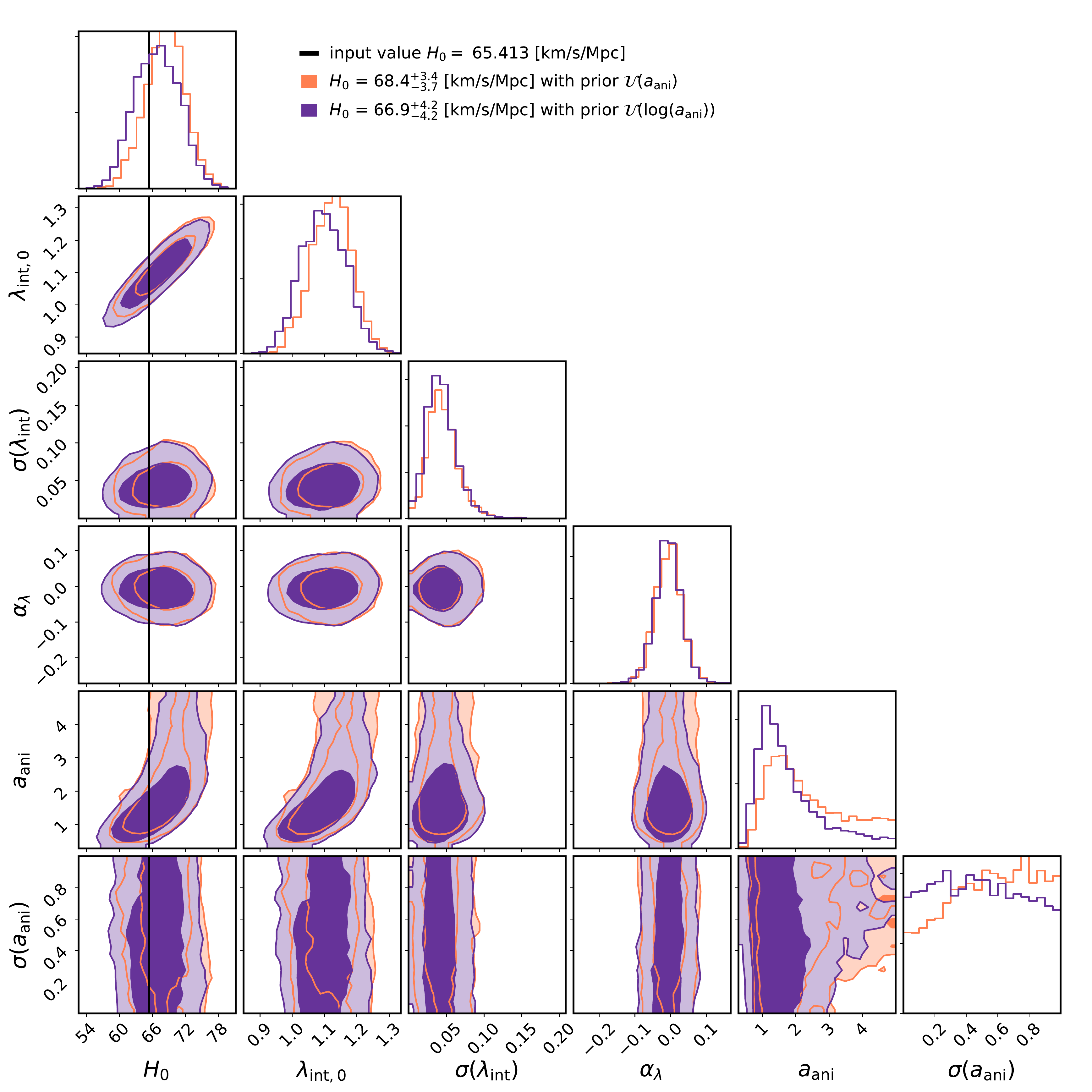}
  \caption{Mock data from the TDLMC Rung3 inference with the parameters and prior specified in Table \ref{table:param_summary_tdlmc}. Orange contours indicate the inference with a uniform prior in $a_{\rm ani}$ while the purple contours indicate the inference with a uniform priors in $\log(a_{\rm ani})$. The thin vertical line indicates the ground truth $H_0$ value in the challenge.  \faGithub\href{https://github.com/TDCOSMO/hierarchy_analysis_2020_public/blob/6c293af582c398a5c9de60a51cb0c44432a3c598/TDLMC/TDLMC_rung3_inference.ipynb}{~source} }
  \label{fig:tdlmc_rung3_omega_m_fixed_om}
\end{figure*}

Figure \ref{fig:tdlmc_rung3_omega_m_fixed_om} shows the posteriors of the hierarchical analysis with the priors specified in Table \ref{table:param_summary_tdlmc}.

We recover the assumed value for the Hubble constant ($H_0 = 65.413$ \Hunit) within the uncertainties of our inference. We find $H_0 = 66.9^{+4.2}_{-4.2}$ \Hunit for the $\mathcal{U}(\log(a_{\rm ani}))$ prior and $H_0 = 68.4^{+3.4}_{-3.7}$ \Hunit for the $\mathcal{U}(a_{\rm ani})$ prior.
We note that a uniform prior in $\log(a_{\rm ani})$ is a slightly less informative prior than a uniform prior in $a_{\rm ani}$ in the same range, as already pointed out by \cite{Birrer:2016}. In the remaining of this work $\mathcal{U}(\log(a_{\rm ani}))$ is the prior of choice in the absence of additional data that constrain the stellar anisotropy of massive elliptical galaxies to provide $H_0$ constraints.
The hierarchical analysis and the additional degree of freedom in the mass profile allows us to accurately correct for the insufficient assumptions in the mass profiles on the simulated galaxies. The kinematics modeling indicates that there is more mass in the central part of the galaxies than is modeled with a single power-law profile and infers $\lambda_{\rm int} > 1$.

We notice a nonzero inferred scatter in the internal MST distribution. One contributing source to this scatter is the fact that the external convergence component was added in post-processing in the TDLMC time delays (Eqn. \ref{eqn:time_delay_mst}). The rescaling was not applied to the velocity dispersion (Eqn. \ref{eqn:kinematics_mst}), leading to an artificial scatter in this relation equivalent to the distribution scatter of $\kappa_{\rm ext}$, $\sigma(\kappa_{\rm ext}) = 0.025$. As the mean in the convergence distribution in the TDLMC is $\langle\kappa_{\rm ext}\rangle=0$, we do not expect biases beyond a scatter to occur.

The velocity dispersion measurements allow us to constrain $\lambda_{\rm int}$ and effectively probe a more flexible mass model family.
Generally, the velocity dispersion estimates have a 5\% relative uncertainty on each individual mock lens.
As an ensemble, the 13 lenses of the EPFL submission in the TDLMC Rung3 provide information to infer $\lambda_{\rm int}$ to 2.8\% precision (see Eqn. \ref{eqn:sigma_v_lambda_error_propagation}) in the limit of a perfect anisotropy model.

The final achieved precision on $H_0$ from the sample of lenses, however, is 8\%, dominated by the uncertainty in $\lambda_{\rm int}$.
The fact that, within our chosen priors, the kinematics cannot constrain $\lambda_{\rm int}$ to better than 8\% comes from the uncertainty in the anisotropy model. More constraining priors on the anisotropy distribution of the stellar orbits in the lensing galaxies are the key to reducing the uncertainty in the $H_0$ inference \citep[see e.g.,][]{Birrer:2016, Shajib:2018, Yildirim:2020}.

\section{TDCOSMO mass profile and $H_0$ inference} \label{sec:tdcosmo_analysis}
Having verified the hierarchical approach introduced in Section \ref{sec:hierarchy} in simultaneously constraining mass profiles and $H_0$ with imaging, kinematics and time-delay observations in the TDLMC (Section \ref{sec:tdlmc}) we employ the inference on the TDCOSMO sample set to measure $H_0$. The inference on the TDCOSMO data is identical to the validation on the TDLMC, apart from some necessary modifications due to the additional complexity in the line-of-sight structure of the real data. In Section \ref{sec:tdcosmo_sample_overview} we summarize the data and individual analyses for each single lens of the TDCOSMO sample. In Section \ref{sec:tdcosmo_hierarchical} we describe the hierarchical analysis and present the results.

\subsection{TDCOSMO sample overview} \label{sec:tdcosmo_sample_overview}
The analysis presented in this work heavily relies on data and analysis products collected and presented in the literature. We give here a detailed list of the references relevant for our work for the seven lenses of the TDCOSMO sample.

\begin{enumerate}
\item \textit{B1608+656:}
The discovery in the Cosmic Lens All-Sky Survey (CLASS) is presented by \cite{Meyers:1995} with the source redshift by \cite{Fassnacht:1996}.
The imaging modeling is presented by \cite{Suyu:2009} and \cite{Suyu:2010}.
The time-delay measurement is presented by \cite{Fassnacht:1999, Fassnacht:2002}.
The velocity dispersion measurement of $260$ km/s presented by \cite{Suyu:2010} is based on Keck-LRIS spectroscopy. The statistical uncertainty is $\pm7.7$ km/s with a systematic spread of $\pm13$ km/s depending on wavelength and stellar template solution. The combined uncertainty is $260\pm15$ km/s. A previous measurement by \cite{Koopmans:2003} with $247\pm35$ km/s with Echellette Spectrograph and Imager (ESI) on Keck-II is consistent with the more recent one by \cite{Suyu:2010}.
The line-of-sight analysis is presented by \cite{Suyu:2010}, based on galaxy number counts by \cite{Fassnacht:2011}.

\item \textit{RXJ1131-1231:}
The discovery is presented by \cite{Suyu:2013} and \cite{Sluse:2003}.
The imaging modeling is presented by \cite{Suyu:2014} (for \textit{HST}) and \cite{Chen:2019} (for Keck Adaptive Optics data). An independent analysis of the \textit{HST} data was performed by \cite{Birrer:2016}.
The time-delay measurement is presented by \cite{Tewes:2012}.
The velocity dispersion measurement of $323\pm20$ km/s presented by \cite{Suyu:2013} is based on Keck-LRIS spectroscopy and includes systematics.
The line-of-sight analysis is presented by \cite{Suyu:2013}.

\item \textit{HE0435-1223:}
The discovery is presented by \cite{Wisotzki:2002}.
The image modeling is presented by \cite{Wong:2017} (for \textit{HST}) and \cite{Chen:2019} (for Keck Adaptive Optics data).
The time-delay measurement is presented by \cite{Bonvin:2016}.
The velocity dispersion measurement of $222\pm15$ km/s presented by \cite{Wong:2017} is based on Keck-LRIS spectroscopy and includes systematic uncertainties. An independent measurement of $222\pm34$ km/s by \cite{Courbin:2011} using VLT is in excellent agreement.
The line-of-sight analysis is presented by \cite{Rusu:2017}.

\item \textit{SDSS1206+4332:}
The discovery is presented by \cite{Oguri:2005}.
The image modeling is presented by \cite{Birrer:2019}.
The time-delay measurement is presented by \cite{Eulaers:2013} with an update by \cite{Birrer:2019}.
The velocity dispersion measurement of $290\pm30$ km/s presented by \cite{Agnello:2016} is based on Keck-DEIMOS spectroscopy and includes systematic uncertainties.
The line-of-sight analysis is presented by \cite{Birrer:2019}.

\item \textit{WFI2033-4723:}
The discovery is presented by \cite{Morgan:2004}, the image modeling by \cite{Rusu:2020} and the time-delay measurement by \cite{Bonvin:2019}.
The velocity dispersion measurement from VLT MUSE is presented by \cite{Sluse:2019} with $250\pm10$ km/s only accounting for statistical error and $250 \pm 19$ km/s including systematic uncertainties.
The line-of-sight analysis is presented by \cite{Rusu:2020}.

\item \textit{DES0408-5354:}
The discovery is presented by \cite{Lin:2017, Diehl:2017}. The imaging modeling is presented by \cite{Shajib0408}. A second team within STRIDES and TDCOSMO is performing an independent and blind analysis using a different modeling code (Yildirim et al in prep).
The time-delay measurement is presented by \cite{Courbin:2018}.
The velocity dispersion measurements are presented by \cite{BuckleyGeer:2020}. We used the values from Table 3 in \cite{Shajib0408}. The measurements are from Magellan with $230\pm37$ km/s (mask A) and $236\pm42$ km/s (mask B), from Gemini with $220\pm21$ km/s and from VLT MUSE with $227\pm9$ km/s. The reported values do not include systematic uncertainties and covariances among the different measurements. Following \cite{Shajib0408} we add a covariant systematic uncertainty of $\pm17$ km/s to the reported values.
The line-of-sight analysis is presented by \cite{BuckleyGeer:2020}.

\item \textit{PG1115+080:}
The discovery is presented by \cite{Weymann:1980}. The image modeling is presented by \cite{Chen:2019} using Keck Adaptive Optics.
The time-delay measurement is presented by \cite{Bonvin:2018}, while the line-of-sight analysis by \cite{Chen:2019}.
The velocity dispersion measurement of $281\pm25$ km/s, presented by \cite{Tonry:1998}, is based on Keck-LRIS spectroscopy.
In this work we add new acquired integral-field spectroscopy obtained with the Multi-Object Survey Explorer (MUSE) on the VLT in March 2019 (0102.A-0600(C), PI Agnello), and we thus go in some detail about the observations. The details and the data will be presented in a forthcoming paper by Agnello et al. (in prep).
At the location of the lens, 3h of total exposure time were obtained, in clear or photometric conditions and nominal seeing of 0.8$^{\prime\prime}$ FWHM.
Due to the proximity of the four quasar images to the main galaxy, a dedicated extraction routine was used in order to optimally deblend all components.
We followed the same procedure as by \cite{Sluse:2019} and \cite{Braibant:2014}, fitting each spectral channel as a superposition of a Sersic profile (for the main lens) and four point sources as identical Moffat profiles. The separation between the individual components is held fixed to the \textit{HST}-NICMOS measurements \citep{Sluse:2012}.

A nearby star in the MUSE field-of-view was used as a reference PSF. From this direct modeling, the FWHM of the PSF was found to be $0^{\prime\prime}.67\pm0^{\prime\prime}.1$, with some variation with wavelength that was accounted for in the model-based deblending.
This procedure produced an optimal subtraction of the quasar spectra, at least within 1$^{\prime\prime}$ from the center of the lens.
The lens galaxy 1D spectra were then extracted in two square apertures ($R<0^{\prime\prime}.6$, $0^{\prime\prime}.6<R<1^{\prime\prime}.0$), and processed with the Penalized PiXel-Fitting (\textsc{ppxf}) code presented in \cite{Cappellari:2004} and further upgraded in \cite{Cappellari:2017} to obtain velocity dispersions.

The velocity dispersion measurement results from a linear combination of stellar template spectra to which a sum of orthogonal polynomials is added to adjust the continuum shape of the templates to the observed galaxy specttrum.
The spectral library used for the fit is the Indo-US spectral library, 1273 stars covering the region from 3460 - 9464 \AA at a spectral resolution of 1.35 \AA FWHM \citep{Valdes:2004}.

We measure for the inner aperture ($R<0.6^{\prime\prime}$) a stellar velocity dispersion value of $277 \pm 6.5$ km/s and for the outer ($0^{\prime\prime}.6<R<1^{\prime\prime}.0$) a value of $241 \pm 8.8$ km/s. The uncertainties only include the statistical errors.
In order to estimate the systematics, we performed a number of \textsc{ppxf} fits on the smaller aperture, changing each time the wavelength range, the degree of the additive polynomial and the number of stellar templates used to fit the galaxy spectra. We obtained a systematic uncertainty of $\pm 23.6$ km/s that, as for the case of DES0408, we treat as fully covariant among the two aperture measurements.
With the spectral resolution of MUSE, systematic uncertainties are within $\approx10\%$ and about three times larger than the nominal, statistical uncertainties thanks to the high signal-to-noise of the spectra.

\end{enumerate}

All the TDCOSMO analyses of lenses used uniform priors on all relevant parameters when performing the inference with a PEMD model \footnote{For the composite models, priors on the mass-concentration relation of the dark matter profiles were imposed.}. Six out of the seven lenses were modeled blindly\footnote{The first lens, B1608+656, and the reanalysis of RXJ1131-1231 with AO data were not executed blindly.}, that is $H_0$ values were never seen by the modeler at any step of the process.

\begin{figure}
  \centering
  \includegraphics[angle=0, width=80mm]{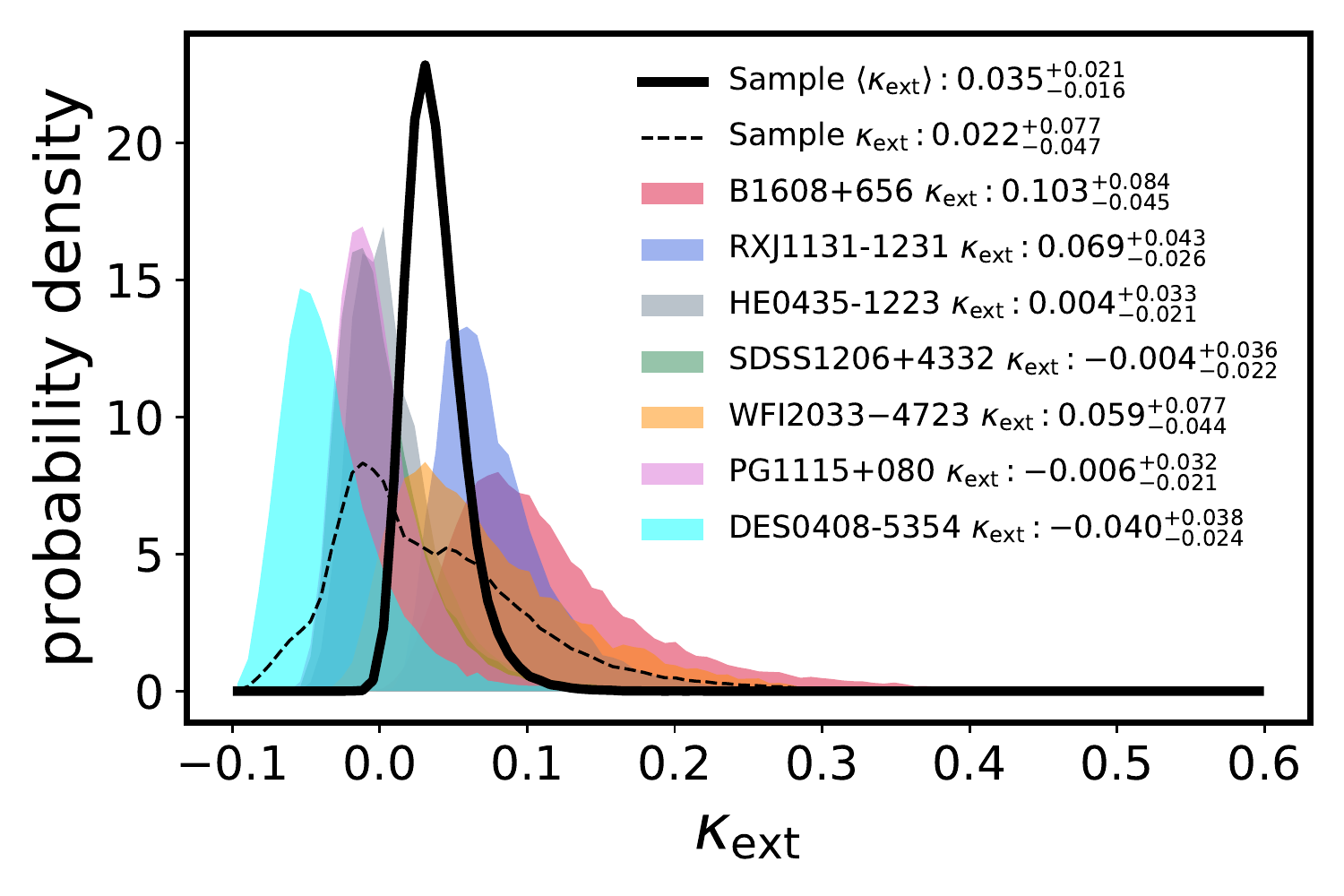}
  \caption{External convergence posteriors for the individual TDCOSMO lenses.
   \faGithub\href{https://github.com/TDCOSMO/Ahierarchy_analysis_2020_public/blob/6c293af582c398a5c9de60a51cb0c44432a3c598/TDCOSMO_sample/tdcosmo_sample.ipynb}{~source} }
  \label{fig:kappa_ext_tdcosmo}
\end{figure}

Detailed line-of-sight analyses for each lens have been performed based on weighted relative number counts of galaxies along the line of sight on deep photometry and spectroscopic campaigns \citep[e.g.,][]{Rusu:2017}.
Furthermore, for a fraction of the lenses, we have used also an external shear constraint inferred by the strong lens modeling to inform the line-of-sight convergence estimate.
The weighted galaxy number count and external shear summary statistics have been applied on the Millenium Simulation \citep{Springel:2005} with ray-tracing \citep{Hilbert:2009} to extract a posterior in $p(\kappa_{\rm ext})$ with the prior from the Millenium Simulation and semi-analytic galaxy evolution model with painted synthetic photometry on top \citep{DeLuciaBlaizot:2007}\footnote{The Millenium Simulation uses the following flat $\Lambda$CDM cosmology: $\Omega_{\rm m} = 0.25$, $\Omega_{\rm b} = 0.045$, $H_0 = 73$ \Hunit, $n=1$, and $\sigma_{8}=0.9$.}. The external convergence and shear values from the Millenium simulation are computed from the observer to the source plane, $\kappa_{\rm ext} \approx \kappa_{\rm s}$. The coupling of the strong lens deflector \citep[e.g.,][]{Barkana:1996, McCully:2014, Birrer:2017} is not included in the calculation of $\kappa_{\rm s}$.
Figure \ref{fig:kappa_ext_tdcosmo} shows the $\kappa_{\rm ext}$ posteriors for the individual lenses. For the overall sample mean, we get $\langle\kappa_{\rm ext}\rangle = 0.035^{+0.021}_{-0.016}$ with a scatter of $\sigma(\kappa_{\rm ext}) = 0.046$ around the mean.
Nearby massive galaxies along the line of sight were included explicitly in the modeling where required, and the external convergence term was adapted accordingly in order to not double count mass structure in the analysis.
Table \ref{table:tdcosmo_sample} presents the redshifts and the relevant lens model posteriors that are used in our analysis.

\begin{table*}
\caption{Overview of the TDCOSMO sample posterior products used in this work. We list lens redshift $z_{\rm lens}$, source redshift $z_{\rm source}$, half-light radius of the deflector $r_{\rm eff}$, Einstein radius of the deflector $\theta_{\rm E}$, power-law slope $\gamma_{\rm pl}$, external convergence $\kappa_{\rm ext}$ and inferred time-delay distance from the power-law model based on imaging data and time delays, not including external convergence or internal MST terms, $D_{\Delta t}^{\rm pl}$.}
\begin{center}
\begin{threeparttable}
\begin{tabular}{l l l l l l l l}
    \hline
    name & $z_{\rm lens}$ & $z_{\rm source}$ & $r_{\rm eff}$ [arcsec] & $\theta_{\rm E}$ [arcsec] & $\gamma_{\rm pl}$ & $\kappa_{\rm ext}$ & $D_{\Delta t}^{\rm pl}$ [Mpc] \\

    \hline \hline
    B1608+656 & 0.6304 & 1.394 & $0.59\pm 0.06$ & $0.81\pm0.02$ & $2.08\pm0.03$ & ${+0.103}_{-0.045}^{+0.084}$ & ${4775}_{-130}^{+138}$\\
	RXJ1131-1231 & 0.295 & 0.654 & $1.85\pm0.05$ & $1.63\pm0.02$ & $1.95\pm0.05$ & ${+0.069}_{-0.026}^{+0.043}$ & ${1947}_{-35}^{+35}$ \\
	HE0435-1223	& 0.4546 & 1.693 & $1.33\pm0.05$ & $1.22\pm0.05$ & $1.93\pm0.02$ & ${+0.004}_{-0.021}^{+0.032}$ & ${2695}_{-157}^{+159}$ \\
	SDSS1206+4332 & 0.745 & 1.789 & $0.34\pm0.05$ & $1.25\pm0.01$ & $1.95\pm0.05$ & ${-0.004}_{-0.021}^{+0.036}$ & ${5846}_{-608}^{+628}$ \\
	WFI2033-4723 & 0.6575 & 1.662 & $1.41\pm 0.05$ & $0.94\pm0.02$ & $1.95\pm0.02$ & ${+0.059}_{-0.044}^{+0.078}$ & ${4541}_{-152}^{+134}$ \\
	PG1115+080 & 0.311 & 1.722 & $0.53\pm0.05$ & $1.08\pm0.02$ & $2.17\pm0.05$ & ${-0.006}_{-0.021}^{+0.032}$ & ${1458}_{-115}^{+117}$ \\
	DES0408-5354 & 0.597 & 2.375 & $1.20\pm0.05$ & $1.92\pm0.01$ & $1.90\pm0.03$ & ${-0.040}_{-0.024}^{+0.037}$ & ${3491}_{-74}^{+75}$\\
    \hline
\end{tabular}
\begin{tablenotes}
\end{tablenotes}
\end{threeparttable}
\end{center}
\label{table:tdcosmo_sample}
\end{table*}

\subsection{TDCOSMO hierarchical inference} \label{sec:tdcosmo_hierarchical}

We use for each lens the individual time-delay distance likelihood according to Equation \ref{eqn:td_image_likelihood} that was derived in previous works of this collaboration from a lens model inference on imaging data and the time-delay measurements from the PEMD inference, not including external convergence or internal MST, $D_{\Delta t}^{\rm pl}$.
We add the same MST transform as a distribution mean $\lambda_{\rm int, 0}$ and scaling $alpha_{\lambda}$ with $r_{\rm eff}/\theta_{\rm E}$, and with Gaussian scatter across the data set, identical to the TDLMC validation in Section \ref{sec:tdlmc}. The individual $p(\kappa_{\rm ext})$ distributions are added for each lens and in the inference combined with the internal MST parameters.

For the kinematic modeling, we make the same assumptions as for the TDLMC sample (Section \ref{sec:tdlmc_individual_lens}) with the anisotropy model of \cite{Osipkov:1979, Merritt:1985} (Eqn. \ref{eqn:r_ani}) with a parameterization of the transition radius relative to the half-light radius (Eqn. \ref{eqn:a_ani}). The approach is consistent with the previous kinematic analysis and sufficiently verified on the TDLMC to the level of accuracy we can expect from this analysis.
We also assume a Hernquist light profile with $r_{\rm eff}$, in conjunction with the power-law lens model posteriors $\theta_{\rm E}$ and $\gamma_{\rm pl}$ to model the dimensionless kinematic quantity $J$ (Eqn. \ref{eqn:sigma_convolved}, \ref{eqn:kinematics_cosmography}), also incorporating the slit mask and seeing conditions of the individual observations.

For each of the lenses in the TDCOSMO sample, we use the distribution $p(\kappa_{\rm ext})$ as derived on the individual blinded analyses and do not invoke an additional population parameter. We leave the hierarchical analysis of the line-of-sight selection to future work. We want to stress that the overall selection bias in this hierarchical approach does not impact the $H_0$ constraints as the kinematics constrains the overall MST (Eqn. \ref{eqn:mst_split}). An overall shift in the distribution of $\kappa_{\rm s}$ will be compensated by $\lambda_{\rm int}$ in the inference, thus leaving the $H_0$ constraints invariant.

\begin{figure*}
  \centering
  \includegraphics[angle=0, width=150mm]{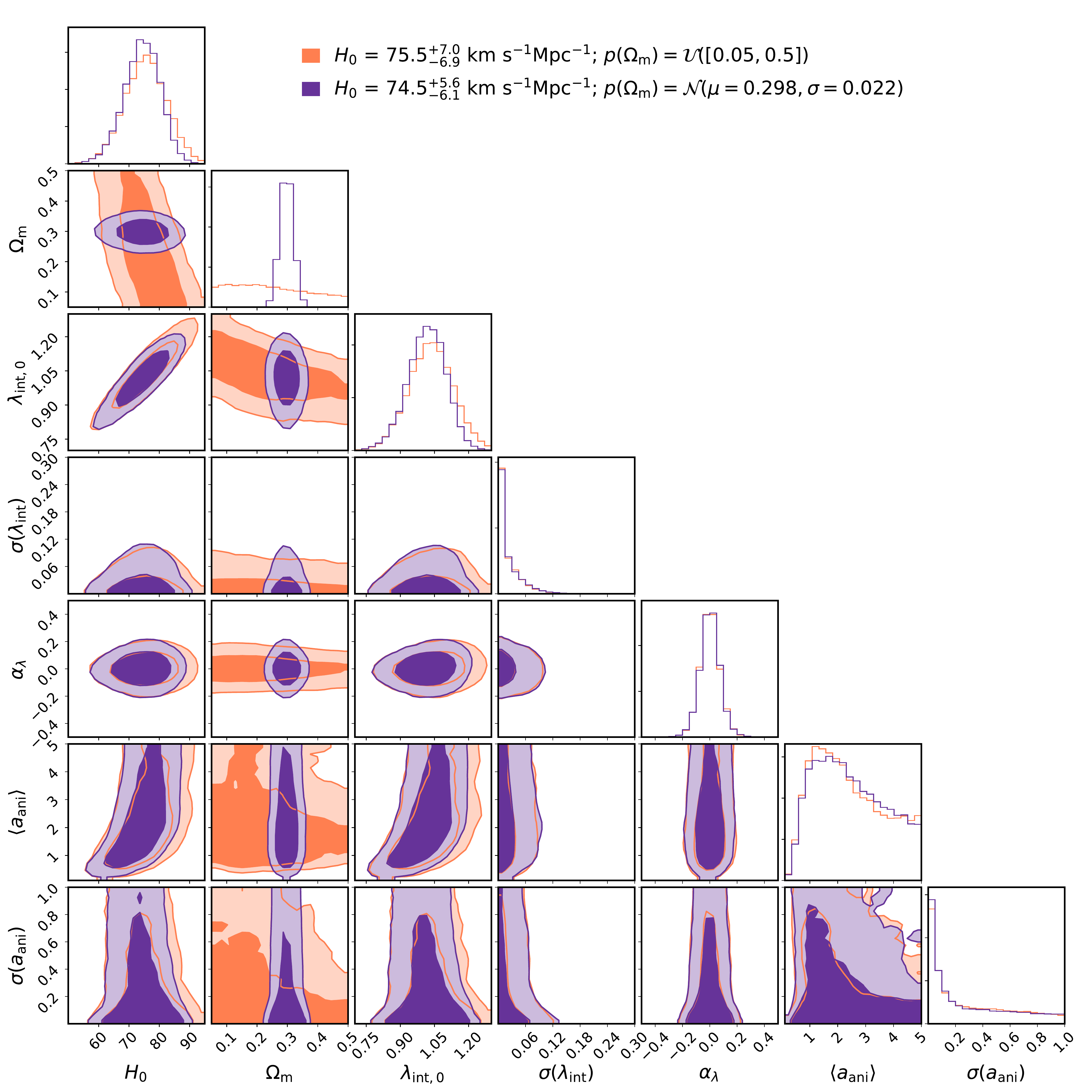}
  \caption{Hierarchical analysis of the TDCOSMO-only sample when constraining the MST with kinematic information. Parameter and priors are specified in Table \ref{table:param_summary_tdcosmo}. Orange contours correspond to the inference with uniform prior on $\Omega_{\rm m}$, $\mathcal{U}([0.05, 0.5])$, while the purple contours correspond to the prior based on the Pantheon sample with $\mathcal{N}(\mu=0.298, \sigma=0.022)$.  \faGithub\href{https://github.com/TDCOSMO/hierarchy_analysis_2020_public/blob/6c293af582c398a5c9de60a51cb0c44432a3c598/TDCOSMO_sample/tdcosmo_sample.ipynb}{~source}}
  \label{fig:tdcosmo_results_only}
\end{figure*}

We assume a flat $\Lambda$CDM cosmology with a uniform prior on $H_0$ in [0, 150] \Hunit.
For $\Omega_{\rm m}$ we chose the prior based on the Pantheon sample \citep{Scolnic:2018}, $\mathcal{N}(\mu=0.298, \sigma=0.022)$. We also perform the inference with a flat prior on $\Omega_{\rm m}$ in [0.05, 0.5] to allow for comparison with the previous work by \cite{Wong:2020} and \cite{Millon:2020} and to illustrate cosmology dependences in the time-delay cosmography inference.
Table \ref{table:param_summary_tdcosmo} summarizes all the hierarchical hyper-parameters sampled in the analysis of this section. The posteriors of the TDCOSMO sample inference are presented in Figure \ref{fig:tdcosmo_results_only}.

For the tight prior on $\Omega_{\rm m}$, we measure $H_0 = $ \Htdcosmo \Hunit. For an unconstrained relative expansion history with a prior on $\Omega_{\rm m}$ uniform in [0.05, 0.5], we measure $H_0 = $ \HtdcosmoFlatOmega \Hunit.
The 9\% precision on $H_0$ is significantly inflated relative to previous studies with the same data set \citep[][]{Wong:2020, Millon:2020}.
The increase in uncertainty with respect to the H0LiCOW analysis is attributed to two main factors: 1) we relaxed the assumption of NFW+stars or power-law mass density profiles; 2) we considered the impact of covariance between lenses when accounting for uncertainties potentially arising from assumptions about mass profile and stellar anisotropy models. As we show in the next sections, however, this uncertainty can be reduced by adding external information to further constrain the mass profile and anisotropy of the deflectors.
The inferred scatter in $\lambda_{\rm int}$, $\sigma(\lambda_{\rm int})$, is consistent with zero. This is a statement on the internally consistent error bars on $H_0$ among the TDCOSMO sample \citep[][]{Wong:2020, Millon:2020}.

\begin{table*}
\caption{Summary of the model parameters sampled in the hierarchical inference on the TDCOSMO sample in Section \ref{sec:tdcosmo_analysis} and posteriors presented in Figure \ref{fig:tdcosmo_results_only}.}
\begin{center}
\begin{threeparttable}
\begin{tabular}{l l l}
    \hline
    name & prior & description \\
    \hline \hline
    Cosmology (Flat $\Lambda$CDM) \\
    $H_0$ [\Hunit] & $\mathcal{U}([0, 150])$  & Hubble constant \\
    $\Omega_{\rm m}$ & $\mathcal{U}([0.05, 0.5])$ or $\mathcal{N}(\mu=0.298, \sigma=0.022)$ & current normalized matter density \\
    \hline
    Mass profile \\
    $\lambda_{\rm int, 0}$ & $\mathcal{U}([0.5, 1.5])$ & internal MST population mean for $r_{\rm eff}/\theta_{\rm E}=1$\\
    $\alpha_{\lambda}$ & $\mathcal{U}([-1, 1])$  & slope of $\lambda_{\rm int}$ with $r_{\rm eff}/\theta_{\rm E}$ of the deflector (Eqn. \ref{eqn:lambda_scaling}) \\
    $\sigma(\lambda_{\rm int})$ & $\mathcal{U}(\log([0.001, 0.5]))$ & 1-$\sigma$ Gaussian scatter in $\lambda_{\rm int}$ at fixed $r_{\rm eff}/\theta_{\rm E}$ \\
    \hline
    Stellar kinematics \\
    $\langle a_{\rm ani}\rangle$ & $\mathcal{U}(\log([0.1, 5]))$ & scaled anisotropy radius (Eqn. \ref{eqn:r_ani}, \ref{eqn:a_ani}) \\
    $\sigma(a_{\rm ani})$ & $\mathcal{U}(\log([0.01, 1]))$  & $\sigma(a_{\rm ani}) \langle a_{\rm ani}\rangle$ is the 1-$\sigma$ Gaussian scatter in $a_{\rm ani}$\\
    \hline
    Line of sight \\
    $\kappa_{\rm ext}$ & $p(\kappa_{\rm ext})$ of individual lenses (Fig. \ref{fig:kappa_ext_tdcosmo}) & external convergence of lenses \\
    \hline
\end{tabular}
\begin{tablenotes}
\end{tablenotes}
\end{threeparttable}
\end{center}
\label{table:param_summary_tdcosmo}
\end{table*}

\section{SLACS analysis of galaxy density profiles} \label{sec:slacs_analysis}

Gravitational lenses with imaging and kinematics data can add valuable information about the mass profiles of the lenses. Even though the kinematics data in the current TDCOSMO sample is limited, an additional sufficiently large data set with precise measurements can significantly improve the precision on the mass profiles of the population and thus on the Hubble constant.
Resolved kinematics observations may in addition provide constraints on the anisotropy distribution of stellar orbits.

When incorporating external data sets as part of the hierarchical framework, it is important that those external lenses are drawn from the same population as the time-delay lenses - unless explicitly marginalized over population differences.
Provided that (i) the lensing sample has a known selection function, (ii) the lens modeling is performed to the same level of precision and with the same model assumptions as the time-delay lenses, (iii) the kinematic modeling assumptions are identical and (iv) the anisotropy uncertainties are mitigated on the population level, we can fold in the extracted likelihood (Eqn. \ref{eqn:spec_image_likelihood}) into the hierarchical analysis, applying the same population dependence on $\lambda_{\rm int}$ and $a_{\rm ani}$.

Selection biases can arise from different aspects. Ellipticity and shear naturally increase the abundance of quadruple lenses relative to double lenses. \cite{Holder:2003} use $N$-body simulations to estimate the level of external shear due to structure near the lens and conclude that the local environment is the dominant contribution that drives the external shear bias in quadruple lenses. \cite{Huterer:2005} investigate the external shear bias and conclude that this effect is not sufficient to explain the observed quadruple-to-double ratio.
\cite{Collett:2016} conclude, based on idealized simulations, that selection based on image brightness and separation leads to significant selection bias in the slope of the mass profiles. In addition, \cite{Collett:2016} also find a line-of-sight selection bias in quadruply lensed quasars relative to the overall population on the level of 0.9\%. The bias is less prominent for doubly imaged quasars.
The specific discovery channel can also lead to selection effects. \cite{Dobler:2008} note that a spectroscopically selected search, as performed for the Sloan Lens ACS (SLACS) survey \citep{Bolton:2006}, can lead to significant biases on the selected velocity dispersion in the resulting sample. However, \cite{Treu:2006} show that, at fixed velocity dispersion, the SLACS sample is indistinguishable from other elliptical galaxies.

In this section we present a hierarchical analysis of the SLACS sample \citep{Bolton:2006, Bolton:2008} following the same hierarchical approach as the TDCOSMO sample, based on the imaging modeling by \cite{Shajib_slacs:2020}.
The SLACS sample of strong gravitational lenses is a sample of massive elliptical galaxies selected from the Sloan Digital Sky Survey (SDSS) by the presence in their spectra of emission lines consistent with a higher redshift.
Follow-up high-resolution observations with \textit{HST} revealed the presence of strongly lensed sources.
The SLACS data set allows us to further constrain the population distribution in the mass profile parameter $\lambda_{\rm int}$ and the anisotropy distribution $a_{\rm ani}$ and, thus, can add significant information to the TDCOSMO sample to be used jointly in Section \ref{sec:joint_analysis} to constrain $H_0$.

In Section \ref{sec:slacs_imaging} we describe the imaging data and lens model inference. In Section \ref{sec:slacs_spectroscopy} we describe the spectroscopic data set used and how we model it, including VLT VIMOS IFU data for a subset of the lenses. We analyze the selection effect of the SLACS sample in Section \ref{sec:slacs_selection} and in Section \ref{sec:slacs_los_convergence} we constrain the line-of-sight convergence for the individual lenses. In Section \ref{sec:slacs_constraints} we present the results of the hierarchical analysis of the SLACS sample in regard to mass profile and anisotropy constraints.

\subsection{SLACS imaging} \label{sec:slacs_imaging}
To include additional lenses in the hierarchical analysis, we must ensure that the quality and the choices made in the analysis are on equal footing with the TDCOSMO sample. \cite{Shajib_slacs:2020} presents a homogeneous lens model analysis of 23 SLACS lenses from \textit{HST} imaging data. The lens model assumptions are a PEMD model with external shear, identical to the derived products we are using from the TDCOSMO sample.
The scaling of the analysis was made possible by advances in the automation of the modeling procedure \citep[e.g.,][]{Shajib:2019} with the \textsc{dolphin} pipeline package. The underlying modeling software is \textsc{lenstronomy} \citep{Birrer_lenstronomy, Birrer:2015} for which we also performed the TDLMC validation (Section \ref{sec:tdlmc}).

\cite{Shajib_slacs:2020} first select 50 SLACS lenses for uniform modeling from the sample of 85 lenses presented by \cite{Auger:2009}. The selection criteria for these lenses are: (i) no nearby satellite or large perturber galaxy within approximately twice the Einstein radius, (ii) absence of multiple source galaxies or complex structures in the lensed arcs that require large computational cost for source reconstruction, and (iii) the main deflector galaxy is not disk-like. These criteria are chosen so that the modeling procedure can be carried out automatically and uniformly without tuning the model settings on a lens-by-lens basis.
Using the \textsc{dolphin} package on top of \textsc{lenstronomy}, a uniform and automated modeling procedure is performed on the 50 selected lenses with V-band data (Advance Camera for Surveys F555W filter, or Wide Field and Planetary Camera 2 F606W filter).

After the modeling, 23 lenses are selected to have good quality models. The criteria for this final selection are: (i) good fitting to data by visually inspecting the residual between the image and the model-based reconstruction, and (ii) the median of the power-law slope does not diverge to unusual values (i.e., $\lesssim1.5$ or $\gtrsim2.5$)\footnote{We note that the prior on the power-law slope $\gamma_{\rm pl}$ is chosen to be uniform in [1, 3] during the Bayesian inference with MCMC.}.
For the TDCOSMO sample, iterative PSF corrections have been performed, based on the presence of the bright quasar images, to guarantee a well matched and reliable PSF in the modeling. For the SLACS lenses, such an iterative correction on the image itself cannot be performed due to the absence of quasars in these systems. Nevertheless, extensive tests with variations of the PSF have been performed by \cite{Shajib_slacs:2020} and the impact on the resulting power-law slope inference was below $\sim$0.005 on the population mean of $\gamma_{\rm pl}$.
The half light radius for the deflector galaxies are taken from \cite{Auger:2009} in V-band (measured along the intermediate axis).

\subsection{SLACS spectroscopy} \label{sec:slacs_spectroscopy}
The constraints on the MST rely on the kinematics observations. In this section we provide details on the data set and reduced products we are using in this work, on top of the already described ones for the TDCOSMO lenses. These include SDSS's Baryon Oscillation Spectroscopic Survey (BOSS) fiber spectroscopy \citep{Dawson:2013} and VLT VIMOS IFU observations.

\subsubsection{SDSS fiber spectroscopy}
All the SLACS lenses have BOSS spectra available as part of SDSS-III. The fiber diameter is 3$^{\prime\prime}$ and the nominal seeing of the observations are 1$^{\prime\prime}$.4 FWHM.
The measurements of the velocity dispersion from the SDSS reduction pipeline were originally presented by \cite{Bolton:2008}.
However, in this work, we use improved measurements of the velocity dispersion, determined using an improved set of templates as described in \cite{Shu:2015}.
The SDSS measurements are in excellent agreement with the subsample measured with VLT X-shooter presented by \cite{Spiniello:2015}.

\subsubsection{VLT VIMOS IFU data} \label{sec:ifu_data_description}
The VLT VIMOS IFU data set is described in \cite{Czoske:2008} and subsequently used in \cite{Barnabe:2009, Barnabe:2011, Czoske:2012}. The VIMOS fibers were in a configuration with spatial sampling of 0.67$^{\prime\prime}$, and the seeing was 0$^{\prime\prime}$.8 FWHM.

The first moment (velocity) and second moment (velocity dispersion) of the individual VIMOS fibers are fit with a single stellar template for each fiber individually and the uncertainties in the measurements are quantified within Bayesian statistics. Templates were chosen by fitting a random sample of IndoUS spectra to the aperture-integrated VIMOS IFU spectra and selecting one of the best-fitting (in the least-squares sense) template candidates \citep[we refer to details to ][]{Czoske:2008}. Marginalization over template mismatch adds another 5--10\% measurement uncertainties. Within this additional error budget, the integrated velocity dispersion measurements of \cite{Czoske:2008} are consistent with the SDSS measured values of \cite{Bolton:2008}.
We bin the fibers in radial bins in steps of 1$^{\prime\prime}$ from the center of the deflector. The binning is performed using luminosity weighting and propagation of the independent errors to the uncertainty estimate per bin.
Where necessary, we exclude fibers that point on satellite galaxies or line-of-sight contaminants.
In this work, we make use of the relative velocity dispersion measurements in radial bins when inferring $H_0$.
We do so by introducing a separate internal MST distribution $\lambda_{\rm ifu}$, effectively replacing $\lambda_{\rm int}$ when evaluating the likelihood of the IFU data. $\lambda_{\rm ifu}$ is entirely constrained by the IFU data. The MST information that propagates in the joint constraints of TDCOSMO+SLACS analysis, $\lambda_{\rm int}$, (Section \ref{sec:joint_analysis}) is derived from the SDSS velocity dispersion measurements only.
In this form, the IFU data informs the anisotropy parameter but not the mass profile directly.
We leave the amplitude calibration and usage of this data set to constrain the MST for future work.

From the original sample of 17 SLACS lenses with VIMOS observations, we drop five objects that are fast rotators (when the first moments dominate the averaged dispersion in the outer radius bin) and one slow rotator with velocity dispersion >380 km/s. This is necessary to match this sample with the TDCOSMO one in velocity dispersion space; the fast rotators are, in fact, all in a lower velocity dispersion range ($\sigma^{\rm P}$ in [185, 233] km s$^{-1}$).
Finally, we excluded one more galaxy for which there is no estimate of the Einstein radius, and thus we cannot combine lensing and dynamics. In this way, we end up with a sample of ten lenses, prior to further local environment selection.

\subsection{SLACS selection function} \label{sec:slacs_selection}

The SLACS lenses were preselected from the spectroscopic database of the SDSS based on the presence of absorption-dominated galaxy continuum at one redshift and nebular emission lines (Balmer series, [OII] 3727 {A}, or [OIII] 5007 {A}) at another, higher redshift. Details on the method and selection can be found in \cite{Bolton:2004, Bolton:2006} and \cite{Dobler:2008}. The lens and source redshifts of the SLACS sample are significantly lower than for the TDCOSMO sample.

\cite{Treu:2009} studied the relation between the internal structure of early-type galaxies and their environment with two statistics: the projected number density of galaxies inside the tenth nearest neighbor ($\Sigma_{10}$) and within a cone of radius one $h^{-1}$~Mpc ($D_1$) based on photometric redshifts. It was observed that the local physical environment of the SLACS lenses is enhanced compared to random volumes, as expected for massive early-type galaxies, with 12 out of 70 lenses in their sample known to be in group/cluster environments.

In this study, we are specifically only looking for lenses whose lensing effect can be described as the mass profile of the massive elliptical galaxy and an uncorrelated line-of-sight contribution. Assuming SLACS and TDCOSMO lenses are galaxies within the same homogeneous galaxy population and with the local environment selection of SLACS lenses, the remaining physical mass components in the deflector model are the same physical components of the lensing effect we model in the TDCOSMO sample. The uncorrelated line-of-sight contribution can be characterized based on large scale structure simulations.

\subsubsection{Deflector morphology and lensing information selection}
Our first selection cut on the SLACS sample is based on \cite{Shajib_slacs:2020}, which excludes a subset of lenses based on their unusual lens morphology (prominent disks, two main deflectors, or complex source morphology) to derive reliable lensing properties using an automated and uniform modeling procedure.

This first cut reduced the total SLACS sample of 85 lenses, presented by \cite{Auger:2009}, to 51 lenses\footnote{To use the IFU data set more optimally, we add the lens SDSSJ0216-0813, which is the remaining lens within the IFU quality sample that was not selected by \cite{Shajib_slacs:2020} from the original SLACS sample.}. Out of these 51 lenses, 23 lenses had good quality models from an automated and uniform modeling procedure as described in Section \ref{sec:slacs_imaging}. Producing good quality models for the rest of the SLACS lenses would require careful treatment on a lens-by-lens basis, which was out of the scope of \citet{Shajib_slacs:2020}.

\subsubsection{Mass proxy selection}
We want to make sure that the deflector properties are as close as possible to the TDCOSMO sample. To do so without introducing biases regarding uncertainties in the velocity dispersion measurements, we chose a cut based on Singular Isothermal Sphere (SIS) equivalent dispersions, $\sigma_{\rm SIS}$, derived from the Einstein radius and the lensing efficiency only. The deflectors of the TDCOSMO sample span a range of $\sigma_{\rm SIS}$ in [200, 350] km s$^{-1}$ and we select the same range for the SLACS sample.

\subsubsection{Local environment selection} \label{sec:slacs_local_selection}
We use the DESI Legacy Imaging Surveys (DLS) \citep{Dey:2019} to characterize the environment of the SLACS lenses. We query the DR7 Tractor source photometry catalog \citep{Lang:2016} removing any object that is morphologically consistent with being a point source convolved with the DLS point spread function. We use the R band data to count objects with $18<R<23$ within 120$^{\prime\prime}$ of the lens galaxy but more than 3$^{\prime\prime}$ from the lens.

We quantify the environment with two numbers: $N_{2^{\prime}}$, the total number of galaxies within 2 arcminutes and an inverse projected distance weighted count $N_{1/r}$ within the same 2 arcminutes aperture, defined as \citep{Greene:2013}
\begin{equation} \label{eqn:n_r_stat}
N_{1/r} \equiv \sum_{i; r < 2'} \frac{1}{r_{i}}.
\end{equation}
$N_{2^{\prime}}$ and $N{1/r}$ are physically meaningful numbers for our analysis as $N_{2^{\prime}}$ should approximately trace the total mass close enough to significantly perturb the lensing \citep[see][]{Collett:2013}, and $N_{1/r}$ should be skewed larger by masses close along the line of sight of the lens which are likely to have the most significant perturbative effect.
We assess the uncertainty on $N_{2^{\prime}}$ and $N_{1/r}$ by taking every object within 120$^{\prime\prime}$ of the lens and bootstrap resampling from their R band magnitude errors, before reapplying the $18<R<23$ cut. Where the SLACS lens is not in the DLS DR7 footprint we queried the DLS DR8 catalog instead.
To put $N_{2^{\prime}}$ and $N_{1/r}$ into context, we perform the same cuts centered on $10^5$ random points within the DLS DR7 footprint. Dividing the SLACS $N_{2^{\prime}}$ and $N_{1/r}$ by the median $\langle N_{2^{\prime}\rangle_{\rm rand}}$ and $\langle N_{1/r}\rangle_{\rm rand}$ of the calibration lines of sight allows us to assess the relative over-density of the SLACS lenses as
\begin{equation}
 \zeta_{N} \equiv \frac{N_{2^{\prime}}}{\langle N_{2^{\prime}}\rangle_{\rm rand}}
\end{equation}
and
\begin{equation}
 \zeta_{1/r} \equiv \frac{N_{1/r}}{\langle N_{1/r}\rangle_{\rm rand}}.
\end{equation}

We compare this metric on our sample with the overlapping sample of \cite{Treu:2009} where local 3-dimensional quantities in the form of $D_1$ are available, and we find good agreement between these two statistics in terms of a rank correlation.

We remove lenses that have $\zeta_{1/r} > 2.10$ within the 2 arcminutes aperture from our sample. This selection cut corresponds to $D_1 = 1.4$ Mpc$^{-3}$ for the subset by \cite{Treu:2009}.
Independently of the $\zeta_{1/r}$ cut, we check and flag all lenses within the \cite{Shajib_slacs:2020} sample that have prominent nearby perturbers present in the \textit{HST} data within 5$^{\prime\prime}$. We do not find any additional lenses with prominent nearby perturbers not already removed by the selection cut of $\zeta_{1/r} > 2.10$.

\subsubsection{Combined sample selection}
\begin{figure*}
  \centering
  \includegraphics[angle=0, width=150mm]{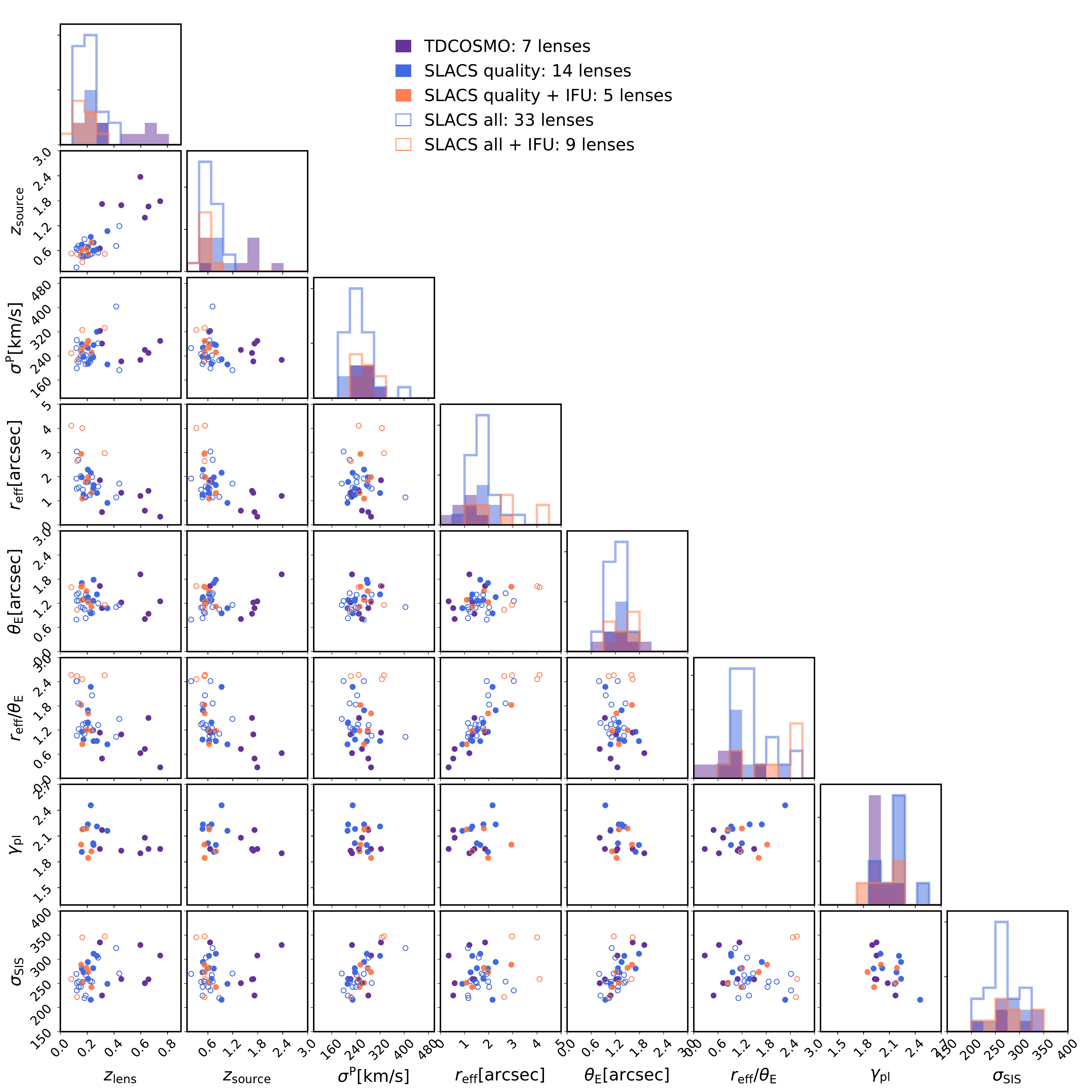}
  \caption{Sample selection of the SLACS lenses being added to the analysis and comparison with the TDCOSMO data set. The comparisons are in lens redshift, $z_{\rm lens}$, source redshift, $z_{\rm source}$, measured velocity dispersion, $\sigma^{\rm P}$, half light radius of the deflector, $r_{\rm eff}$, Einstein radius of the deflector, $\theta_{\rm E}$, the ratio of half light radius to Einstein radius, $r_{\rm eff}/\theta_{\rm E}$, and the SIS equivalent velocity dispersion estimated from the Einstein radius and a fiducial cosmology, $\sigma_{\rm SIS}$. Open dots correspond to lenses included in our selection without quality lens models. Red points correspond to SLACS lenses which have VIMOS IFU data.  \faGithub\href{https://github.com/TDCOSMO/hierarchy_analysis_2020_public/blob/6c293af582c398a5c9de60a51cb0c44432a3c598/JointAnalysis/sample_selection.ipynb}{~source}}
  \label{fig:sample_selection}
\end{figure*}

With the combined selection on the SLACS sample based on the morphology, mass proxy, local environment, and for the IFU lenses also rotation, we end up with 33 SLACS lenses of which nine lenses have IFU data. 14 lenses out of the sample have quality lens models by \cite{Shajib_slacs:2020}, including five lenses with IFU data.
Figure \ref{fig:sample_selection} shows how the individual lenses among the different samples, TDCOSMO, SLACS and the subset with IFU data are distributed in key parameters of the deflector.

We discuss possible differences between the SLACS and TDCOSMO samples and the possibility of trends within the samples impacting our analysis in a systematic way in Section \ref{sec:discussion_population_systematics} after presenting the results of the hierarchical analysis of the joint sample.
We list all the relevant measured values and uncertainties of the 33 SLACS lenses in Appendix \ref{app:slacs_data}.

\subsection{Line of sight convergence estimate} \label{sec:slacs_los_convergence}

We compute the probability for the external convergence given the relative number counts, $P(\kappa_\mathrm{ext}|\zeta_1,\zeta_{1/r})$, following \cite{Greene:2013} \citep[see e.g.,][]{Rusu:2017,Rusu:2020,Chen:2019,BuckleyGeer:2020}. In brief, we select from the Millennium Simulation \citep[MS;][]{Springel:2005} line of sights which satisfy the relative weighted number density constraints measured above, in terms of both number counts and $1/r$ weighting (Eqn. \ref{eqn:n_r_stat}). While the MS consists only of dark matter halos, we use the catalog of galaxies painted on top of these halos following the semi-analytical models of \citet{DeLuciaBlaizot:2007}. We implement the same magnitude cut, aperture radius etc. which were employed in measuring the relative weighted number densities for the SLACS lenses, in order to compute $\zeta_1,\zeta_{1/r}$ corresponding to each line of sight in the MS. We then use the $\kappa$ maps computed by \citet{Hilbert:2009} and read off the values corresponding to the location of the selected line of sight, thus constructing the $p(\kappa_\mathrm{ext}|\zeta_1,\zeta_{1/r})$  probability density function (PDF). The \citet{Hilbert:2009} maps were computed for a range of source redshift planes. Over the range spanned by the source redshifts of the SLACS lenses, there are 17 MS redshift planes, with spacing $\Delta z \sim 0.035$ - $0.095$. We used the maps best matching the source redshift of each SLACS lens. For 23 of the SLACS lenses there are available external shear measurements by \cite{Shajib_slacs:2020}, which we used, optionally, as a third constraint. Compared to previous inferences of $p(\kappa)$ for the TDCOSMO lenses, we made two computational simplifications to our analysis, in order to be able to scale our technique to the significantly larger number of lenses: 1) We did not resample from the photometry of the MS galaxies, taking into account photometric uncertainties similar to those in the observational data. A toy simulation showed that this step results in negligible differences. 2) We use only 1/8 of the lines of sight in the MS. We then checked that this results in $\Delta\kappa \lesssim 0.001$ offsets, negligible for the purpose of our analysis.

\begin{figure}
  \centering
  \includegraphics[angle=0, width=80mm]{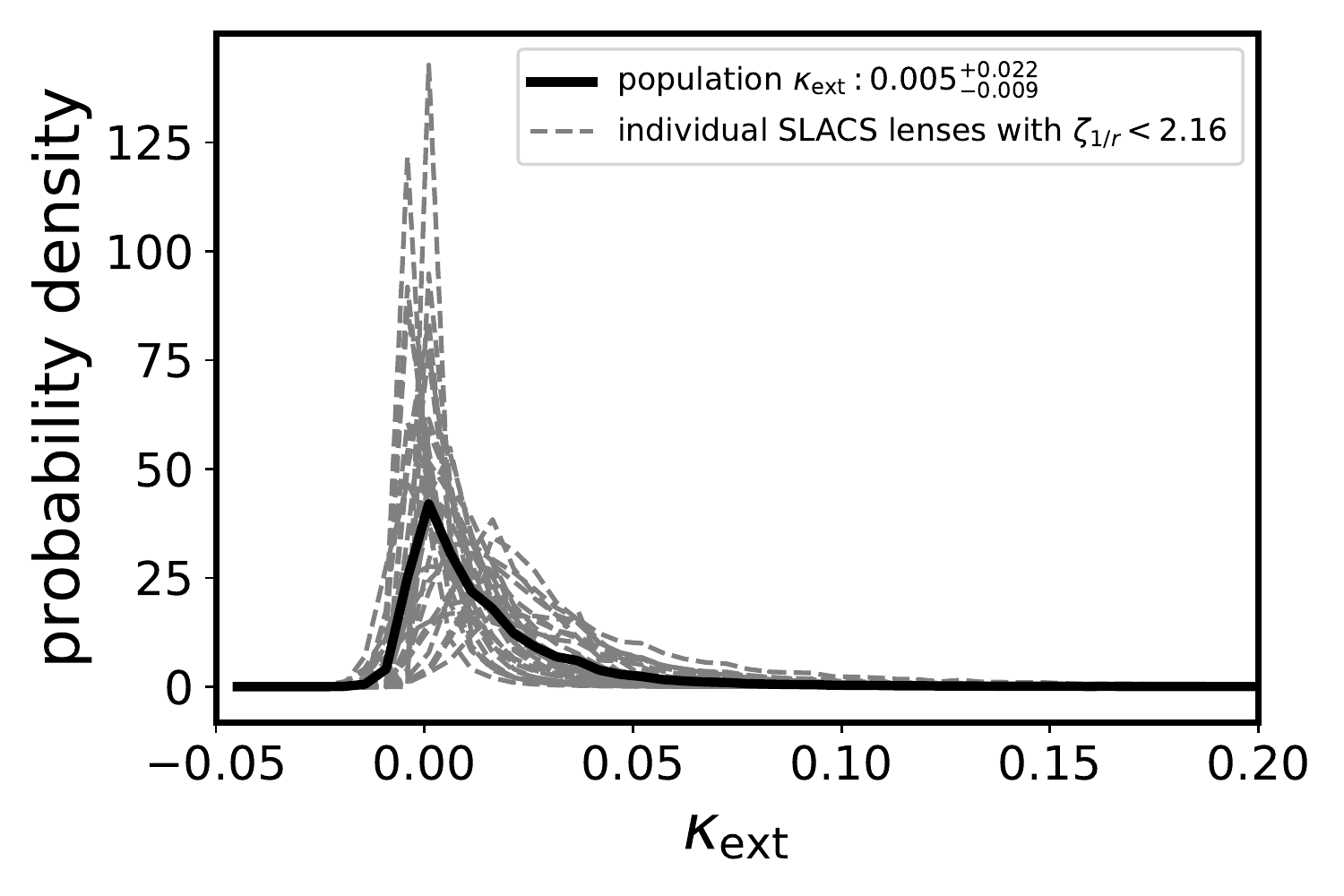}
  \caption{External convergence posteriors of the 33 SLACS lenses that pass our morphology and local environment selection cut based on the weighted number counts (gray dashed lines) and the population distribution (black solid line).
   \faGithub\href{https://github.com/TDCOSMO/hierarchy_analysis_2020_public/blob/6c293af582c398a5c9de60a51cb0c44432a3c598/JointAnalysis/sample_selection.ipynb}{~source} }
  \label{fig:kappa_ext_slacs}
\end{figure}

Figure \ref{fig:kappa_ext_slacs} shows the $p(\kappa_\mathrm{ext}|\zeta_1,\zeta_{1/r})$ distributions for the sub-selected sample based on morphology and local environment. As expected from the significantly lower source redshifts of the SLACS sample compared to the TDCOSMO lenses, most of $p(\kappa)$ PDFs for the individual lenses are very narrow and peak at $\sim$zero, with dispersion $\sim0.01$. This is because the volume is smaller and thus there are relatively few structures in the MS at these low redshifts to contribute. In fact, the relative weighted number density constraints have relatively little impact on most of the $p(\kappa)$ distributions, which resemble the PDFs for all lines of sight. Finally, we note that, while our approach to infer $p(\kappa)$ for the SLACS lenses is homogeneous, this is not the case for the TDCOSMO lenses. This is by necessity, as the environmental data we used for the TDCOSMO lenses has varied in terms of depth, number of filters and available targeted spectroscopy. Nonetheless, as we have shown through simulations by \citet{Rusu:2017,Rusu:2020}, such differences do not bias the $p(\kappa)$ inference.

\subsection{SLACS inference} \label{sec:slacs_constraints}
Here we present the hierarchical inference on the mass profile and anisotropy parameters from the selected sample of the SLACS lenses. We remind the reader that we use 33 SLACS lenses, of which 14 have imaging modeling constraints on the power-law slope $\gamma_{\rm pl}$. Nine of the lenses in our final sample have also VLT VIMOS IFU constraints in addition to SDSS spectroscopy (five of which have imaging modeling constraints on the power-law slope).
The separate inference presented in this section is meant to provide consistency checks and to gain insights into how the likelihood of the SLACS data set is going to impact the constraints on the mass profiles, and thus $H_0$, when combining with the TDCOSMO data set.

\begin{figure}
  \centering
  \includegraphics[angle=0, width=80mm]{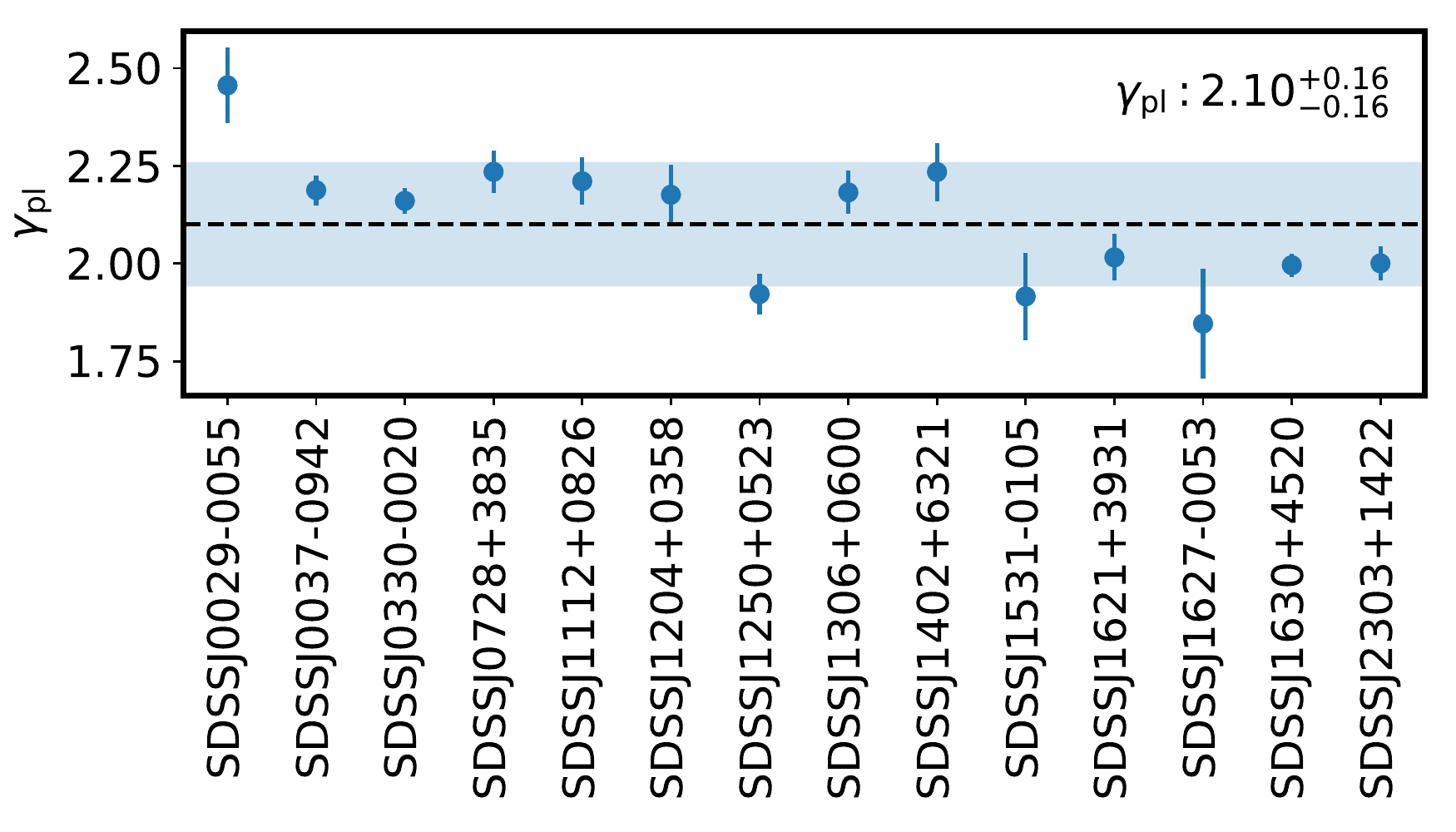}
  \caption{Power-law slope $\gamma_{\rm pl}$ inferences obtained from \textit{HST} imaging data for the 14 SLACS lenses within our selection cut from Shajib et al. 2020. We derive a population distribution from these lenses to be applied on the subset of lenses without measured $\gamma_{\rm pl}$ from imaging data. The black dashed line indicates the population mean and the blue band the 1-sigma population width based on the 14 individual measurements.
   \faGithub\href{https://github.com/TDCOSMO/hierarchy_analysis_2020_public/blob/6c293af582c398a5c9de60a51cb0c44432a3c598/JointAnalysis/sample_selection.ipynb}{~source} }
  \label{fig:pl_slope_slaces}
\end{figure}

We are making use of the marginalized posteriors in the lens model parameters of \cite{Shajib_slacs:2020} in the same way as for the TDLMC and TDCOSMO sample.
For SLACS lenses that do not have a model and parameter inference by \cite{Shajib_slacs:2020}, we use the Einstein radii measured by \cite{Auger:2009} derived from a singular isothermal ellipsoid (SIE) lens model.
For the power-law slopes of those lenses we apply the inferred Gaussian population distribution prior on $\gamma_{\rm pl}$ from the selected sample which has measured values, with $\gamma_{\rm pl, pop} = 2.10\pm0.16$. Figure \ref{fig:pl_slope_slaces} presents the imaging data inferred $\gamma_{\rm pl}$ for the 14 quality lenses selected in our sample by \cite{Shajib_slacs:2020}.

Table \ref{table:param_summary_slacs} presents the parameters and priors used in the hierarchical inference of this section. In particular, we fix the cosmology to assess constraining power and consistency with the TDCOSMO data set. We separate the inference on $\lambda_{\rm int, 0}$ of the VIMOS IFU data set from the SDSS measurements to assess systematic differences between the two data products. Further more, we use a uniform prior in $a_{\rm ani}$, $\mathcal{U}(a_{\rm ani})$, rather than a logarithmic prior $\mathcal{U}(\log(a_{\rm ani}))$, to assess and illustrate the information on the anisotropy parameter from the IFU data set.

For the analysis of the SLACS-only sample in this section, we fix the cosmological model. The cosmological dependence folds in the prediction of the velocity dispersion through the distance ratio $D_{\rm s}/D_{\rm ds}$ (Eqn. \ref{eqn:kinematics_cosmography}). This ratio is not sensitive to $H_0$ and the SLACS-only data set is not constraining $H_0$. When combining the SLACS and TDCOSMO sample in the next section, the cosmology dependence is fully taken into account.

\begin{table*}
\caption{Summary of the model parameters sampled in the hierarchical inference on the SLACS sample of Section \ref{sec:slacs_analysis}. The SLACS-only analysis is for the purpose of illustrating the constraining power on the mass profile and to assess consistencies with the TDCOSMO sample. For this purpose, we fix the cosmology to a fiducial value in the SLACS-only inference.}
\begin{center}
\begin{threeparttable}
\begin{tabular}{l l l}
    \hline
    name & prior & description \\
    \hline \hline
    Cosmology (Flat $\Lambda$CDM) \\
    $H_0$ [\Hunit] & $= 73$  & Hubble constant \\
    $\Omega_{\rm m}$ & $=0.3$ & current normalized matter density \\
    \hline
    Mass profile \\
	$\lambda_{\rm int, 0}$ & $\mathcal{U}([0.5, 1.5])$ & internal MST population mean for $r_{\rm eff}/\theta_{\rm E}=1$\\
    $\alpha_{\lambda}$ & $\mathcal{U}([-1, 1])$  & slope of $\lambda_{\rm int}$ with $r_{\rm eff}/\theta_{\rm E}$ of the deflector (Eqn. \ref{eqn:lambda_scaling}) \\
    $\sigma(\lambda_{\rm int})$ & $\mathcal{U}([0, 0.5])$ & 1-$\sigma$ Gaussian scatter in the internal MST from SDSS \\
    \hline
    Stellar kinematics \\
    $\langle a_{\rm ani}\rangle$ & $\mathcal{U}([0.1, 5])$ & scaled anisotropy radius (Eqn. \ref{eqn:r_ani}, \ref{eqn:a_ani}) \\
    $\sigma(a_{\rm ani})$ & $\mathcal{U}([0, 1])$  & $\sigma(a_{\rm ani}) \langle a_{\rm ani}\rangle$ is the 1-$\sigma$ Gaussian scatter in $a_{\rm ani}$\\
    \hline
    Normalization of IFU data \\
    $\lambda_{\rm ifu}$ & $\mathcal{U}([0.5, 1.5])$ & internal MST population constraint from IFU data \\
    $\sigma(\lambda_{\rm ifu})$ & $\mathcal{U}([0, 0.5])$ & 1-$\sigma$ Gaussian scatter in $\lambda_{\rm ifu}$  \\
    \hline
    Line of sight \\
    $\kappa_{\rm ext}$ & $p(\kappa_{\rm ext})$ of individual lenses (Fig. \ref{fig:kappa_ext_slacs}) & external convergence of lenses \\
    \hline
\end{tabular}
\begin{tablenotes}
\end{tablenotes}
\end{threeparttable}
\end{center}
\label{table:param_summary_slacs}
\end{table*}

\begin{figure*}
  \centering
  \includegraphics[angle=0, width=160mm]{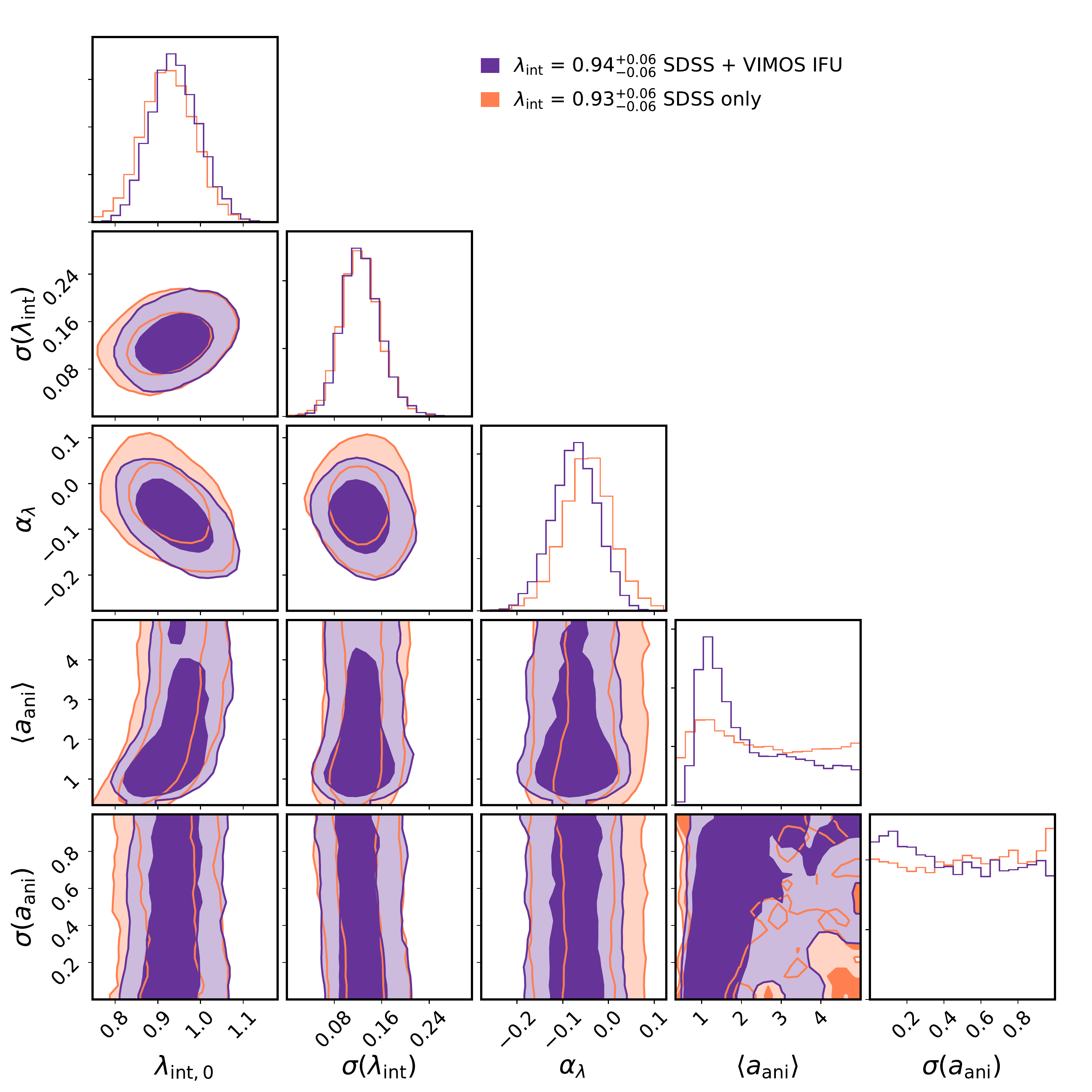}
  \caption{Posterior distribution for the SLACS sample with priors according to Table \ref{table:param_summary_slacs}. Orange: Inference with the SDSS spectra. Purple: Inference with SDSS spectra and VIMOS IFU data set. The posterior of $\lambda_{\rm int, 0}$ was blinded during the analysis.  \faGithub\href{https://github.com/TDCOSMO/hierarchy_analysis_2020_public/blob/6c293af582c398a5c9de60a51cb0c44432a3c598/SLACS_sample/SLACS_constraints.ipynb}{~source}}
\label{fig:results_slacs}
\end{figure*}

We perform two posterior inferences: one with the SDSS velocity dispersion data only, and one combining SDSS and VIMOS IFU binned dispersions. Figure \ref{fig:results_slacs} shows the two different posteriors. The constraints on $\lambda_{\rm int}$ (parameters $\lambda_{\rm int, 0}$, $\alpha_{\lambda}$, $\sigma(\lambda_{\rm int})$) come for all three cases entirely from the kinematics of the SDSS measurements.

All the parameters are statistically consistent with each other and the TDCOSMO analysis of Section \ref{sec:tdcosmo_analysis} except the posterior in the scatter of the internal MST, $\sigma(\lambda_{\rm int})$. The TDCOSMO constraints of $\sigma(\lambda_{\rm int})$ are consistent with zero scatter in the mass profile parameter and 2-sigma bound at 0.1, while the inference of the SLACS sample results in a larger scatter.
An underestimation of uncertainties in the velocity dispersion measurements, if not accounted for in the analysis,  will directly translate to an increase in $\sigma(\lambda_{\rm int})$.
We point out the excellent agreement of the anisotropy distribution with the TDLMC Rung3 hydrodynamical simulations (Section \ref{sec:tdlmc}).

\section{Hierarchical analysis of TDCOSMO+SLACS} \label{sec:joint_analysis}
We describe now the final and most stringent analysis of this work, obtained by combining the analysis of the TDCOSMO lenses, presented in Section \ref{sec:tdcosmo_analysis}, and that of the SLACS sample, presented in Section \ref{sec:slacs_analysis}.
The parameterization and priors have been validated on the TDLMC mock data set in Section \ref{sec:tdlmc}. We remind the reader that the choices of the analyses are identical and thus we can combine the TDCOSMO and SLACS sample on the likelihood level. We define the parameterization and priors of our hierarchical model in Section \ref{sec:joint_param} and present the result and the $H_0$ measurement in Section \ref{sec:joint_results}.

\subsection{Parameterization and priors} \label{sec:joint_param}
For our final $H_0$ measurement, we assume a flat $\Lambda$CDM cosmology with uniform prior in $H_0$ in [0, 150] \Hunit and a narrow prior on $\Omega_{\rm m}$ with $\mathcal{N}(\mu=0.298, \sigma=0.022)$ from the Pantheon SNIa sample \citep[][see Section \ref{sec:cosmo_param}]{Scolnic:2018}.
For $\lambda_{\rm int}$, we assume an identical distribution for the selected population of the SLACS lenses and the TDCOSMO sample for the scaling in $r_{\rm eff}/\theta_{\rm E}$ (Eqn. \ref{eqn:lambda_scaling}). We also assume the same stellar anisotropy population distributions for the SLACS and TDCOSMO lenses.
To account for potential systematics in the VIMOS IFU measurement (see Section \ref{sec:ifu_data_description}), we introduce a separate a separate internal MST distribution $\lambda_{\rm ifu}$, effectively replacing $\lambda_{\rm int}$ when fitting the IFU data. This approach allows us to use the anisotropy constraints from the IFU data while not requiring a perfect absolute calibration of the measurements. For the external convergence we use the individual $p(\kappa_{\rm ext})$ distributions from the two samples.

As discussed in Section \ref{sec:slacs_analysis}, there is an inconsistency in the inferred spread in the $\lambda_{\rm int}$ distribution between the SLACS and TDCOSMO sample. We attribute this inconsistency to uncertainties that were not accounted for in the velocity dispersion measurements of the SDSS data products. In our joint analysis, we add a parameter that describes an additional relative uncertainty in the velocity dispersion measurements, $\sigma_{\sigma^{\rm P}, {\rm sys}}$, such that the total uncertainty in the velocity dispersion measurements is the square of the quoted measurement uncertainty plus this unaccounted term,
\begin{equation} \label{eqn:sigma_sys}
 \sigma^2_{\sigma^{\rm P}, {\rm tot}} = \sigma^2_{\sigma^{\rm P}, {\rm measurement}}+ (\sigma^{P}\sigma_{\sigma^{\rm P}, {\rm sys}})^2.
\end{equation}
$\sigma_{\sigma^{\rm P}, {\rm sys}}$ is the same for all the SDSS measured velocity dispersions.
Table \ref{table:param_summary_joint} presents all the parameters being fit for, including their priors, in our joint analysis of the SLACS and TDCOSMO sample. \footnote{The notebooks are publicly available and we facilitate the use of different priors and cosmological models. All choices presented here are made blindly in regard to $H_0$.}

\begin{table*}
\caption{Summary of the model parameters sampled in the hierarchical inference on the TDCOSMO+SLACS sample.}
\begin{center}
\begin{threeparttable}
\begin{tabular}{l l l}
    \hline
    name & prior & description \\
    \hline \hline
    Cosmology (Flat $\Lambda$CDM) \\
    $H_0$ [\Hunit] & $\mathcal{U}([0, 150])$  & Hubble constant \\
    $\Omega_{\rm m}$ & $\mathcal{N}(\mu=0.298, \sigma=0.022)$ & current normalized matter density \\
    \hline
    Mass profile \\
    $\lambda_{\rm int, 0}$ & $\mathcal{U}([0.5, 1.5])$ & internal MST population mean \\
    $\alpha_{\lambda}$ & $\mathcal{U}([-1, 1])$  & slope of $\lambda_{\rm int}$ with $r_{\rm eff}/\theta_{\rm E}$ of the deflector (Eqn. \ref{eqn:lambda_scaling}) \\
    $\sigma(\lambda_{\rm int})$ & $\mathcal{U}(\log([0.001, 0.5]))$ & 1-$\sigma$ Gaussian scatter in the internal MST \\
    \hline
    Normalization of IFU data \\
    $\lambda_{\rm ifu}$ & $\mathcal{U}([0.5, 1.5])$ & internal MST population constraint from IFU data \\
    $\sigma(\lambda_{\rm ifu})$ & $\mathcal{U}(\log([0.01, 0.5]))$ & 1-$\sigma$ Gaussian scatter in $\lambda_{\rm ifu}$  \\
    \hline
    Stellar kinematics \\
    $\langle a_{\rm ani}\rangle$ & $\mathcal{U}(\log(a_{\rm ani}))$ for $a_{\rm ani}$ in [0.1, 5] & scaled anisotropy radius (Eqn. \ref{eqn:r_ani}, \ref{eqn:a_ani}) \\
    $\sigma(a_{\rm ani})$ & $\mathcal{U}(\log([0.01, 1]))$  & $\sigma(a_{\rm ani}) \langle a_{\rm ani}\rangle$ is the 1-$\sigma$ Gaussian scatter in $a_{\rm ani}$\\
    $\sigma_{\sigma^{\rm P}, {\rm sys}}$ & $\mathcal{U}(\log([0.01, 0.5]))$ & systematic uncertainty on $\sigma^{\rm P}_{\rm SDSS}$ measurements (Eqn. \ref{eqn:sigma_sys}) \\
    \hline
    Line of sight \\
    $\kappa_{\rm ext}$ & $p(\kappa_{\rm ext})$ of individual lenses (Fig. \ref{fig:kappa_ext_tdcosmo} \& \ref{fig:kappa_ext_slacs}) & external convergence of lenses \\
    \hline
\end{tabular}
\begin{tablenotes}
\end{tablenotes}
\end{threeparttable}
\end{center}
\label{table:param_summary_joint}
\end{table*}

\subsection{Results} \label{sec:joint_results}
Here we present the posteriors of the joint hierarchical analysis of 33 SLACS lenses (nine of which have IFU data) and the seven quasar time-delay TDCOSMO lenses for the parameterization and priors described in Table \ref{table:param_summary_joint}.
To trace back information to specific data sets, we sample different combinations of the TDCOSMO and SLACS data sets under the same priors.
The TDCOSMO-only inference was already presented in Section \ref{sec:tdcosmo_analysis} and results in $H_0 = $ \Htdcosmo~\Hunit. 
Besides the TDCOSMO-only result, we perform the inference for the TDCOSMO+SLACS$_{\rm IFU}$ data set, effectively allowing anisotropy constraints being used on top of the TDCOSMO data set, resulting in $H_0 = $ \Htdcosmoifu~\Hunit; the TDCOSMO+SLACS$_{\rm SDSS}$ data set, using the SLACS lenses with their SDSS spectroscopy to inform the analysis, results in $H_0 = $ \HtdcosmoSLACS~\Hunit. For our final inference of this work of the joint data sets of TDCOSMO+SLACS$_{\rm SDSS + IFU}$, we measure $H_0 = $ \Hjoint~\Hunit.

\begin{figure*}
  \centering
  \includegraphics[angle=0, width=160mm]{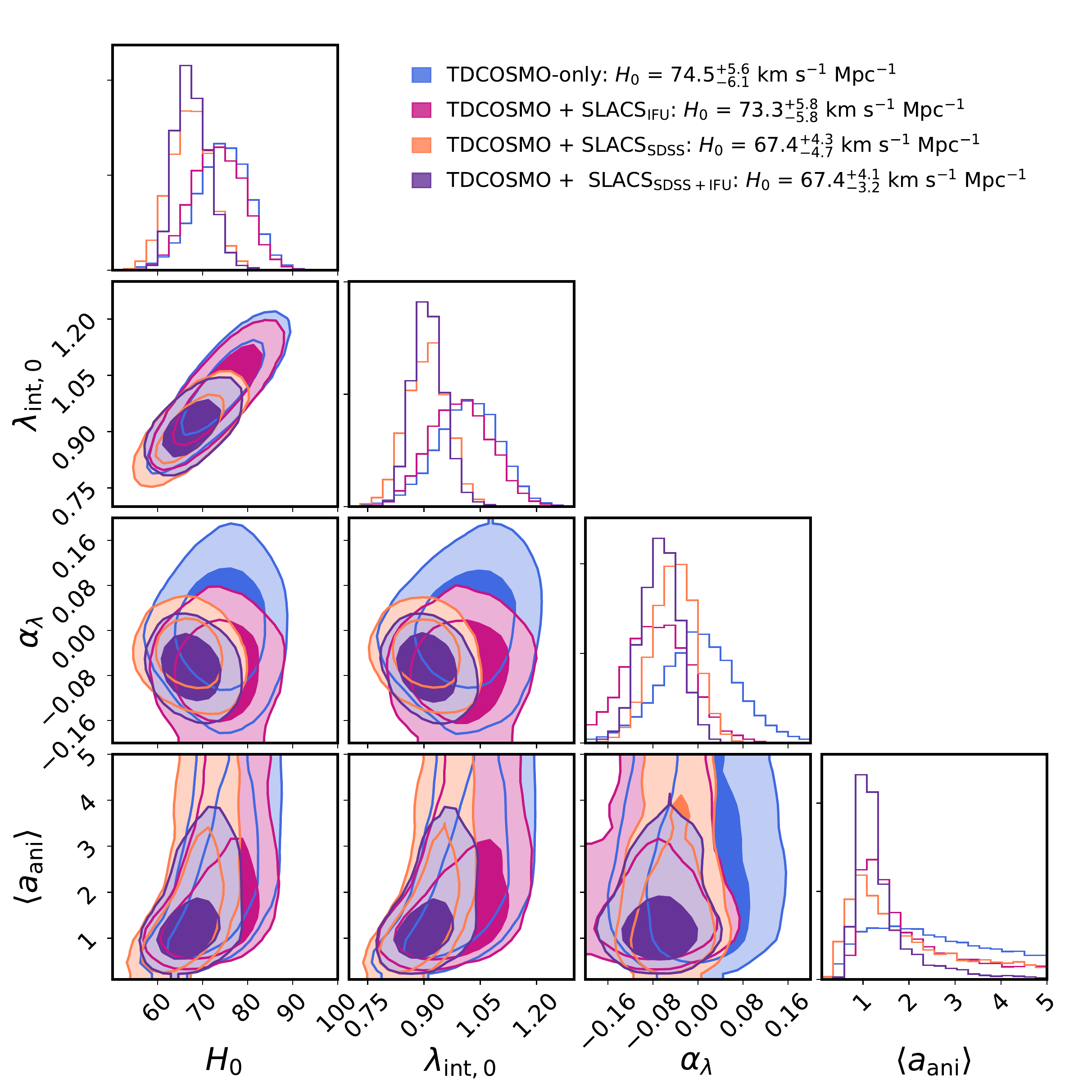}
  \caption{Posterior distributions of the key parameters for the hierarchical inference. Blue: constraints from the TDCOSMO-only sample. Violet: constraints with the addition of IFU data of nine SLACS lenses to inform the anisotropy prior on the TDCOSMO sample, TDCOSMO+SLACS$_{\rm IFU}$. Orange: constraints with a sample of 33 additional lenses with imaging and kinematics data (\textit{HST} imaging + SDSS spectra) from the SLACS sample, TDCOSMO+SLACS$_{\rm SDSS}$. Purple: Joint analysis of TDCOSMO and 33 SLACS lenses with SDSS spectra of which nine have VIMOS IFU data, TDCOSMO+SLACS$_{\rm SDSS+IFU}$. Priors are according to Table \ref{table:param_summary_joint}. The 68th percentiles of the 1D marginalized posteriors are presented in Table \ref{table:results}. The posteriors in $H_0$ and $\lambda_{\rm int, 0}$ were held blinded during the analysis.  \faGithub\href{https://github.com/TDCOSMO/hierarchy_analysis_2020_public/blob/6c293af582c398a5c9de60a51cb0c44432a3c598/JointAnalysis/joint_inference.ipynb}{~source}}
\label{fig:results}
\end{figure*}

Figure \ref{fig:results} presents the key parameter posteriors of the TDCOSMO-only, TDCOSMO+SLACS$_{\rm IFU}$, TDCOSMO+SLACS$_{\rm SDSS}$, and the TDCOSMO+SLACS$_{\rm SDSS+IFU}$ analyses. Not shown on the plot are the $\Omega_{\rm m}$ posteriors (effectively identical to the prior), the $\sigma_{\sigma^{\rm P}, {\rm sys}}$ posteriors for the SDSS kinematics measurements, the distribution scatter parameters $\sigma(\lambda_{\rm int}$ and $\sigma(a_{\rm ani})$, and the IFU calibration nuisance parameter $\lambda_{\rm ifu}$. All the one-dimensional marginalized posteriors, except for the nuisance parameter $\lambda_{\rm ifu}$, of the different combinations of the data sets are provided in Table \ref{table:results}.

\begin{table*}
\caption{Marginalized posteriors of our hierarchical Bayesian cosmography inference based on the priors and parameterization specified in Table \ref{table:param_summary_joint} for a flat $\Lambda$CDM cosmology.}
\begin{center}
\begin{threeparttable}
\begin{tabular}{l l l l l l l l }
    \hline
    Data sets & $H_0$ [\Hunit]  & $\lambda_{\rm int, 0}$ & $\alpha_{\lambda}$ & $\sigma(\lambda_{\rm int})$ & $a_{\rm ani}$ & $\sigma(a_{\rm ani})$  & $\sigma_{\sigma^{\rm P}, {\rm sys}}$\\
    \hline \hline
    TDCOSMO-only & ${74.5}_{-6.1}^{+5.6}$ & ${1.02}_{-0.09}^{+0.08}$ & ${0.00}_{-0.07}^{+0.07}$ & ${0.01}_{-0.01}^{+0.03}$ & ${2.32}_{-1.17}^{+1.62}$ & ${0.16}_{-0.14}^{+0.50}$ & -  \\ 
TDCOSMO + SLACS$_{\rm IFU}$ & ${73.3}_{-5.8}^{+5.8}$ & ${1.00}_{-0.08}^{+0.08}$ & ${-0.07}_{-0.06}^{+0.06}$ & ${0.07}_{-0.05}^{+0.09}$ & ${1.58}_{-0.54}^{+1.58}$ & ${0.15}_{-0.13}^{+0.47}$ & -  \\ 
TDCOSMO + SLACS$_{\rm SDSS}$ & ${67.4}_{-4.7}^{+4.3}$ & ${0.91}_{-0.06}^{+0.05}$ & ${-0.04}_{-0.04}^{+0.04}$ & ${0.02}_{-0.01}^{+0.04}$ & ${1.52}_{-0.70}^{+1.76}$ & ${0.28}_{-0.25}^{+0.45}$ & ${0.06}_{-0.02}^{+0.02}$  \\ 
TDCOSMO + SLACS$_{\rm SDSS + IFU}$ & ${67.4}_{-3.2}^{+4.1}$ & ${0.91}_{-0.04}^{+0.04}$ & ${-0.07}_{-0.04}^{+0.03}$ & ${0.06}_{-0.04}^{+0.08}$ & ${1.20}_{-0.27}^{+0.70}$ & ${0.18}_{-0.15}^{+0.50}$ & ${0.06}_{-0.02}^{+0.02}$  \\ 
    \hline
\end{tabular}
\begin{tablenotes}
\end{tablenotes}
\end{threeparttable}
\end{center}
\label{table:results}
\end{table*}

\begin{figure*}%
    \centering
    \subfloat[Fit to the time-delay distance]{{\includegraphics[width=5cm]{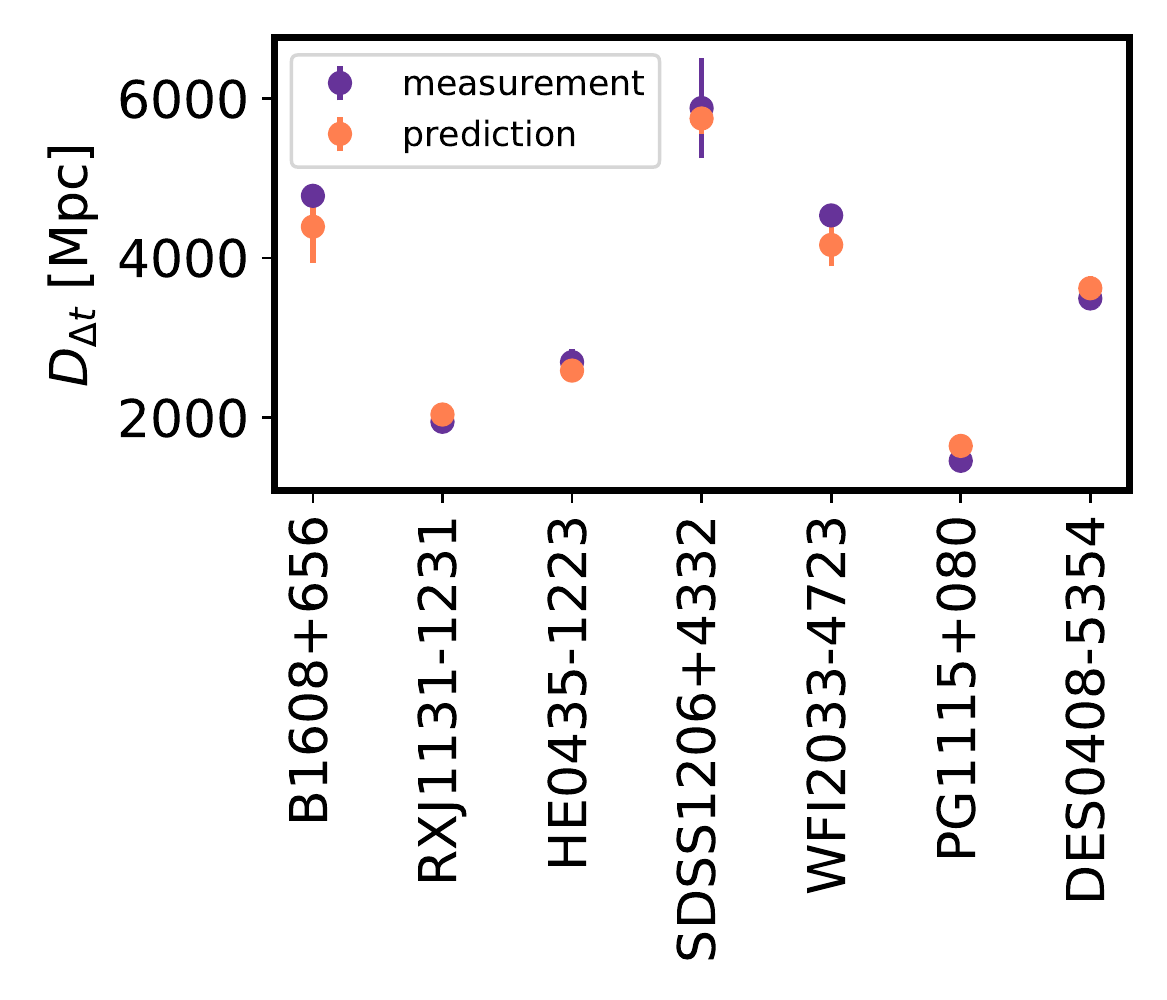} }}%
    \qquad
    \subfloat[Fit of velocity dispersion]{{\includegraphics[width=5cm]{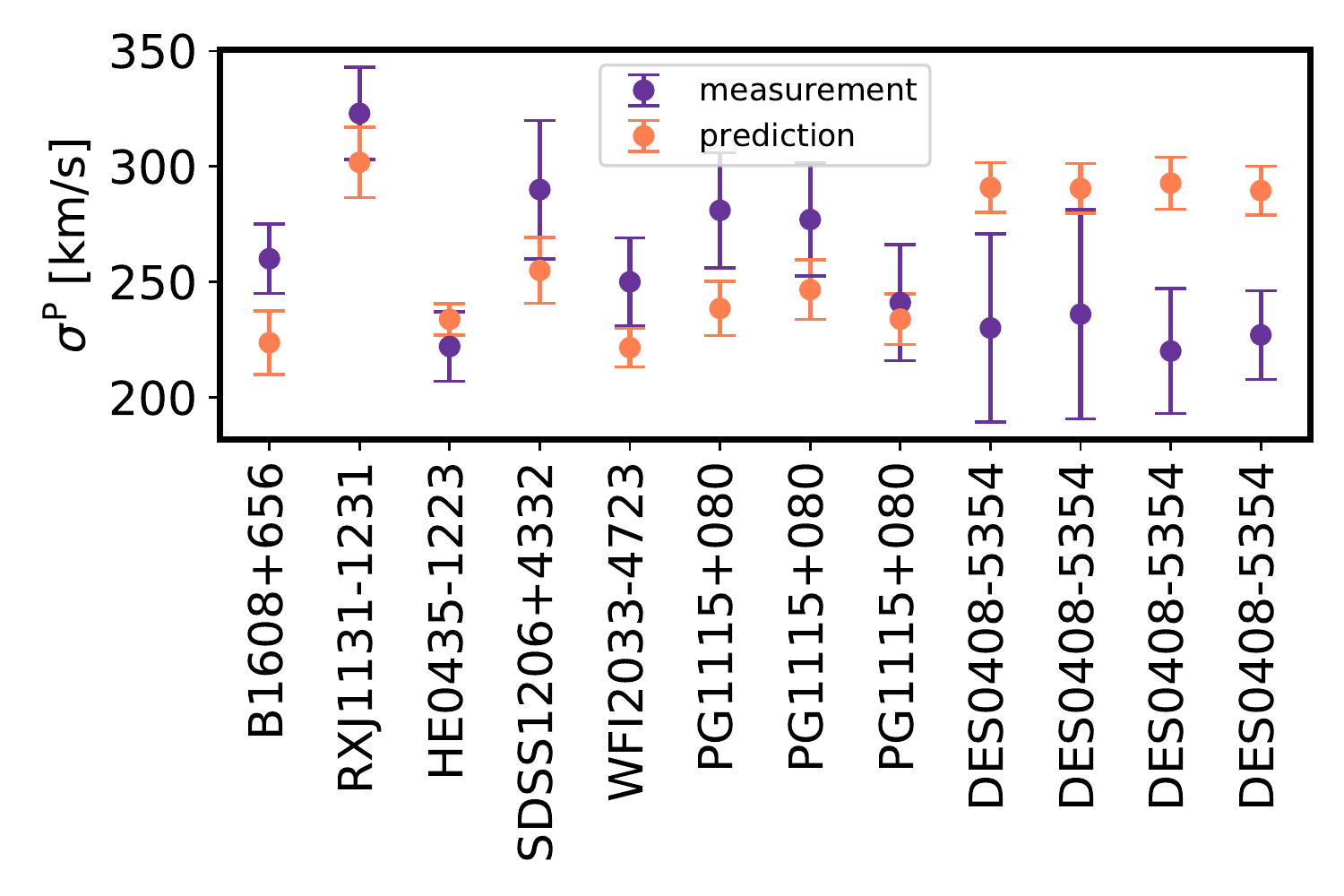} }}%
    \caption{Illustration of the goodness of the fit of the maximum likelihood model of the joint analysis in describing the TDCOSMO data set. Blue points are the measurements with the diagonal elements of the measurement covariance matrix. Orange points are the model predictions with the diagonal elements of the model covariance uncertainties. \textbf{Left:} Comparison of measured time-delay distance from imaging data and time delays compared with the predicted value from the cosmological model, the internal and external MST (and their distributions). \textbf{Right:} Comparison of the velocity dispersion measurements and the predicted values. In addition to the MST terms, the uncertainty in the model also includes the uncertainty in the anisotropy distribution $a_{\rm ani}$. For lenses with multiple velocity dispersion measurements, the diagonal terms in the error covariance are illustrated.  \faGithub\href{https://github.com/TDCOSMO/hierarchy_analysis_2020_public/blob/6c293af582c398a5c9de60a51cb0c44432a3c598/JointAnalysis/joint_inference.ipynb}{~source}}%
    \label{fig:tdcosmo_best_fit}%
\end{figure*}

\begin{figure*}
  \centering
  \includegraphics[angle=0, width=160mm]{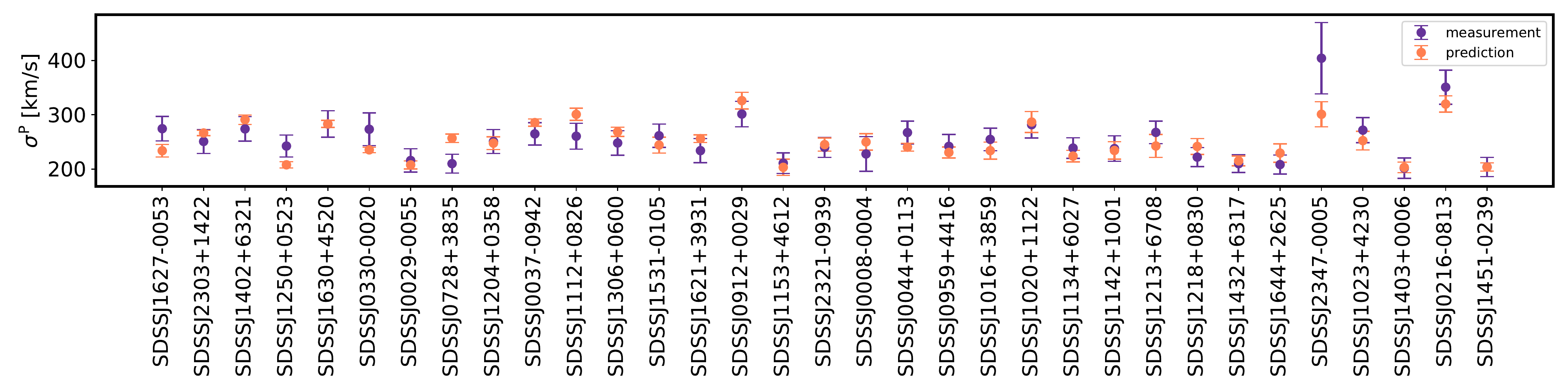}
  \caption{Illustration of the goodness of the fit of the maximum likelihood model of the joint analysis in describing the SDSS velocity dispersion measurements of the 34 SLACS lenses in our sample. Blue points are the measurements with the diagonal elements of the measurement covariance matrix. Orange points are the model predictions with the diagonal elements of the model covariance uncertainties. The measurement uncertainties include the uncertainties in the quoted measurements and the additional uncertainty of $\sigma_{\sigma^{\rm P}, {\rm sys}}$. The model uncertainties include the lens model uncertainties and the marginalization over the $\lambda_{\rm int}$ and $a_{\rm ani}$ distribution.  \faGithub\href{https://github.com/TDCOSMO/hierarchy_analysis_2020_public/blob/6c293af582c398a5c9de60a51cb0c44432a3c598/JointAnalysis/joint_inference.ipynb}{~source}}
\label{fig:sdss_best_fit}
\end{figure*}

\begin{figure*}
  \centering
  \includegraphics[angle=0, width=160mm]{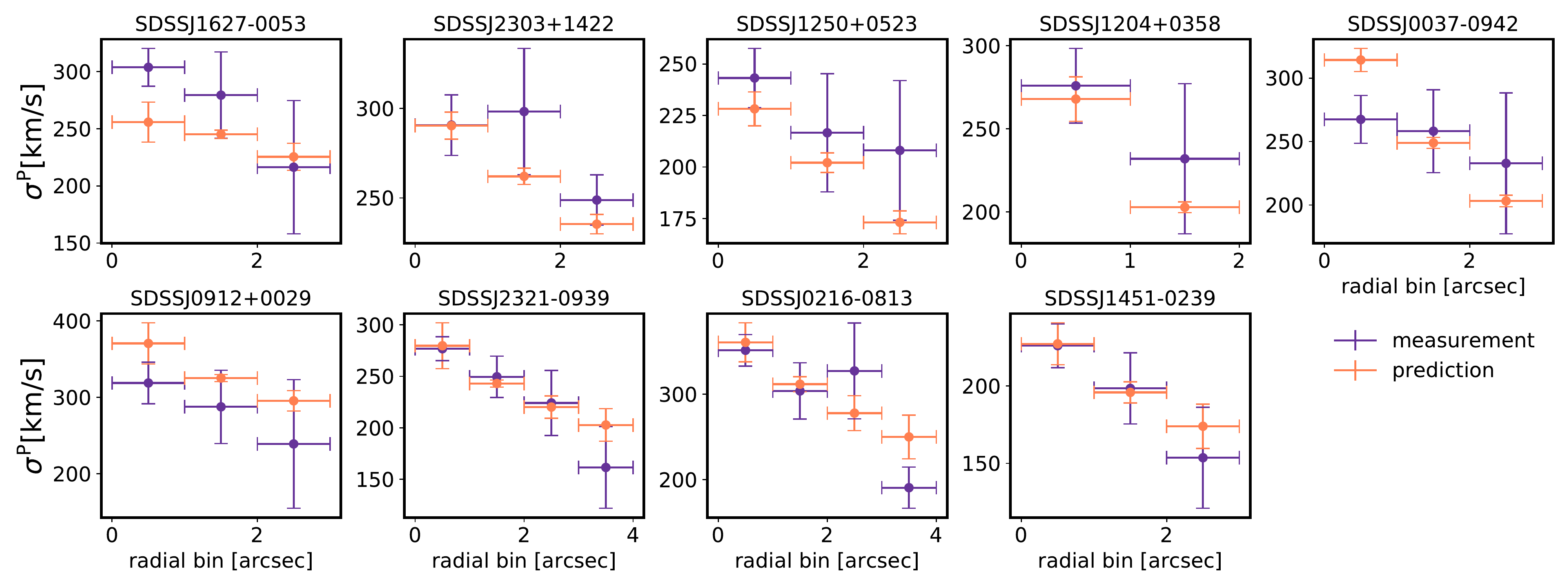}
  \caption{Illustration of the goodness of the fit of the maximum likelihood model of the joint analysis in describing the VIMOS radially binned IFU velocity dispersion measurements of the nine SLACS lenses with VIMOS data in our sample. Blue points are the measurements with the diagonal elements of the measurement covariance matrix. Orange points are the model predictions with the diagonal elements of the model covariance uncertainties. The measurement uncertainties include the uncertainties in the quoted measurements and the additional uncertainty of $\sigma_{\sigma^{\rm P}, {\rm sys}}$. The model uncertainties include the lens model uncertainties and the marginalization over the $\lambda_{\rm int}$ and $a_{\rm ani}$ distribution.  \faGithub\href{https://github.com/TDCOSMO/hierarchy_analysis_2020_public/blob/6c293af582c398a5c9de60a51cb0c44432a3c598/JointAnalysis/joint_inference.ipynb}{~source}}
\label{fig:ifu_best_fit}
\end{figure*}

We compare the best fit model prediction of the joint TDCOSMO+SLACS$_{\rm SDSS+IFU}$ inference to the time-delay distance and kinematics of the TDCOSMO data set in Figure \ref{fig:tdcosmo_best_fit}, to the SDSS velocity dispersion measurements in Figure \ref{fig:sdss_best_fit} and to the IFU data set in Figure \ref{fig:ifu_best_fit}. The model prediction uncertainties include the population distributions in $\lambda_{\rm int}$ and $a_{\rm ani}$ and the measurement uncertainty in the SDSS and VIMOS velocity dispersion uncertainties include the inferred $\sigma_{\sigma^{\rm P}, {\rm sys}}$ uncertainty.

\begin{figure*}
  \centering
  \includegraphics[angle=0, width=160mm]{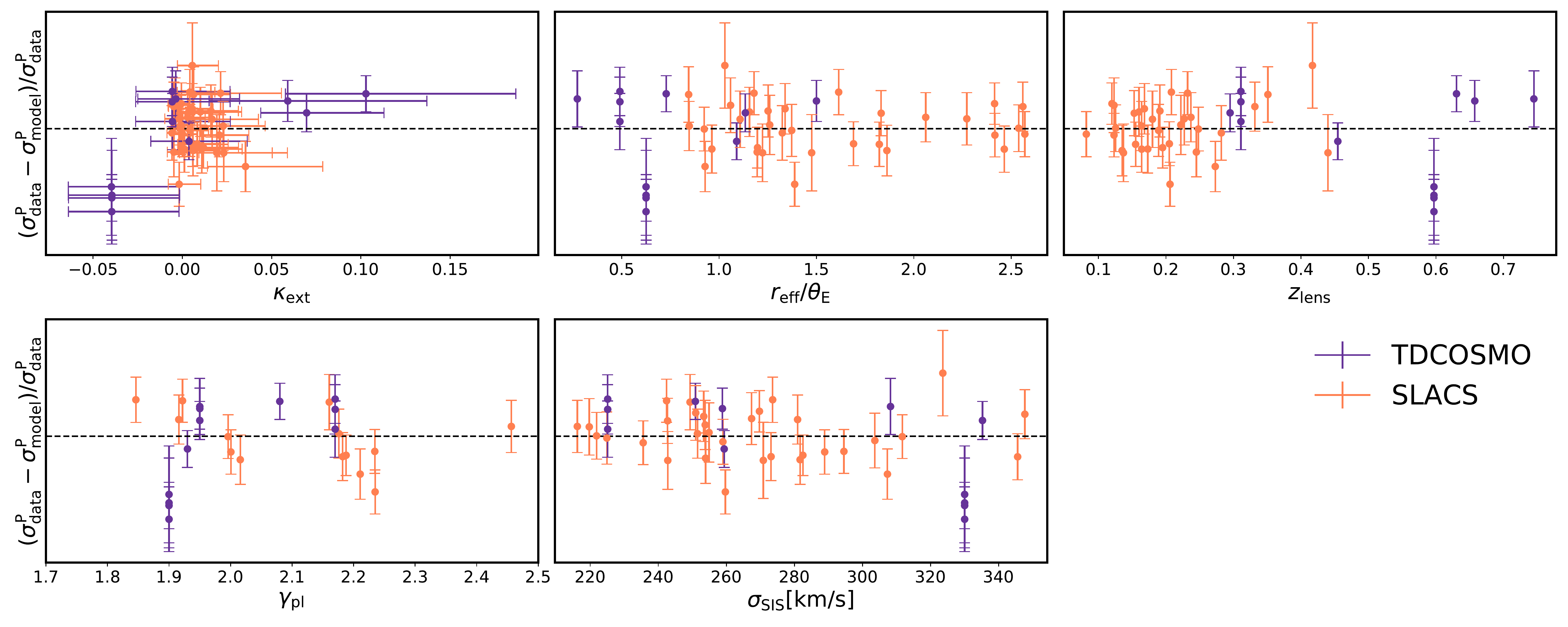}
  \caption{Relative difference of the measured vs the predicted velocity dispersion for the SLACS and TDCOSMO sample as a function of different parameters associated with the line-of-sight and the lensing galaxy. In particular, these are the relative inverse distance weighted over density $\zeta_{1/r}$, the ratio of half-light radius to Einstein radius $r_{\rm eff}/\theta_{\rm E}$, lens redshift $z_{\rm lens}$, and SIS equivalent velocity dispersion $\sigma_{\rm SIS}$.  \faGithub\href{https://github.com/TDCOSMO/hierarchy_analysis_2020_public/blob/6c293af582c398a5c9de60a51cb0c44432a3c598/JointAnalysis/joint_inference.ipynb}{~source}}
\label{fig:kin_fit_trends}
\end{figure*}

In Figure \ref{fig:kin_fit_trends} we assess trends in the fit of the kinematic data in regards to lensing deflector properties. We see that with the $r_{\rm eff}/\theta_{\rm E}$ scaling by $\alpha_{\lambda}$ (Eqn. \ref{eqn:lambda_scaling}) we can remove systematic trends in model predictions. We do not find statistically significant remaining trends in our data set beyond the ones explicitly parameterized and marginalized over.

\section{Discussion \protect\footnote{This section, with the exception of Section \ref{sec:tension_status}, was written before the results of the combined TDCOSMO+SLACS analysis were known to the authors and, thus, reflect the assessment of uncertainties present in our analysis agnostic to its outcome.}} \label{sec:discussion}

In this section, we discuss the interpretation of our measurement of $H_0$, the robustness of the uncertainties, and present an avenue for further improvements in the precision while maintaining accuracy.
We first summarize briefly the key assumptions of this work, and give a physical interpretation of the results (Section \ref{sec:meaning}). Second, we estimate the contribution of each individual assumption and dataset to the total error budget of the current analysis on $H_0$ (Section \ref{sec:joint_error_budget}). Third, we discuss specific aspects of the analysis that need further investigations to maintain accuracy with increased precision in Section \ref{sec:systematics_discussion}. Fourth, in Section \ref{sec:future} we present the near future prospects for collecting data sets and revising the analysis to increase further the precision on $H_0$ with strong lensing time-delay cosmography. Finally, in Section \ref{sec:tension_status}, we compare and discuss the $H_0$ measurement of this work with previous work by the TDCOSMO collaboration.

\subsection{Physical interpretation of the result}
\label{sec:meaning}
While consistent with the results of \citet{Wong:2020, Millon:2020}, our inference of $H_0$ has significantly lower precision for the TDCOSMO sample, even with the addition of external datasets from SLACS. The larger uncertainty was expected and is a direct result of relaxing the assumptions on the mass profile.
By introducing a mass-sheet degeneracy parameter, we add the maximal degree of freedom in $H_0$ while having minimal constraining power by lensing data on their own. This is the most conservative approach when adding a single degree of freedom in our analysis.
While mathematically this result is clearly understood, it is worth discussing the physical interpretation of this choice.

If we had perfect cosmological numerical simulations or perfect knowledge of the internal mass distribution within elliptical galaxies, we would not have to worry about the internal MST. The approach chosen by our collaboration \citep{Wong:2020,Shajib:2019,Millon:2020} was to assume physically motivated mass profiles with degrees of freedom in their parameters. In particular, the collaboration used two different mass profiles, a power-law elliptical mass profile, and a composite mass profile separating the luminous component (with fixed mass-to-light ratio) and a dark component described as a NFW profile.
The good fit to the data, the small pixellated corrections on the profiles from the first lens system \citep{Suyu:2010}, and the good agreement of $H_0$ inferred with the two mass profiles was a positive sanity check on the result \citep{Millon:2020}.

In this paper we have taken a different viewpoint, and asked how much can the mass profiles depart from a power-law and still be consistent with the data. By phrasing the question in terms of the MST we can conveniently carry out the calculations, because the MST leaves the lensing observables unchanged and therefore it corresponds to minimal constraints and assumptions, and thus maximal uncertainties with one additional degree of freedom. However, after the inference, one has to examine the inferred MST transformed profile and evaluate it in comparison with existing and future data to make sure it is realistic. We know that the exact MST cannot be the actual answer because profiles have to go zero density at large radii, but the approximate MST discussed in Section~\ref{sec:individual_lens} provides a convenient interpretation with the addition of a cored mass component.

\begin{figure*}
  \centering
  \includegraphics[angle=0, width=160mm]{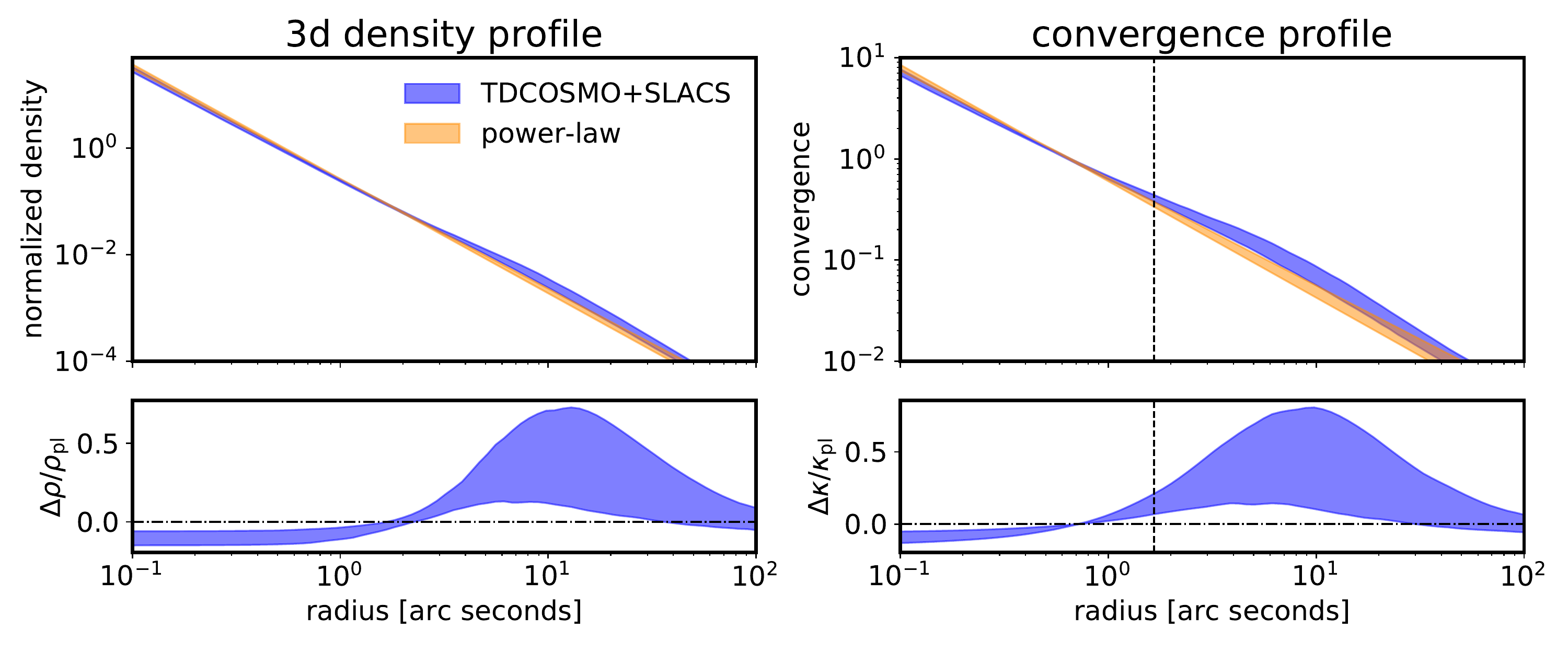}
  \caption{Illustration of the inferred mass profile of the joint TDCOSMO+SLACS$_{\rm SDSS+IFU}$ analysis. A pure power-law with $\gamma_{\rm pl}=2.10 \pm 0.05$ is shown in orange. In blue is the result of this work of $\lambda_{\rm int} = 0.91\pm 0.045$ when interpreted as a cored mass component with $R_{\rm c}$ uniform in $[3^{\prime\prime}, 10^{\prime\prime}]$. Three dimensional density are illustrated on the left and the lensing convergence on the right. The dashed vertical line on the right panels indicates the Einstein radius. Relative difference in respect to the power-law model are presented in the bottom panels.  \faGithub\href{https://github.com/TDCOSMO/hierarchy_analysis_2020_public/blob/6c293af582c398a5c9de60a51cb0c44432a3c598/MST_impact/MST_pl_cored.ipynb}{~source}}
\label{fig:pl_mst_inferred}
\end{figure*}

Figure \ref{fig:pl_mst_inferred} illustrates a cored mass component approximating the MST inferred from this work, $\lambda_{\rm int} = 0.91\pm 0.04$, in combination with a power-law model inferred from the population mean of the SLACS analysis by \cite{Shajib_slacs:2020}.
The analysis presented here guarantees that the inferred mass profile is consistent with the properties of TDCOSMO and SLACS lenses. We discuss below how additional data may allow us to constrain the models even further and thus reduce the overall uncertainty while keeping the assumptions at a minimum.

\subsection{Statistical error budget and known systematics} \label{sec:joint_error_budget}
The total error budget of 5\% on $H_0$ in our combined TDCOSMO+SLACS analysis can be traced back to specific aspects of the data and the uncertainties in the model components/assumptions. Fixing $\lambda_{\rm int}$ to a single-valued number (i.e., $\lambda_{\rm int} = 1$) is equivalent to assuming a power-law profile
and leads to an uncertainty in $H_0$ of 2\% \citep{Millon:2020}. By subtracting in quadrature 2\% from our total uncertainty, we estimate that the total error contribution of the MST ($\lambda_{\rm int}$) to the error budget is 4.5\%.
Once the MST is introduced, the uncertainty in the mass profile is dominated by uncertainties in the measurement and modeling assumptions of the velocity dispersion. 
The statistical constraints on the combined velocity dispersion measurements of 33 SLACS lenses with SDSS spectroscopy, accounting for the $\sigma_{\sigma^{\rm P}, {\rm sys}}$ contribution, and the TDCOSMO spectroscopic data set contribute 3\% to the total error budget. The remaining 3.5\% error contribution (in quadrature) to the total $H_0$ error budget arises in equal parts from the uncertainty in the anisotropy prior distribution ($\langle a_{\rm ani}\rangle$, $\sigma(a_{\rm ani})$) and the MST dependence with $r_{\rm eff}/\theta_{\rm E}$ ($\alpha_{\lambda}$).
The uncertainty in the line-of-sight selection effect of the SLACS sample contributes a statistical uncertainty smaller than 0.5\%. We note that an overall unaccounted-for shared $\kappa_{\rm ext}$ term of the ensemble of lenses in our sample would be mitigated through our MST parameterization and thus not affect our $H_0$ inference.

\subsection{Unaccounted-for systematics} \label{sec:systematics_discussion}

Our framework is conservative in the sense that it imposes minimal assumptions of the mass profile in regards to $H_0$.
Furthermore, the methods presented here have been internally reviewed and validated on the hydrodynamical simulations used in the TDLMC \citep{Ding_tdlmc2018, Ding:2020} (Section \ref{sec:tdlmc}). Despite the known limitations of current numerical simulations at the sub-kpc scale, the blind validation on external data corroborates our methodology. In this section, we discuss aspects of our analysis that are not part of our validation scheme. In particular, we discuss uncertainties and potential systematics in the kinematics measurements and selection effects of the different lens samples used in this work. At the current level of precision, these are all subdominant effects, but they may be relevant as we further increase the precision.

\subsubsection{Uncertainties in the kinematics measurement and modeling}
Under the assumptions of this analysis, aperture stellar kinematic measurements drive the overall precision by providing the information needed to mitigate the MST. Given its crucial role, we highlight here the limitations of our kinematic treatment, in order to point the way to further improvements. First, we used a heterogeneous set of stellar velocity dispersions. The TDCOSMO measurements are based on large telescope high-quality data and were the subject of extensive tests to assess systematic measurements, sometimes through repeated measurements. The nominal uncertainties are thus accurate, resulting in the internal consistency of all the TDCOSMO systems with a scatter on $\lambda_{\rm int}$ consistent with zero\footnote{This statement has been tested with a flat prior on $\sigma(\lambda_{\rm int})$.}.

The SLACS-only analysis with the reported uncertainties of the stellar velocity dispersions leads to an inferred scatter in $\lambda_{\rm int}$ of about 10\%.
Assuming the same scatter in $\lambda_{\rm int}$ among the TDCOSMO and SLACS lenses, the discrepancy in the inferred $\sigma(\lambda_{\rm int})$ between the two samples indicates that the reported uncertainties of the stellar velocity dispersions of the SLACS lenses do not reflect the total uncertainty.
For the present analysis, we have addressed this issue by adding additional terms of uncorrelated errors. However, future work should aim to improve the determination of systematics going back to the original data (or acquiring better data), and contemplate the possibility of correlated calibration errors, as due for example to the choice of stellar library or instrumental setup.
Second, our analysis is based on spherical Jeans models, assuming anisotropy of the Osipkov--Merritt form. These approximations are sufficient given the current uncertainties and constraints, but future work should consider at least axis-symmetric Jeans modeling \citep[e.g.,][]{Cappellari:2008, Barnabe:2012, Posacki:2015, Yildirim:2020}, and consider alternate parameterizations of anisotropy. Another possibility is the use of axisymmetric modeling of the phase-space distribution function with a two-integral Schwarzschild method by \cite{Cretton:1999, VerolmeZeeuw:2002} as performed by \cite{Barnabe:2007, Barnabe:2009}.

The addition of more freedom to the kinematic models will require the addition of more empirical information that can be obtained by spatially resolved data on distant lens galaxies, or from high-quality data (including absorption line shapes) of appropriately selected local elliptical galaxies.

\subsubsection{Selection effects of different lens samples} \label{sec:discussion_population_systematics}
One key pillar in this analysis to improve the precision on the $H_0$ measurement from the TDCOSMO sample is the information on the mass profiles of the SLACS sample.
The SLACS sample differs in terms of the redshift distribution and $r_{\rm eff}/\theta_{\rm E}$ relative to the TDCOSMO sample. Beyond our chosen explicit parameterized dependence of the MST parameter $\lambda_{\rm int}$ as a function of $r_{\rm eff}/\theta_{\rm E}$ we do not find trends in the predicted vs measured velocity dispersion within the SLACS sample.
However, we do find differences in the external shear contributions between the SLACS and TDCOSMO sample \citep{Shajib_slacs:2020}. This is expected because of selection effects. The TDCOSMO sample is composed of quads at higher redshift than SLACS. So it is not surprising that the TDCOSMO lenses tend to be more elongated (to increase the size of the quad cross section) and be more impacted by mass structure along the line of sight than SLACS.
Nonetheless, based on previous studies, we have no reason to suspect that the deflectors themselves are intrinsically different between SLACS and TDCOSMO.
Complex angular structure of the lenses might also affect the inference in the power-law slope $\gamma_{\rm pl}$, as the angular degree of freedoms in our model assumptions are, to some degree, limited \citep{Kochanek:2020b}.
A study with more lenses and particularly sampling the redshift range of the TDCOSMO sample (see Fig. \ref{fig:kin_fit_trends}) would allow us to better test our current underlying assumption and in case of a significant redshift evolution to correct for it.

\subsubsection{Line-of-sight structure}
The investigation of the line-of-sight structure of strong gravitational lenses of the TDCOSMO and the SLACS sample follows a specific protocol to provide an individual PDF of the external convergence, $p(\kappa_{\rm ext})$. In our current analysis, the statistical uncertainty of the SLACS line-of-sight structure is subdominant.

In the future -- as the other terms of the error budget shrink and this one becomes more relevant -- the following steps will be necessary. First, the specific choice
of $N$-body simulation and semi-analytic galaxy evolution model will need to be revisited. Second, it will be necessary to investigate how to improve the comparison with simulation products in order to further mitigate uncertainties. For instance, beyond galaxy number count statistics, weak gravitational lensing observations can also add information on the line-of-sight structure \citep[][]{Tihhonova:2018, Tihhonova:2020}.

Ideally, we aim for a validation based on simulations in the full cosmological context. These future simulations should include the presence of the strong lensing deflector, to further quantify nonlinear effects from the line-of-sight structure on the main deflector modeling as well as the main deflector impact on the line-of-sight light path differences \citep[see e.g.,][]{Li:2020}. Meeting the line-of-sight goal will require large box simulations, and for the main deflector this demands a very high fidelity and resolution at the 10-100pc scales dominated by baryons in the form of stars and gas.

\subsubsection{More flexible lens models and extended hierarchical analysis}
Getting the uncertainties right requires careful judgment in the use of theoretical assumptions, validated as much as possible by empirical data. Previous work by TDCOSMO assumed that galaxies were described by power laws or stars plus an NFW profile, leading to a given precision. In this work, we relax this assumption, with the goal to study the impact of the MST. As part of this investigation, we introduce the MST parameter $\lambda_{\rm int}$ in our hierarchical framework and use a PEMD + shear model as baseline. We demonstrate, based on simulations, that these choices are sufficient to the level of precision currently achieved. It is not, however, the end of the story. Additional information will enable better constraints on the mass density profiles. As the precision improves on $H_0$, it will be necessary to keep revisiting our assumptions and validating on a sufficiently large and realistic mock data set.

In the future, additional model flexibility may demand a treatment of more lens model parameters in the full hierarchical context of the inference. Currently, our baseline model is constrained sufficiently by the imaging data of the lensing sample.

However, the development of a hierarchical treatment of additional lensing parameters may also allow us to incorporate lenses with fewer constraints on the lensing nature, such as doubly lensed quasars, or lenses with missing high resolution imaging, or other partially incomplete data products. By pursuing further this development in hierarchical lens modeling, the total number of usable systems can improve, thus, in turn improving the constraints on the Hubble constant.

Substructure adds 0.6\%-2\% of uncorrelated and un-biased uncertainties on the $D_{\Delta t}$ inference\citep{Gilman:2020} for individual lenses. Thus, substructure adds a 0.5\% uncertainty in quadrature on the  combined $H_0$ constraints from the seven TDCOSMO lenses. This effect is highly subdominant to other sources of uncertainties related to the MST in our work and we note that this effect might partially be encapsulated in the scatter in $\lambda_{\rm int}$, $\sigma(\lambda_{\rm int})$, as inferred to be few percent.

\subsection{A pathway forward for time-delay cosmography} \label{sec:future}

After having discussed current limitations on the precision and accuracy of our new proposed hierarchical framework applied to time-delay cosmography, we summarize here the key steps to take in the near future, in terms of improvements on the analysis and addition of data, to improve both precision and accuracy in the $H_0$ measurements.
Given the new hierarchical context, our largest statistical uncertainty on $H_0$ arises from the stellar anisotropy modeling assumptions and the precision on the velocity dispersion measurements. Multiple and spatially resolved high signal-to-noise velocity dispersion measurements of gravitational lenses are able to further constrain the stellar anisotropy distribution. This can be provided by a large VLT-MUSE and Keck-KCWI campaign of multiple lenses and we expect significant constraining power from JWST \citep[][]{Yildirim:2020}.
A complementary approach of studying the mass profile and kinematic structure of the deflector galaxies, is to study the local analogs of those galaxies with high signal-to-noise ratio resolved spectroscopy. Assumptions about potential redshift evolution need to be mitigated and assessed within a lensing sample covering a wide redshift range.

A more straightforward approach in extending our analysis is by incorporating more galaxy-galaxy lenses, in particular lenses that populate a similar distribution to the lensed quasar sample. Such a targeted large sample can reduce potential systematics of our self-similarity assumptions, as well as increase the statistical precision on the mass profiles.
Recent searches for strong gravitational lenses in current and ongoing large area imaging surveys, such as the Dark Energy Survey (DES) and the Hyper-Supreme-Cam survey (HSC) have resulted in hundreds of promising galaxy-galaxy scale candidate lenses \citep[see e.g.,][]{Jacobs:2019, Sonnenfeld:2020} and dozens of lensed quasars \citep[see e.g.,][]{Agnello:2018, Delchambre:2019, Lemon:2020}.

With the next generation large ground and space based surveys (Rubin Observatory LSST, Euclid, Nancy Grace Roman Space Telescope), of order $10^{5}$ galaxy-galaxy lenses and of order $10^3$ quasar-galaxy lenses will be discovered \citep{OM10, Collett:2015}.
Limited follow-up capabilities with high resolution imaging and spectroscopy will be a key limitation and needs to be mitigated with strategic prioritization of targets to maximize resulting precision and accuracy.
We refer to \cite{BirrerTreu:2020} for a forcast based on the precision on $H_0$ we can expect for a current and future lensing sample with spatially resolved kinematics measurement based on the analysis framework presented in this work.

Beyond the addition of external data sets, we emphasize the further demand on the validation of the modeling approach, both in the imaging analysis as well as the stellar anisotropy modeling.
Detailed investigation and data challenges based on realistic data with the same complexity level as the real analysis are a useful tool to make progress. To ensure that the requirements are met in the modeling of the deflector galaxy and the local and line-of-sight environment, validation on realistic simulations in the full cosmological context, including selection effects and ray-tracing through the line-of-sight cone of a cosmological box are required.
Moreover, we also stress that assessing and tracking systematics at the percent level and the mitigation thereof on the joint inference on $H_0$ would be much facilitated by an automatized and homogenized analysis framework encapsulating all relevant aspects of the analysis of individual lenses.

Finally, a decisive conclusion on the current Hubble tension demands for a rigorous assessment of results by different science collaborations. We stress the importance of conducting the analysis blindly in regard to $H_0$ and related quantities to prevent experimenter bias, a procedure our collaboration has incorporated and followed rigorously.
In addition, all measurements of $H_0$ contributing to a decisive conclusion of the tension must guarantee reproducibility. In this work, we provide all software as open-source and release the value-added data products and analysis scripts to the community to facilitate the needed reproducibility.

\subsection{Post-blind discussion of the results and comparison with previous time-delay cosmography work \protect\footnote{This section was written after the results were known to the authors.}} \label{sec:tension_status}

\begin{figure*}
  \centering
  \includegraphics[angle=0, width=160mm]{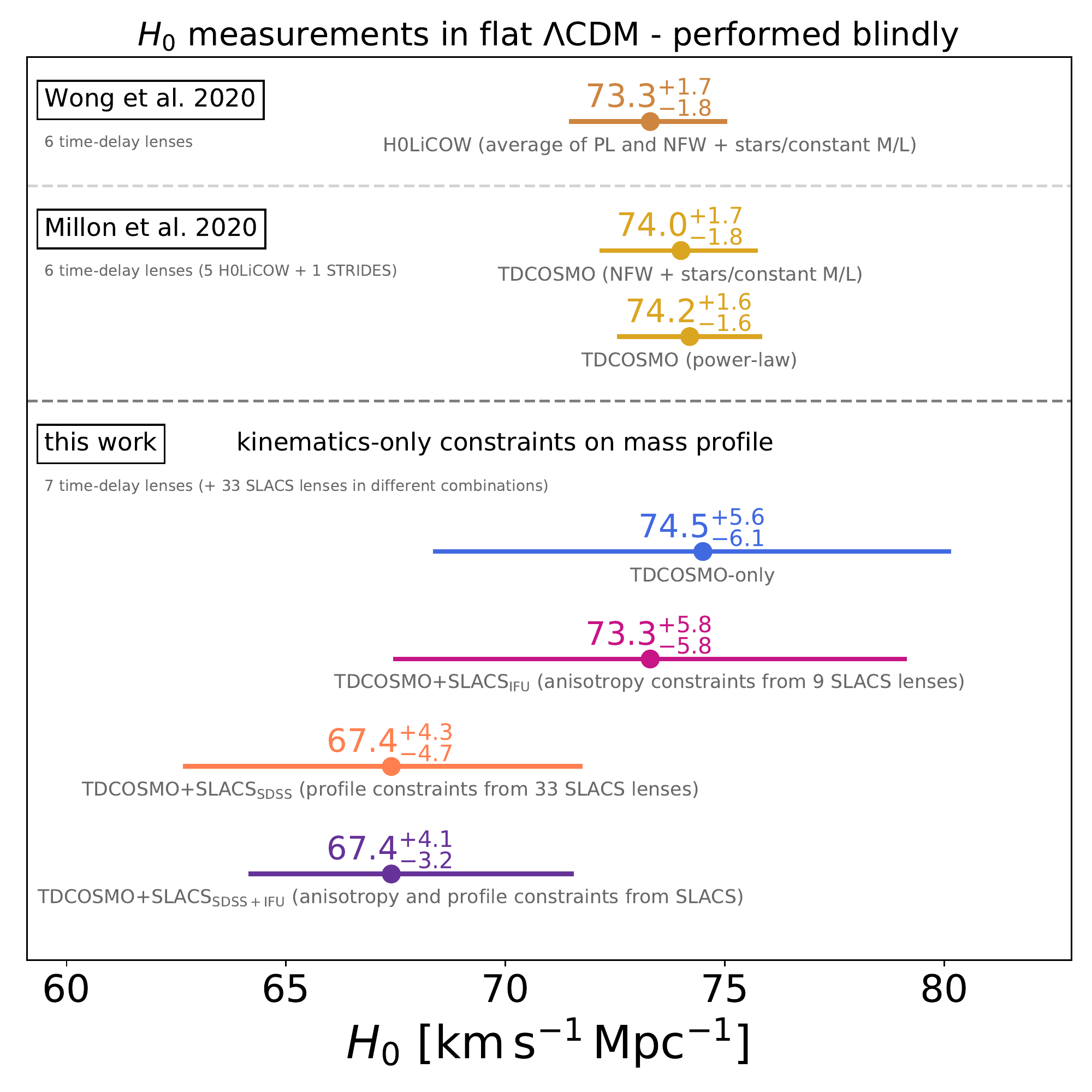}
  \caption{Comparison of different {\bf blind} $H_0$ measurements by the TDCOSMO collaboration, based on different mass profile assumptions and data sets incorporated. All measurements presented on this plot were performed blindly with regard to the inference of $H_0$.
  The measurement on top is the combined H0LiCOW six lenses constraints presented by \cite{Wong:2020}, when averaging power-law and composite NFW plus stars (with constant mass-to-light ratio) on a lens-by-lens basis without correlated errors among the lenses.
  The next two measurements are from \cite{Millon:2020} of six TDCOSMO time-delay lenses (five H0LiCOW lenses\protect\footnotemark and one STRIDES lens by \cite{Shajib0408}), when performing the inference assuming either a composite NFW plus stars (with constant mass-to-light ratio) or the power-law mass density profile for the galaxy acting as a lens.
In the lower panel, we show the results from this work. The main difference with respect to previous work is that we have made virtually no assumption on the radial mass density profile of the lens galaxy, and taken into account the covariance between the lenses. The analysis in this work is constrained only by the stellar kinematics and fully accounts for the uncertainty related to the mass sheet transformation (MST). In this framework, we obtain four measurements according to the datasets considered. The TDCOSMO-only inference is based on the same set of seven lenses as those jointly included by \cite{Millon:2020} and \cite{Wong:2020}. The inferred median value is the same, indicating no bias, and the uncertainties, as expected, are larger. The next three measurements rely on external datasets from the SLACS survey, by making the assumption that the lens galaxies in the two surveys are drawn from the same population. The TDCOSMO+SLACS$_{\rm IFU}$ measurements uses, in addition to the TDCOSMO sample, nine lenses from the SLACS sample with IFU observations to inform the anisotropy prior applied on the TDCOSMO lenses. The TDCOSMO+SLACS$_{\rm SDSS}$ measurement comes from the joint analysis of the TDCOSMO sample and 33 SLACS lenses with SDSS spectroscopy. The TDCOSMO+SLACS$_{\rm SDSS + IFU}$ presents the joint analysis of all three data sets, again assuming self-similar distributions of the mass profiles and stellar anisotropy. The TDCOSMO-only and TDCOSMO+SLACS$_{\rm IFU}$ analyses do not rely on self-similar mass profiles of the SLACS and TDCOSMO sample while the TDCOSMO+SLACS$_{\rm SDSS}$ and TDCOSMO+SLACS$_{\rm SDSS + IFU}$ measurements (orange and purple) do. All the measurements shown in this plot are in statistical agreement with each other. See Section \ref{sec:tension_status} for a discussion and physical interpretation of the results.
   \faGithub\href{https://github.com/TDCOSMO/hierarchy_analysis_2020_public/blob/6c293af582c398a5c9de60a51cb0c44432a3c598/JointAnalysis/tdcosmo_comparison_plot.ipynb}{~source}}
\label{fig:h0_comparison}
\end{figure*}

In this Section we discuss how the measurement presented in this paper related to previous work by members of this collaboration as part of the H0LiCOW, STRIDES, and SHARP projects. We then discuss the relationship between the multiple measurements obtained within the hierarchical framework introduced in this paper. All the relevant measurements are summarized in Figure~\ref{fig:h0_comparison} for quick visualization.

The result of our hierarchical TDCOSMO-only analysis is fully consistent with the assumptions on the mass profiles made in previous H0LiCOW/STRIDES/SHARP work \citep[see e.g.,][]{Wong:2020, Shajib0408,Millon:2020}. The consistency is reinforced by \citep{Yang:2020} who concluded that the combination of kinematics and time-delay constraints are consistent with General Relativity, an underlying assumptions of time-delay cosmography. The only difference with respect to the H0LiCOW/STRIDES/SHARP analysis is that the uncertainty has significantly increased. This was expected, because we have virtually eliminated the assumptions on the radial mass profile of elliptical galaxies and, due to the MST, the only source of information left to enable a $H_0$ measurement is the stellar kinematics. Without lensing information, due to the well known mass-anisotropy degeneracy, unresolved kinematics has limited power to constrain the mass profiles. Since our parametrization is maximally degenerate with H$_0$ and our assumptions are minimal, this 9\% error budget accounts for potential effects of the MST.

Another set of results is obtained within the hierarchical framework with the addition of external information. Under the additional assumption that the galaxies in the external datasets are drawn from the same population as the TDCOSMO deflectors, these results achieve higher precision than TDCOSMO alone. Adding the SLACS dataset shrinks the uncertainty to 5\% and shifts the mean inferred $H_0$ to a value about 6 \Hunit lower than the TDCOSMO-only analysis. This shift is consistent within the uncertainties achieved by the TDCOSMO-only analysis and can be traced back to two factors: (i) the anisotropy constraints prefer a lower $a_{\rm ani}$ value and this moves $H_0$ down relative to the chosen prior on $a_{\rm ani}$. The VIMOS+IFU inference is about 2 \Hunit lower than the equivalent TDCOSMO-only inference.
(ii) The SLACS lenses prefer an overall lower -- but statistically consistent -- $\lambda_{\rm int, 0}$ value for a given anisotropy model by about 8\%. The negative trend of $\lambda_{\rm int}$ with $r_{\rm eff}/\theta_{\rm E}$ ($\alpha_{\lambda}$) partially mitigates an even lower $\lambda_{\rm int}$ value preferred by the SLACS sample relative to the TDCOSMO sample.

The shift between the TDCOSMO and TDCOSMO+SLACS results can have two possible explanations (if it is not purely a statistical fluctuation). One option is that elliptical galaxies are more radially anisotropic (and therefore have a flatter mass density profile to reproduce the same velocity dispersion profile) than the prior used to model the TDCOSMO galaxies. The alternative option is that the TDCOSMO and SLACS galaxies are somehow different. Within the observables at disposal, one that may be indicative of a different line of sight anisotropy is the higher ellipticity of the surface brightness and of the projected total mass distribution \citep{Shajib_slacs:2020} of the TDCOSMO deflectors in comparison to the SLACS deflectors. As mentioned in Section \ref{sec:slacs_selection}, this is understood to be a selection effect because ellipticity increases the cross section for quadruple images and TDCOSMO is a sample of mostly quads (six out of seven), while SLACS is mostly doubles \citep{Treu:2009}. Departure from spherical symmetry in elliptical galaxies can arise from rotation or anisotropy. If flattening arises from rotation (which we have neglected in our study) more flattened systems are more likely to be seen edge-on. If it arises from anisotropy, the observed flattening could be due to tangential anisotropy that is not included in our models, or to a smaller degree of radial anisotropy than for other orientations. These two options result in different predictions that can be tested with spatially resolved kinematics of the TDCOSMO lens galaxies. If the shift is just due to an inconsistency between the TDCOSMO prior and the SLACS likelihood, spatially resolved kinematics will bring them in closer alignment. If it is due to intrinsic differences, spatially resolved kinematics will reveal rotation or tangential (less radial) anisotropy. In addition, spatially resolved kinematics of the TDCOSMO sample will reduce the uncertainties of both measurement, and thus resolve whether the shift is a fluctuation or significant.

The other potential way to elucidate the marginal differences between the TDCOSMO and SLACS sample is to obtain precise measurements of mass at scales well beyond the Einstein radius. As seen in Figure \ref{fig:pl_mst_inferred}, a pure power law and the transformed profile differ by up to 50\% in that region (depending on the choice of $R_{\rm c}$). Satellite kinematics or weak lensing would help reduce the freedom of the MST, provided they reach sufficient precision.

\footnotetext{Excluding B1608+656 as this lens was only analyzed with a power-law model and not with a composite model and thus not part of the model comparison analysis. Additional lensing potential perturbations on top of the power-law profile lead to only small amounts of corrections \cite{Suyu:2010}.}

\section{Conclusion} \label{sec:conclusion}

The precision of time-delay cosmography has improved significantly in the past few years, driven by improvement in the quality of the data and methodology. As the precision improves it is critical to revisit assumptions and explore potential systematics, while charting the way forward.

In this work, we relaxed previous assumptions on the mass-profile parameterization and introduced an efficient way to explore potential systematics associated to the mass-sheet degeneracy in a hierarchical Bayesian analysis. In this new approach, the mass density profile of the lens galaxies is only constrained by basic information on stellar kinematics. It thus provides a conservative estimate of how much the mass profile can depart from a power law, and how much the error budget can grow as a result. Based on the consistent results of the power law and stars plus NFW profiles in the inference on $H_0$ \citep{Millon:2020}, we expect very similar conclusions had we performed this analysis with a stars plus NFW profile.

We validated our approach on the Time-Delay Lens Modeling Challenge sample of hydrodynamical simulations. We then applied the formalism and assumptions to the TDCOSMO data set in a blind fashion. Based on the TDCOSMO data set alone we infer $H_0 = $ \Htdcosmo~\Hunit. The uncertainties on $H_0$ are dominated by the precision of the spectroscopic data and the modeling uncertainties therein.
To further increase our precision, we added self-consistently to our analysis a set of SLACS lenses with imaging modeling and independent kinematic constraints. We characterized the candidate lenses to be added and explicitly selected only lenses that do not have significantly enhanced local environments. In total, we were able to add 33 additional lenses with no time delay information of which nine have additional 2D kinematics with VIMOS IFU data that allowed us to further constrain uncertainties in the anisotropy profile of the stellar orbits.
Our most constrained measurement of the Hubble constant is $H_0 = $ \Hjoint~\Hunit from the joint TDCOSMO+SLACS analysis, assuming that the two samples are drawn from the same population.

The 5\% error budget reported in this work addresses conclusively concerns about the MST \citep{SchneiderSluse:2013,Sonnenfeld:2018,Kochanek:2020,Kochanek:2020b}. If the mass density profiles of lens galaxies are not well described by power-laws or stars plus NFW halos, this is the appropriate uncertainty to associate with current time-delay cosmography.
Additional effects are very much subdominant for now as compared with the effect of the MST.
For example, the small level of pixelated corrections to the elliptical power-law model obtained in our previous work suggests that the departure from ellipticity is not required by the data.

Based on the methodology presented and the results achieved, we lay out a roadmap for further improvements to ultimately enable a 1\% precision measurement of the Hubble constant, which is a clear target both for resolving the Hubble tension and to serve as a prior on dark energy studies \citep{Weinberg:2013}.
The key ingredients required to reduce the statistical uncertainties are i) spatially resolved high signal-to-noise kinematic measurements; ii) an increase in the sample size of both lenses with measured time-delays and lenses with high-resolution imaging and precise kinematic measurements.
Potential sources of systematic that should be investigated further to maintain accuracy at the target precision are those arising from: (i) measurements of the stellar velocity dispersion; 
(ii) characterization of the selection function and local environment of all the lenses included in the inference;
(iii) mass profile modeling assumptions beyond the MST and stellar anisotropy modeling assumptions.

Upcoming deep, wide-field surveys (such as those enabled by Vera Rubin Observatory, Euclid and the Nancy Grace Roman Observatory) will discover many thousands of lenses of which several hundred will have accurate time delay measurements \citep[see e.g.,][]{OM10, Collett:2015, Huber:2019}. The analysis framework presented in this work will serve as a baseline for the analysis of these giant samples of lenses; simultaneously enabling precise and accurate constraints on the Hubble constant and the astrophysics of strong lensing galaxies.

\section*{Acknowledgments}
SB thanks Kfir Blum, Veronica Motta, Timo Anguita, Sampath Mukherjee, Elizabeth Buckley-Geer for useful discussions, Hyungsuk Tak for participating in the TDLMC, the TDLMC team for setting up the challenge and Yiping Shu for providing access to the SDSS velocity dispersion measurements.
AJS was supported by the Dissertation Year Fellowship from the UCLA Graduate Division.
TT and AJS acknowledge support from NSF through NSF grant NSF-AST-1906976, from NASA through grant HST-GO-15320 and from the Packard Foundation through a Packard Research Fellowship to TT.
AA was supported by a grant from VILLUM FONDEN (project number 16599). This project is partially funded by the Danish council for independent research under the project ``Fundamentals of Dark Matter Structures'', DFF--6108-00470.
SB and MWA acknowledges support from the Kavli Foundation.
TC is funded by a Royal Astronomical Society Research Fellowship.
C.D.F. and G.C.-F.C. acknowledge support for this work from the National Science Foundation under grant no. AST-1907396 and from NASA through grant HST-GO-15320.
This project has received funding from the European Research Council (ERC) under the European Union’s Horizon 2020 research and innovation program (COSMICLENS:grant agreement No 787886).
CS is supported by a Hintze Fellow at the Oxford Centre for Astrophysical Surveys, which is funded through generous support from the Hintze Family Charitable  Foundation.
SHS thanks the Max Planck Society for support through the Max Planck Research Group.
This work was supported by World Premier International Research Center Initiative (WPI Initiative), MEXT, Japan.  This work was supported by JSPS KAKENHI Grant Number JP20K14511.
SB acknowledges the hospitality of the Munich Institute for Astro- and Particle Physics (MIAPP) of the Excellence Cluster "Universe".
Based on observations collected at the European Southern Observatory under ESO program 0102.A-0600 (PI Agnello), 075.B-0226 (PI Koopmans), 177.B-0682 (PI Koopmans).

This work made use of the following public software packages: \textsc{hierArc} (this work), \textsc{lenstronomy} \citep{Birrer:2015, Birrer_lenstronomy}, \textsc{dolphin} \citep{Shajib_slacs:2020}, \textsc{emcee} \citep{emcee}, \textsc{corner} \citep{corner}, \textsc{ppxf} \citep{pPXF}, \textsc{astropy} \citep{astropy:2013, astropy:2018}, \textsc{FASTELL} \citep{fastell} and standard Python libraries.




\bibliographystyle{mnras}
\bibliography{BibdeskLib}{}


\newpage
\appendix

\section{Internal MST + PEMD}\label{app:mst_pemd}
\begin{figure*}
  \centering
  \includegraphics[angle=0, width=160mm]{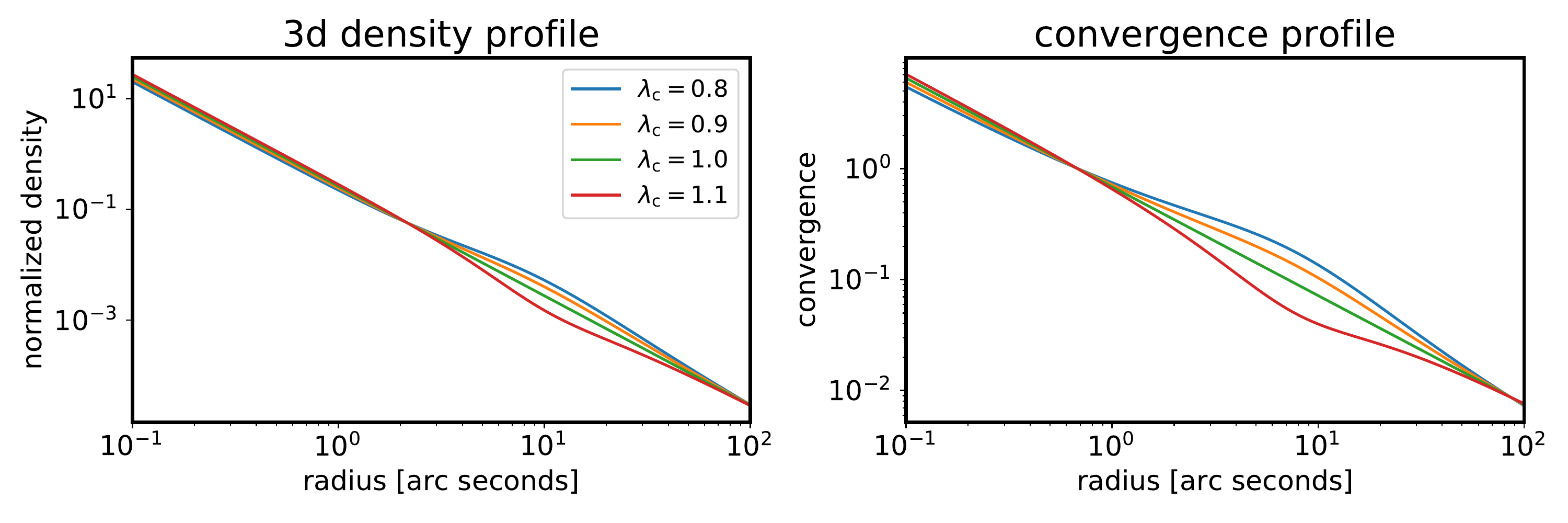}
  \caption{Illustration of the power-law profile (Eqn. \ref{eqn:pl_profile}) in three dimensions (left panel) and in projection (right panel) under an approximate MST with a cored mass component (Eqn. \ref{eqn:core_profile_projected}). The transforms presented here were indistinguishable by the mock imaging data of Figure \ref{fig:mock_lens}.  \faGithub\href{https://github.com/TDCOSMO/hierarchy_analysis_2020_public/blob/6c293af582c398a5c9de60a51cb0c44432a3c598/MST_impact/MST_pl_cored.ipynb}{~source} }
\label{pl_mst_profile}
\end{figure*}

\begin{figure}
  \centering
  \includegraphics[angle=0, width=80mm]{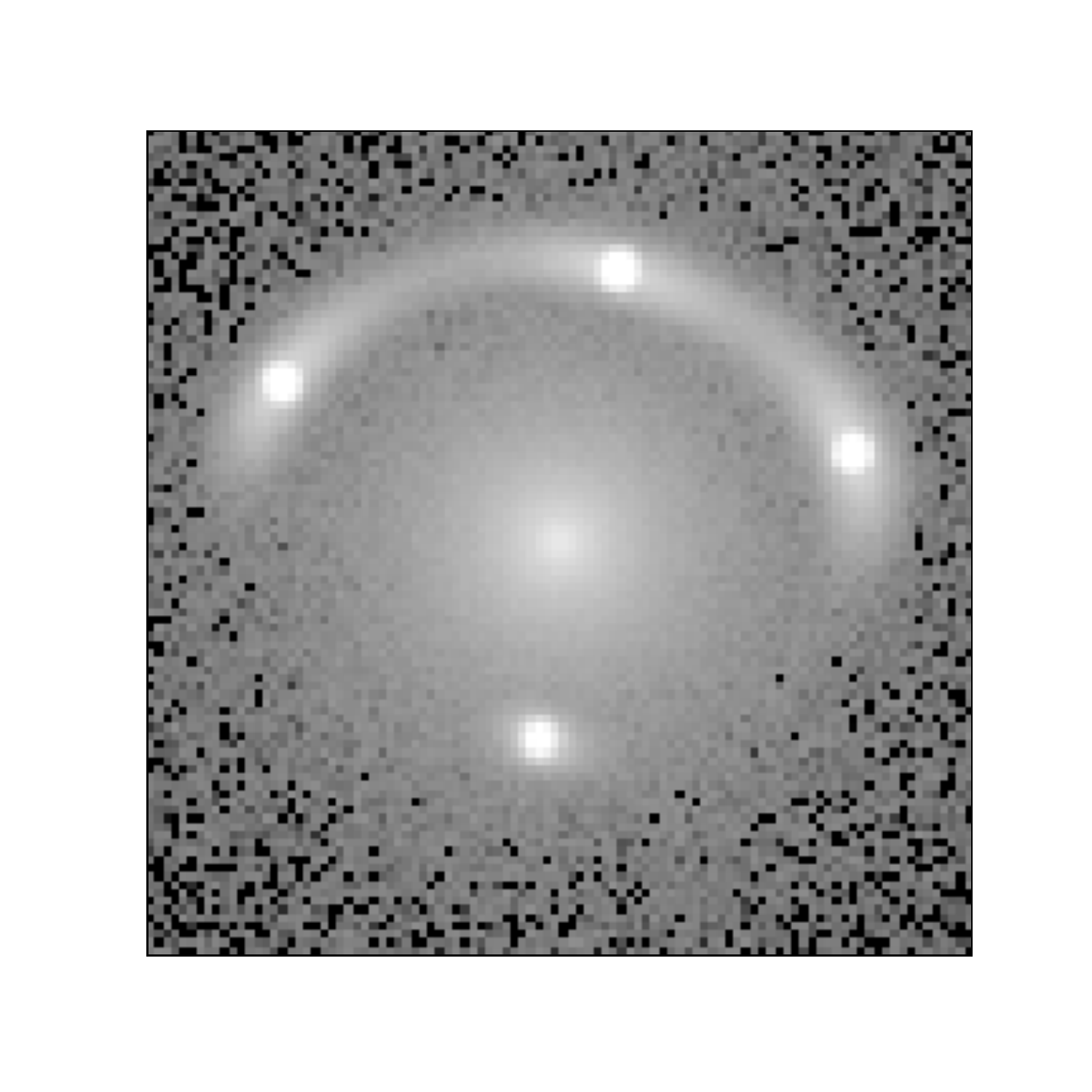}
  \caption{Mock \textit{HST} image with a power-law mass profile for which we perform the inference on the detectability of an approximate MST.  \faGithub\href{https://github.com/TDCOSMO/hierarchy_analysis_2020_public/blob/6c293af582c398a5c9de60a51cb0c44432a3c598/MST_impact/MST_pl_cored.ipynb}{~source} }
  \label{fig:mock_lens}
\end{figure}

Figure \ref{pl_mst_profile} shows different approximate MST's with a core radius of 10 arcseconds on top of a power-law profile \citep[see also][]{Blum:2020}.
Figure \ref{fig:mock_lens} shows the mock lens used in Section \ref{sec:imaging_constraints} to perform the imaging modeling inference on the lens model parameters, including the cored component resembling the MST.

\section{Mass-anisotropy degeneracy} \label{app:anisotropy}

\begin{figure*}
  \centering
  \includegraphics[angle=0, width=160mm]{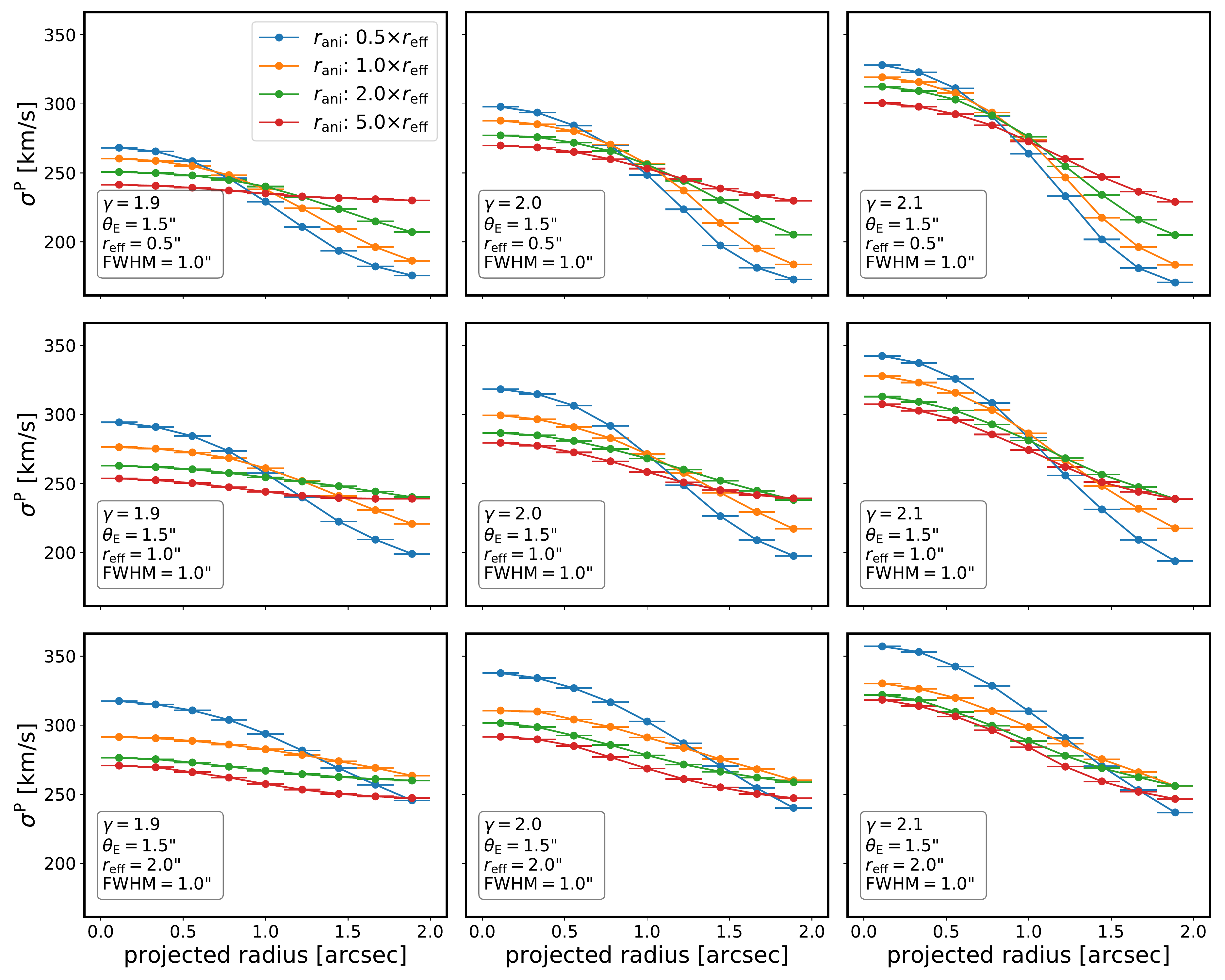}
  \caption{Radial dependence on the projected velocity dispersion measurement for an Osipkov--Merritt anisotropy profile (Eqn. \ref{eqn:r_ani}). \textbf{Top to bottom:} Increase in the half light radius of the deflector. \textbf{Left to right:} Change in the mass profile slope.  \faGithub\href{https://github.com/TDCOSMO/hierarchy_analysis_2020_public/blob/6c293af582c398a5c9de60a51cb0c44432a3c598/MST_impact/anisotropy_ifu.ipynb}{~source}}
  \label{fig:anisotropy_ifu_r_ani}
\end{figure*}

\begin{figure*}
  \centering
  \includegraphics[angle=0, width=160mm]{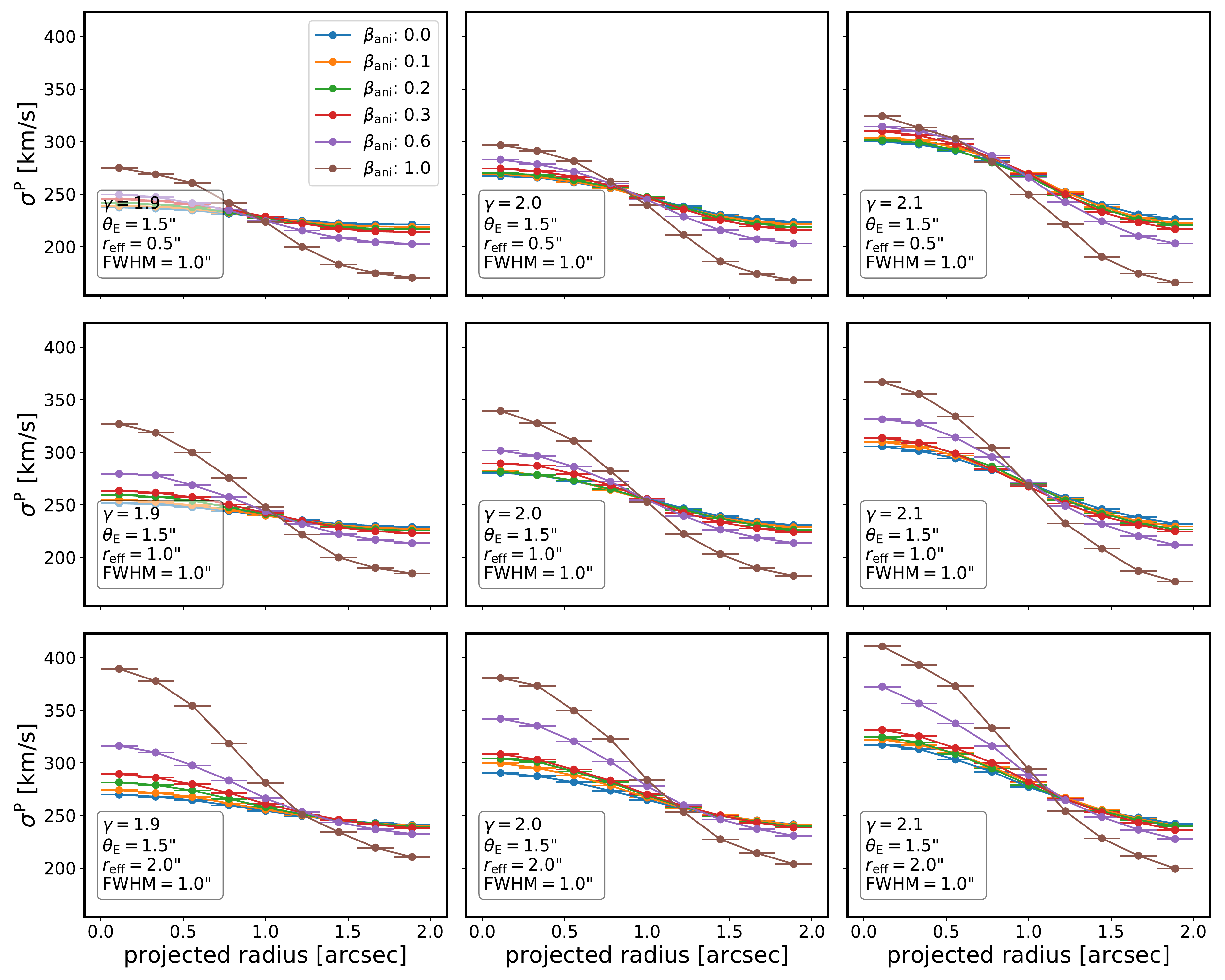}
  \caption{Radial dependence on the projected velocity dispersion measurement for a constant anisotropy $\beta_{\rm ani}$. \textbf{Top to bottom:} Increase in the half light radius of the deflector. \textbf{Left to right:} Change in the mass profile slope.  \faGithub\href{https://github.com/TDCOSMO/hierarchy_analysis_2020_public/blob/6c293af582c398a5c9de60a51cb0c44432a3c598/MST_impact/anisotropy_ifu.ipynb}{~source}}
  \label{fig:anisotropy_ifu_const}
\end{figure*}

\begin{figure*}
  \centering
  \includegraphics[angle=0, width=160mm]{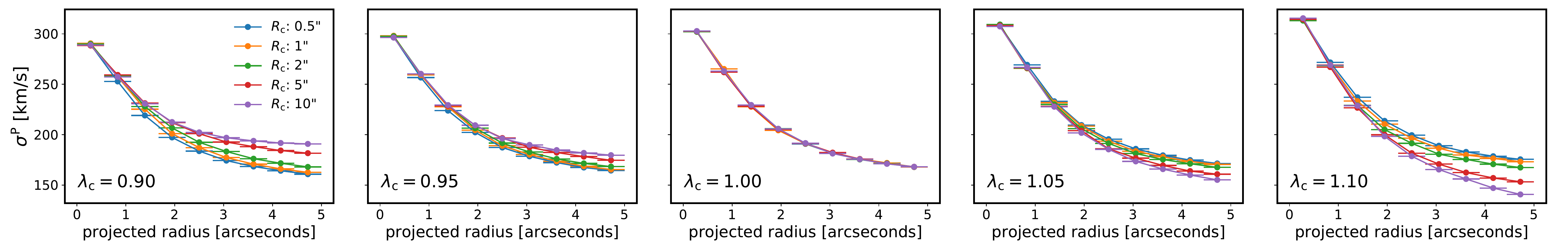}
  \caption{Radial dependence on the projected velocity dispersion measurement for different cored components \ref{eqn:core_profile_projected} on top of a PEMD profile approximating a pure MST, with normalization $\lambda_{\rm c}$ and core radii, $R_{\rm c}$. The projected radius from the center of the galaxy is extended to 5 arcseconds to visibly see the impact on the kinematic of larger cored components.  \faGithub\href{https://github.com/TDCOSMO/hierarchy_analysis_2020_public/blob/6c293af582c398a5c9de60a51cb0c44432a3c598/MST_impact/anisotropy_ifu.ipynb}{~source}}
  \label{fig:anisotropy_ifu_mst}
\end{figure*}

Figure \ref{fig:anisotropy_ifu_r_ani} shows the predicted projected velocity dispersions (Eqn. \ref{eqn:sigma_convolved}) in radial bins form the center for PEMD profiles with different logarithmic mass-profile slopes and half-light radii. We chose a fiducial seeing of FWHM=1$^{\prime\prime}$.0.
Alternatively, we display the results assuming a constant anisotropy $\beta_{\rm ani}(r)={\rm const}$ in Figure \ref{fig:anisotropy_ifu_const}.
In Figure \ref{fig:anisotropy_ifu_mst} we plot, without seeing and under fixed anisotropy model, the predicted radial change in the velocity dispersion for different core masses, $\lambda_{\rm c}$, and core radii, $R_{\rm c}$.

\section{Likelihood calculation} \label{app:likelihood}
In this Section, we provide the specifics of the likelihood calculation for individual lenses and how we efficiently evaluate the likelihood in the hierarchical context. This includes the imaging likelihood (Section \ref{app:imaging_likelihood}), time-delay likelihood (Section \ref{app:td_likelihood}) and velocity dispersion likelihood (Section \ref{app:spec_likelihood}). Section \ref{app:cov_likelihood} describes our formalism to track covariances and the marginalization as implemented in \textsc{hierArc}.

\subsubsection{Imaging likelihood} \label{app:imaging_likelihood}
The likelihood and the lens model inference is not prominently featured in this work, as we are making use of products being derived by our collaboration presented in other work. Nevertheless, the high resolution imaging data and lens model inferences on the likelihood level are essential parts of the analysis.

Given a lens model with parameters $\boldsymbol{\xi}_{\rm mass}$ and surface brightness model with parameters $\boldsymbol{\xi}_{\rm ligth}$, a model of the imaging data can be constructed, $\boldsymbol{d}_{\rm model}$. The likelihood is computed at the individual pixel level accounting for the noise properties from background and other noise properties, such as read-out, as well as the Poisson contribution from the sources. The imaging likelihood is given by
\begin{multline}
  p(\mathcal{D}_{\rm img} | \boldsymbol{\xi}_{\rm mass}, \boldsymbol{\xi}_{\rm ligth}) \\
   = \frac{\exp \left[-\frac{1}{2}\left(\boldsymbol{d}_{\rm data} - \boldsymbol{d}_{\rm model}\right)^{\rm T} \boldsymbol{\Sigma}_{\rm pixel}^{-2}\left(\boldsymbol{d}_{\rm data} - \boldsymbol{d}_{\rm model}\right)\right]}{\sqrt{(2 \pi)^k {\rm det}(\boldsymbol{\Sigma}_{\rm pixel}^{2})}},
\end{multline}
where $k$ is the number of pixels used in the likelihood and $\boldsymbol{\Sigma}_{\rm pixel}$ is the error covariance matrix. Current analyses assume uncorrelated noise properties in the individual pixels and the covariance matrix becomes diagonal. The model of the surface brightness of the lensed galaxy requires high model flexibility. The surface brightness components can be captured with linear components and solved for and marginalized over analytically. TDCOSMO uses pixelized grids as well as smooth basis sets \citep[see e.g.,][for the current methods in use]{Suyu:2006, Birrer:2015}.

\subsubsection{Time-delay likelihood} \label{app:td_likelihood}
The likelihood of the time delay data $\mathcal{D}_{\rm td}$ given a model prediction is
\begin{multline}
  p(\mathcal{D}_{\rm td} | \boldsymbol{\xi}_{\rm mass}, \boldsymbol{\xi}_{\rm ligth}, D_{\Delta t}/\lambda) \\
   = \frac{\exp \left[-\frac{1}{2}\left(\boldsymbol{\Delta t}_{\rm data} - \boldsymbol{\Delta t}_{\rm model}\right)^{\rm T} \boldsymbol{\Sigma}_{\Delta t{\rm data}}^{-2}\left(\boldsymbol{\Delta t}_{\rm data} - \boldsymbol{\Delta t}_{\rm model}\right)\right]}{\sqrt{(2 \pi)^k {\rm det}(\boldsymbol{\Sigma}_{\Delta t{\rm data}}^{2})}},
\end{multline}
with $\boldsymbol{\Delta t}_{\rm data}$ is the data vector of relative time delays, $\boldsymbol{\Sigma}_{\Delta t{\rm data}}^{2}$ is the measurement covariance between the relative delays and
\begin{equation} \label{eqn:td_likelihood}
	\boldsymbol{\Delta t}_{\rm model} = \lambda \frac{D_{\Delta t}}{c} \boldsymbol{\Delta \phi}_{\rm Fermat}(\boldsymbol{\xi}_{\rm mass}, \boldsymbol{\xi}_{\rm light})
\end{equation}
is the model predicted time-delay vector (Eqn. \ref{eqn:time_delay}) with $\boldsymbol{\Delta \phi}_{\rm Fermat}$ is the relative Fermat potential vector (Eqn. \ref{eqn:fermat_potential}).
Effectively, the time-delay distance posterior transform according to Equation \ref{eqn:ddt_mst} under an MST.

\subsubsection{Velocity dispersion likelihood} \label{app:spec_likelihood}

The model prediction of the velocity dispersion transforms under MST according to Equation (\ref{eqn:kinematics_mst}) and cosmological distance ratio relevant for the kinematics is $D_{\rm s}/D_{\rm ds}$ and scales according to Equation (\ref{eqn:kinematics_cosmography}).
We can write the likelihood of the spectroscopic data, $\mathcal{D}_{\rm spec}$, given a model as
\begin{multline}
  p(\mathcal{D}_{\rm spec} | \boldsymbol{\xi}_{\rm mass}, \boldsymbol{\xi}_{\rm light}, \boldsymbol{\beta}_{\rm ani}, D_{\rm s}/D_{\rm ds}, \lambda) \\
   = \frac{\exp \left[-\frac{1}{2}\left(\boldsymbol{\sigma}^{\text{P}}_{\rm data} - \boldsymbol{\sigma}^{\text{P}}_{\rm model}\right)^{\rm T} \boldsymbol{\Sigma}_{\sigma{\rm data}}^{-2}\left(\boldsymbol{\sigma}^{\text{P}}_{\rm data} - \boldsymbol{\sigma}^{\text{P}}_{\rm model}\right)\right]}{\sqrt{(2 \pi)^k {\rm det}(\boldsymbol{\Sigma}_{\sigma{\rm data}}^{2})}} ,
\end{multline}
where $\boldsymbol{\sigma}^{\text{P}}_{\rm data}$ is a vector of velocity dispersion measurements, $\boldsymbol{\Sigma}_{\sigma{\rm data}}^{2}$ is the measurement error covariance between the measurements (including, for example, stellar template fitting, calibration systematics etc.) and
\begin{equation} \label{eqn:kin_model_likelihood}
	\left(\boldsymbol{\sigma}^{\text{P}}_{\rm model}\right)^2 = \lambda c^2 \frac{D_{\rm s}}{D_{\rm ds}} J_{\mathcal{A}_j}(\boldsymbol{\xi}_{\rm mass}, \boldsymbol{\xi}_{\rm light}, \boldsymbol{\beta}_{\rm ani})
\end{equation}
is the model prediction.
The impact of the anisotropy distribution depends on the specific lens and light configuration. We can compute numerically the change in the model predicted dimensionless velocity dispersion component for each individual aperture $\mathcal{A}_j$, $J_{\mathcal{A}_j}(\boldsymbol{\xi}_{\rm mass}, \boldsymbol{\xi}_{\rm light}, \boldsymbol{\beta}_{\rm ani})$
\begin{equation} \label{eqn:kin_anisotropy_transform}
    J_{\mathcal{A}_j}(\boldsymbol{\xi}_{\rm mass}, \boldsymbol{\xi}_{\rm light}, \boldsymbol{\beta}_{\rm ani}) = \phi_{\mathcal{A}_j}(\boldsymbol{\beta}_{\rm ani}) \times J_{\mathcal{A}_j 0}(\boldsymbol{\xi}_{\rm mass}, \boldsymbol{\xi}_{\rm light}).
\end{equation}

\subsubsection{Marginalization and covariances} \label{app:cov_likelihood}

The marginalization over $\boldsymbol{\xi}_{\rm mass}$ and $\boldsymbol{\xi}_{\rm light}$ (Eqn. \ref{eqn:marginal_lens_light}) affects the relative Fermat potential $\boldsymbol{\Delta\phi}_{\rm Fermat}$ in the time-delay likelihood (Eqn. \ref{eqn:td_likelihood}) and the dimensionless factors $\sqrt{J_{\mathcal{A}_j}}$ (Eqn. \ref{eqn:kin_model_likelihood}, \ref{eqn:kin_anisotropy_transform}).
We can compute the marginalized likelihood over $\boldsymbol{\xi}_{\rm mass}$ and $\boldsymbol{\xi}_{\rm light}$ under the assumption that the posteriors in $\boldsymbol{\xi}_{\rm mass}$ and $\boldsymbol{\xi}_{\rm light}$ transform to covariant Gaussian distributions in $\boldsymbol{\Delta\phi}_{\rm Fermat}$ and $\sqrt{J_{\mathcal{A}_j}}$ as a model addition to the error covariances, such that
\begin{equation}
  \boldsymbol{\Sigma}^2_{\rm marg} = \Sigma^2_{\rm data} + \Sigma^2_{\rm model}.
\end{equation}
The model covariance matrix for the time delays can be expressed as
\begin{equation}
  \boldsymbol{\Sigma}^2_{\Delta t{\rm model}} = \text{cov}\left(\boldsymbol{\Delta \phi}_{\rm Fermat}, \boldsymbol{\Delta \phi}_{\rm Fermat} \right) \left(\lambda \frac{D_{\Delta t}}{c}\right)^2,
\end{equation}
the covariance matrix on the kinematics as
\begin{equation}
  \Sigma^2_{\sigma {\rm model}} = \text{cov}\left( \sqrt{J_{\mathcal{A}_i 0}},\sqrt{J_{\mathcal{A}_j 0}}  \right) c^2 \frac{D_{\rm s}}{D_{\rm ds}} \lambda \sqrt{\phi_{\mathcal{A}_i}(\boldsymbol{\beta}_{\rm ani}) \phi_{\mathcal{A}_j}(\boldsymbol{\beta}_{\rm ani})}
\end{equation}
and the cross-covariance between the kinematics and the time delays as
\begin{equation} \label{eqn:cross_covariance}
  \Sigma^2_{\Delta t\sigma {\rm model}} = \text{cov}\left(\boldsymbol{\Delta \phi}_{\rm Fermat}, \sqrt{J_{\mathcal{A}_j 0}}  \right)  D_{\Delta t} \sqrt{\frac{D_{\rm s}}{D_{\rm ds}}} \lambda^{3/2} \sqrt{\phi_{\mathcal{A}_j}(\boldsymbol{\beta}_{\rm ani})}.
\end{equation}
In this form, the model covariances are explicitly dependent on the anisotropy model, the MST and the cosmology.

The covariance between the kinematics and the time delays, $\Sigma^2_{\Delta t\sigma {\rm model}}$, above in Equation (\ref{eqn:cross_covariance}) is primarily impacted by the average density slope parameter $\gamma$ of the mass model. $\gamma$ affects both the kinematics and the Fermat potential and uncertainty in $\gamma$ can lead to covariances. However, if the density slope parameter is well constrained by imaging data (modulo explicit MST), the covariance in Equation (\ref{eqn:cross_covariance}) becomes subdominant relative to the uncertainty in the measurement of the kinematics.

When setting $\Sigma^2_{\Delta t\sigma {\rm model}}=0$, we can separate the inference of $D_{\Delta t}/\lambda$ from the kinematics likelihood and can work directly on the $D_{\Delta t}/\lambda$ posteriors from the inference from the image data, $D_{\rm image}$, and the time-delay measurement, $D_{\rm td}$,
\begin{multline} \label{eqn:td_image_likelihood}
	p(D_{\rm td}, D_{\rm image} | D_{\Delta t}/\lambda) =
	\int  p(D_{\rm image} | \boldsymbol{\xi}_{\rm mass}, \boldsymbol{\xi}_{\rm light}) \\
	\times p(D_{\rm td} | \boldsymbol{\xi}_{\rm mass}, D_{\Delta t}/\lambda)
	p(\boldsymbol{\xi}_{\rm mass}, \boldsymbol{\xi}_{\rm light}) d\boldsymbol{\xi}_{\rm mass} d\boldsymbol{\xi}_{\rm light}.
\end{multline}
This allows us to use individually sampled angular diameter distance posteriors (expression \ref{eqn:single_lens_inference}) without sampling an additional MST and then transform them in post-processing. This is applicable for both, external convergence and internal MST and we effectively evaluate the likelihood on the one-dimensional posterior density in $D_{\Delta t}/\lambda$.

In the same way as for the time-delay likelihood, we can perform the marginalization of the kinematics likelihood over the imaging data constraints
\begin{multline} \label{eqn:spec_image_likelihood}
	p(\mathcal{D}_{\rm spec}, \mathcal{D}_{\rm img} | \boldsymbol{\beta}_{\rm ani}, D_{\rm s}/D_{\rm ds}, \lambda) = \\
	\int  p(\mathcal{D}_{\rm img} | \boldsymbol{\xi}_{\rm mass}, \boldsymbol{\xi}_{\rm light}) p(D_{\rm spec} | \boldsymbol{\xi}_{\rm mass}, \boldsymbol{\xi}_{\rm light}, \boldsymbol{\beta}_{\rm ani}, D_{\rm s}/D_{\rm ds}, \lambda) \\
	\times p(\boldsymbol{\xi}_{\rm mass}, \boldsymbol{\xi}_{\rm light}) d\boldsymbol{\xi}_{\rm mass} d\boldsymbol{\xi}_{\rm light}.
\end{multline}

\section{TDLMC inference with more general anisotropy models} \label{app:tdlmc_anisotropy}
In this work, we presented inferences based on the anisotropy parameterization by \cite{Osipkov:1979, Merritt:1985} (Eqn. \ref{eqn:r_ani}). In this Appendix we perform the inference on the TDLMC with a more general anisotropy parameterization.
\cite{Agnello:2014} introduced a generalization of the Osipkov--Merritt profile with an asymptotic anisotropy value, $\beta_{\infty}$, different than radial

\begin{equation} \label{eqn:gom_anisotropy}
  \beta_{\text{ani}}(r) = \beta_{\infty} \frac{r^2}{r_{\text{ani}}^2+r^2}.
\end{equation}

\begin{figure}
  \centering
  \includegraphics[angle=0, width=80mm]{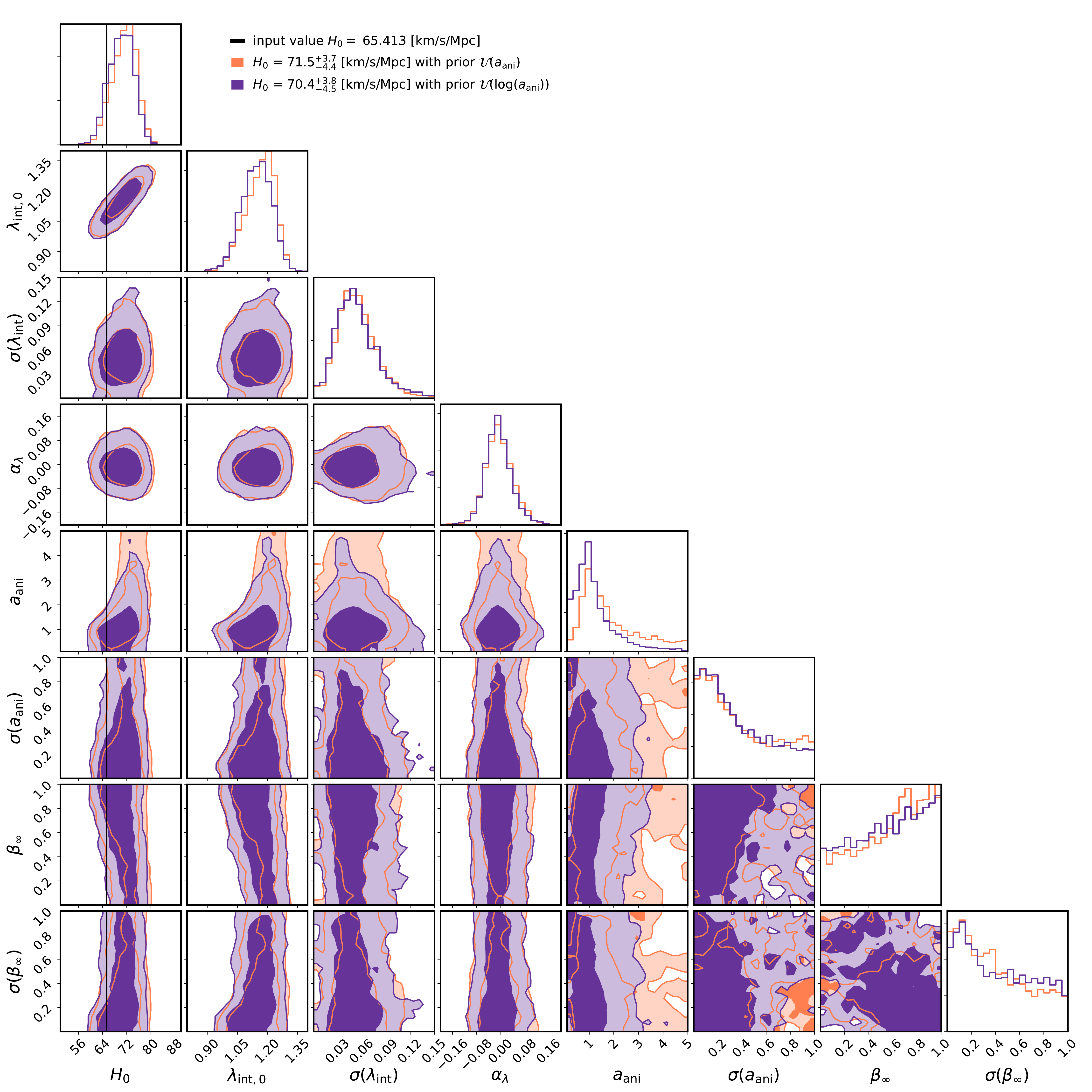}
  \caption{TDLMC Rung3 inference with fixed $\Omega_{\rm m}$ to the correct value and a generalized Osipkov--Merritt anisotropy profile (Eqn. \ref{eqn:gom_anisotropy}). Blue contours indicate the inference with a uniform prior in $a_{\rm ani}$ while the red contours indicate the inference with uniform priors in $\log(a_{\rm ani})$. The thin vertical line indicates the ground truth $H_0$ value in the challenge.  \faGithub\href{https://github.com/TDCOSMO/hierarchy_analysis_2020_public/blob/6c293af582c398a5c9de60a51cb0c44432a3c598/TDLMC/TDLMC_rung3_inference.ipynb}{~source}}
  \label{fig:tdlmc_rung3_omega_m_fixed_gom}
\end{figure}

We perform the identical analysis as presented in Section \ref{sec:tdlmc} except for the addition of one free parameter, $\beta_{\infty}$.
Table \ref{table:param_summary_tdlmc_gom} presents the parameters and priors used in the hierarchical analysis on the TDLMC data set. Figure \ref{fig:tdlmc_rung3_omega_m_fixed_gom} shows the results of this inference for the two different priors in $a_{\rm ani}$. The additional degree of freedom in the anisotropy is not constrained by the mock data and leads to a prior-volume effect. The constraining power on the mass profile relies on the mean anisotropy in the orbits within the aperture of the measurement, and not particularly on the parameterization of the radial dependence \citep[see also e.g.,][]{Agnello:2014b}. It is more challenging to find uninformative priors in higher dimension. As we found an uninformative prior in a simpler parameterization that leads to a consistent result on the TDLMC data set, we do not explore more degrees of freedom in the anisotropy parameterization in this work.

\begin{figure}
  \centering
  \includegraphics[angle=0, width=80mm]{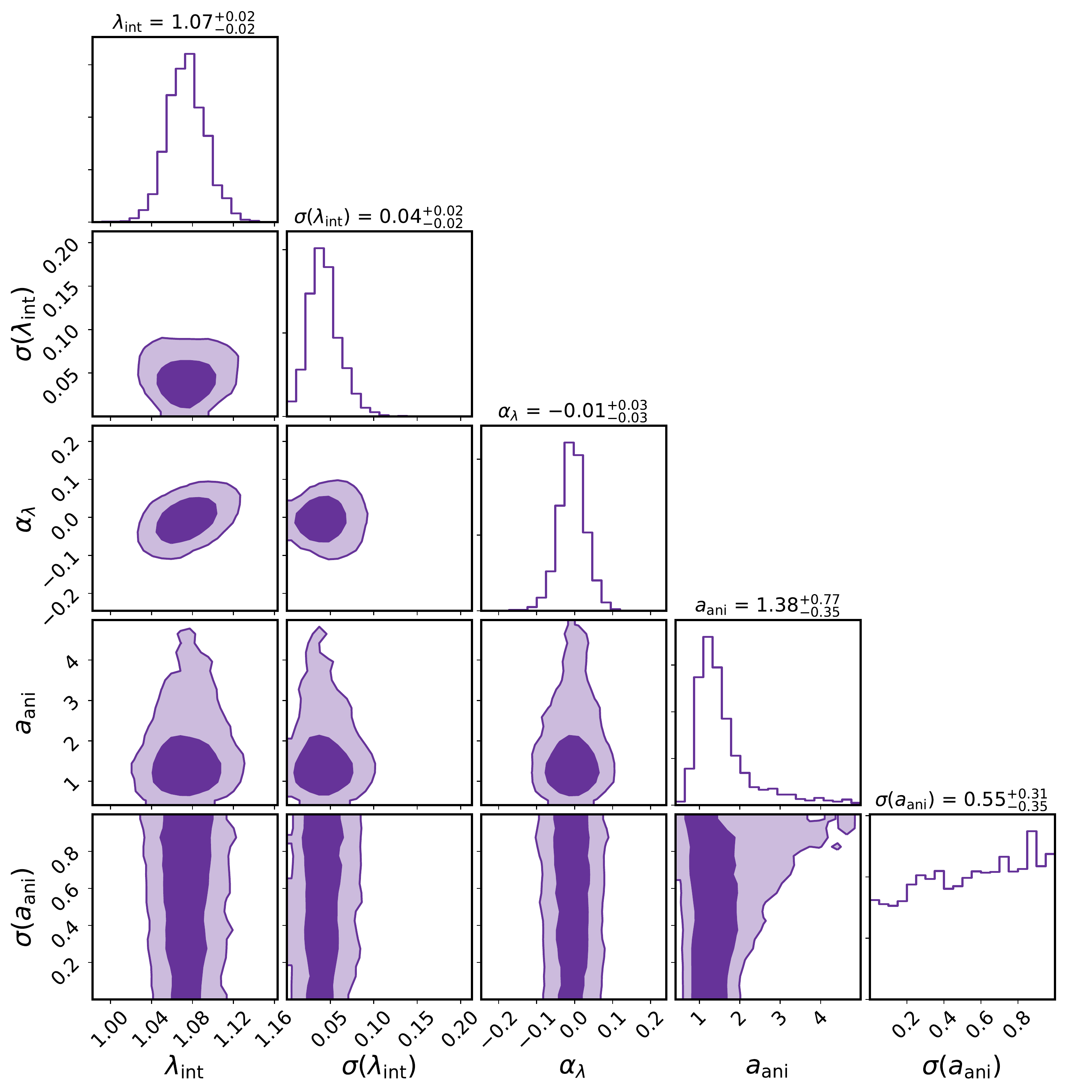}
  \caption{TDLMC Rung3 inference on the profile and anisotropy parameter when assuming the correct cosmology.  \faGithub\href{https://github.com/TDCOSMO/hierarchy_analysis_2020_public/blob/6c293af582c398a5c9de60a51cb0c44432a3c598/TDLMC/TDLMC_rung3_inference.ipynb}{~source} }
  \label{fig:tdlmc_rung3_cosmo_fixed_om}
\end{figure}

\begin{figure}
  \centering
  \includegraphics[angle=0, width=80mm]{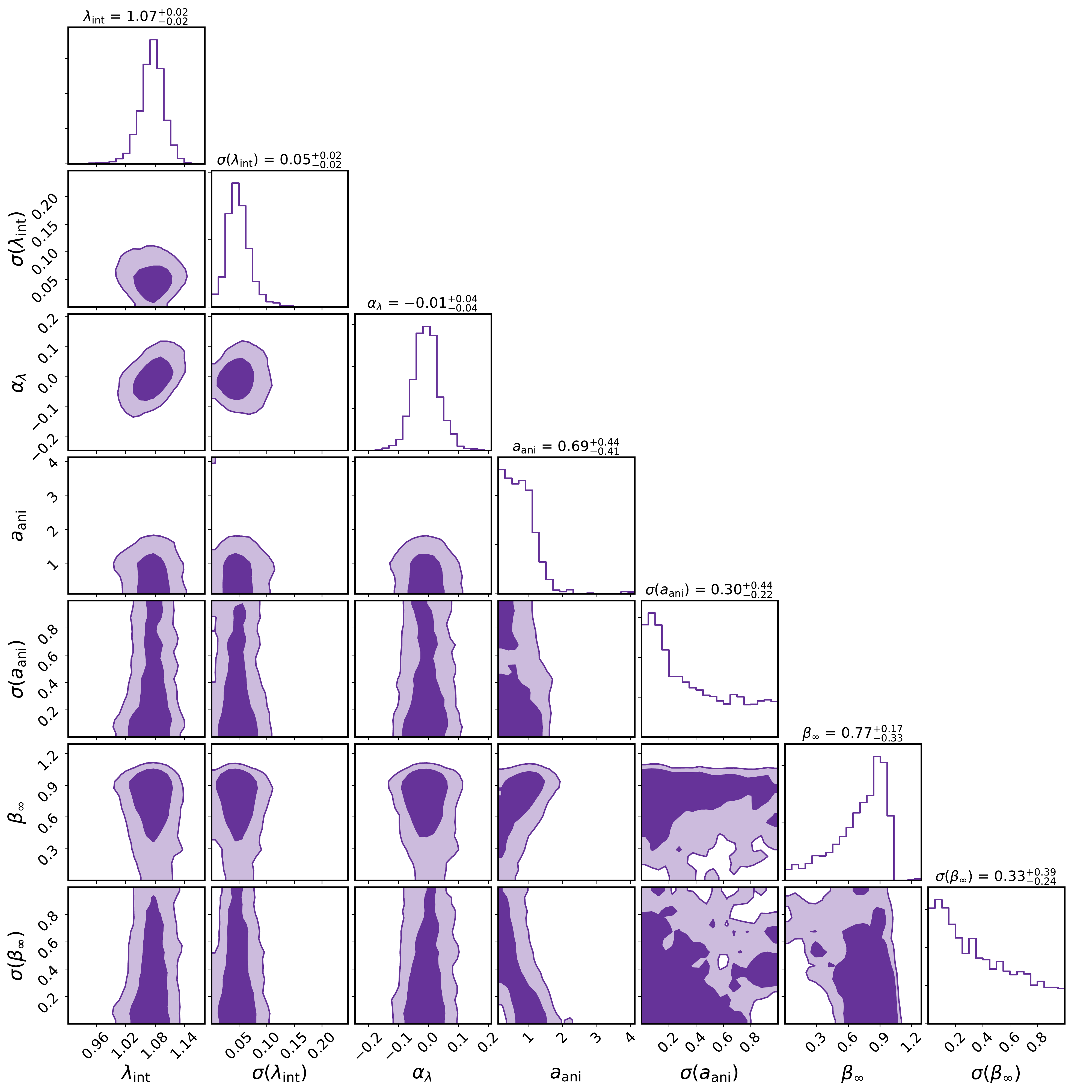}
  \caption{TDLMC Rung3 inference on the profile and anisotropy parameter when assuming the correct cosmology for a generalized Osipkov--Merritt anisotropy profile (Eqn. \ref{eqn:gom_anisotropy}).  \faGithub\href{https://github.com/TDCOSMO/hierarchy_analysis_2020_public/blob/6c293af582c398a5c9de60a51cb0c44432a3c598/TDLMC/TDLMC_rung3_inference.ipynb}{~source}}
  \label{fig:tdlmc_rung3_cosmo_fixed_gom}
\end{figure}

On the mock data with known input cosmology, we can also reverse the problem and ask which anisotropy parameter configurations result in statistically consistent cosmologies. To do so, we fix the cosmology to the input values and only perform the inference on the anisotropy parameters.
Figure \ref{fig:tdlmc_rung3_cosmo_fixed_om} presents the results for the Osipkov--Merritt model of Section \ref{sec:tdlmc} and Figure \ref{fig:tdlmc_rung3_cosmo_fixed_gom} presents the results for the generalized Osipkov--Merritt profile of this Appendix.
The posterior on the anisotropy parameter can be interpreted as an informative prior on the anisotropy model parameters from the hydrodynamical simulations of the TDLMC. We do not make use of such a prior in this work but note the consistent inference of the anisotropy parameters for the TDCOSMO+SLACS analysis with this exercise performed on the TDLMC.

\begin{table*}
\caption{Summary of the model parameters sampled in the hierarchical inference on TDLMC Rung3 with the anisotropy model of Equation \ref{eqn:gom_anisotropy}.}
\begin{center}
\begin{threeparttable}
\begin{tabular}{l l l}
    \hline
    name & prior & description \\
    \hline \hline
    Cosmology \\
    $H_0$ [\Hunit] & $\mathcal{U}([0, 150])$  & Hubble constant \\
    $\Omega_{\rm m}$ & $=0.27$ & current normalized matter density \\
    \hline
    Mass profile \\
    $\lambda_{\rm int}$ & $\mathcal{U}([0.8, 1.2])$ & internal MST population mean \\
    $\sigma(\lambda_{\rm int})$ & $\mathcal{U}([0, 0.2])$ & 1-$\sigma$ Gaussian scatter in the internal MST \\
    \hline
    Stellar kinematics \\
    $\langle a_{\rm ani}\rangle$ & $\mathcal{U}([0.1, 5])$ or $\mathcal{U}(\log([0.1, 5]))$ & scaled anisotropy radius (Eqn. \ref{eqn:r_ani}, \ref{eqn:a_ani}) \\
    $\sigma(a_{\rm ani})$ & $\mathcal{U}([0, 1])$  & $\sigma(a_{\rm ani}) \langle a_{\rm ani}\rangle$ is the 1-$\sigma$ Gaussian scatter in $a_{\rm ani}$\\
    $\beta_{\infty}$ & $\mathcal{U}([0, 1])$ & anisotropy at infinity (Eqn. \ref{eqn:gom_anisotropy}) \\
    $\sigma(\beta_{\infty})$  & $\mathcal{U}([0, 1])$  & 1-$\sigma$ Gaussian scatter in $\beta_{\infty}$ distribution  \\
    \hline
    Line of sight \\
    $\langle\kappa_{\rm ext}\rangle$ & $=0$  & population mean in external convergence of lenses \\
    $\sigma(\kappa_{\rm ext})$ & $=0.025$  & 1-$\sigma$ Gaussian scatter in $\kappa_{\rm ext}$ \\
    \hline
\end{tabular}
\begin{tablenotes}
\end{tablenotes}
\end{threeparttable}
\end{center}
\label{table:param_summary_tdlmc_gom}
\end{table*}

\section{SLACS sample details} \label{app:slacs_data}
In this Appendix we provide the detailed numerical numbers used in this analysis for the SLACS lenses. Table \ref{table:slacs_data_summary} lists the data derived from external works that are used in our analysis for the 33 lenses of the SLACS sample. Redshifts are from SDSS presented by \cite{Auger:2009}, Einstein radii from \cite{Auger:2009} and \cite{Shajib_slacs:2020} (where available), half-light radii, $r_{\rm eff}$, from \cite{Auger:2009}, power-law slopes from \cite{Shajib_slacs:2020} (where available) and velocity dispersions are based on \cite{Bolton:2008} and \cite{Shu:2015}. Local environment statistics $\zeta_{1/r}$ and external shear $\kappa_{\rm ext}$ are derived in this work (see Section \ref{sec:slacs_selection} and Section \ref{sec:slacs_los_convergence}).

\begin{table*}
\caption{Summary of the parameters being used of the individual 33 SLACS lenses selected in Section \ref{sec:slacs_analysis} to infer mass profile constraints in combination of imaging and kinematics. Aside the name, lens and source redshift, the Einstein radius $\theta_{\rm E}$ , half-light radius of the deflector $r_{\rm eff}$, imaging data-only inference on the power-law slope $\gamma_{\rm pl}$ (where available), 1/r weighted galaxy number count $\zeta_{1/r}$, external convergence $\kappa_{\rm ext}$, measured velocity dispersion $\sigma_{\rm SDSS}$ and whether VIMOS IFU data is available are provided.}
\begin{center}
\begin{threeparttable}
\begin{tabular}{l l l l l l l l l l}
    \hline
    name  & $z_{\rm lens}$ & $z_{\rm source}$ & $\theta_{\rm E}$ [arcsec] & $r_{\rm eff}$ [arcsec] & $\gamma_{\rm pl}$ & $\zeta_{1/r}$ & $\kappa_{\rm ext}$ & $\sigma_{\rm SDSS}$[km/s] & IFU \\
    \hline \hline
SDSSJ0008-0004 & 0.44 & 1.192 & 1.159$\pm$0.020 & 1.710$\pm$0.060 & - & 1.47 & ${+0.019}_{-0.021}^{+0.040}$ & 228$\pm$27 & no \\
SDSSJ0029-0055 & 0.227 & 0.931 & 0.951$\pm$0.004 & 2.160$\pm$0.076 & 2.46$\pm$0.10 & 1.14 & ${-0.002}_{-0.008}^{+0.015}$ & 216$\pm$15 & no \\
SDSSJ0037-0942 & 0.195 & 0.632 & 1.503$\pm$0.017 & 1.800$\pm$0.063 & 2.19$\pm$0.04 & 1.60 & ${+0.012}_{-0.010}^{+0.020}$ & 265$\pm$8 & yes \\
SDSSJ0044+0113 & 0.12 & 0.197 & 0.795$\pm$0.020 & 1.920$\pm$0.067 & - & 1.68 & ${-0.001}_{-0.002}^{+0.005}$ & 267$\pm$9 & no \\
SDSSJ0216-0813 & 0.3317 & 0.5235 & 1.160$\pm$0.020 & 2.970$\pm$0.200 & - & 0.83 & ${-0.005}_{-0.003}^{+0.005}$ & 351$\pm$19 & yes \\
SDSSJ0330-0020 & 0.351 & 1.071 & 1.079$\pm$0.012 & 0.910$\pm$0.032 & 2.16$\pm$0.03 & 1.32 & ${+0.006}_{-0.013}^{+0.021}$ & 273$\pm$23 & no \\
SDSSJ0728+3835 & 0.206 & 0.688 & 1.282$\pm$0.006 & 1.780$\pm$0.062 & 2.23$\pm$0.06 & 1.12 & ${-0.002}_{-0.006}^{+0.012}$ & 210$\pm$8 & no \\
SDSSJ0912+0029 & 0.164 & 0.324 & 1.627$\pm$0.020 & 4.010$\pm$0.140 & - & 1.71 & ${+0.001}_{-0.004}^{+0.010}$ & 301$\pm$9 & yes \\
SDSSJ0959+4416 & 0.237 & 0.531 & 0.961$\pm$0.020 & 1.980$\pm$0.069 & - & 1.41 & ${+0.003}_{-0.006}^{+0.012}$ & 242$\pm$13 & no \\
SDSSJ1016+3859 & 0.168 & 0.439 & 1.090$\pm$0.020 & 1.460$\pm$0.051 & - & 1.58 & ${+0.005}_{-0.007}^{+0.012}$ & 255$\pm$10 & no \\
SDSSJ1020+1122 & 0.282 & 0.553 & 1.200$\pm$0.020 & 1.590$\pm$0.056 & - & 0.54 & ${-0.006}_{-0.003}^{+0.005}$ & 282$\pm$13 & no \\
SDSSJ1023+4230 & 0.191 & 0.696 & 1.414$\pm$0.020 & 1.770$\pm$0.062 & - & 1.65 & ${+0.016}_{-0.010}^{+0.016}$ & 272$\pm$12 & no \\
SDSSJ1112+0826 & 0.273 & 0.629 & 1.422$\pm$0.015 & 1.320$\pm$0.046 & 2.21$\pm$0.06 & 1.96 & ${+0.035}_{-0.021}^{+0.043}$ & 260$\pm$15 & no \\
SDSSJ1134+6027 & 0.153 & 0.474 & 1.102$\pm$0.020 & 2.020$\pm$0.071 & - & 1.49 & ${+0.003}_{-0.006}^{+0.012}$ & 239$\pm$8 & no \\
SDSSJ1142+1001 & 0.222 & 0.504 & 0.984$\pm$0.020 & 1.240$\pm$0.043 & - & 1.18 & ${-0.001}_{-0.005}^{+0.008}$ & 238$\pm$16 & no \\
SDSSJ1153+4612 & 0.18 & 0.875 & 1.047$\pm$0.020 & 1.160$\pm$0.041 & - & 1.55 & ${+0.017}_{-0.014}^{+0.026}$ & 211$\pm$11 & no \\
SDSSJ1204+0358 & 0.164 & 0.631 & 1.287$\pm$0.009 & 1.090$\pm$0.038 & 2.18$\pm$0.08 & 1.89 & ${+0.023}_{-0.013}^{+0.023}$ & 251$\pm$12 & yes \\
SDSSJ1213+6708 & 0.123 & 0.64 & 1.416$\pm$0.020 & 1.500$\pm$0.052 & - & 1.00 & ${-0.004}_{-0.004}^{+0.008}$ & 267$\pm$7 & no \\
SDSSJ1218+0830 & 0.135 & 0.717 & 1.450$\pm$0.020 & 2.700$\pm$0.095 & - & 1.40 & ${+0.006}_{-0.008}^{+0.014}$ & 222$\pm$7 & no \\
SDSSJ1250+0523 & 0.232 & 0.795 & 1.119$\pm$0.029 & 1.320$\pm$0.046 & 1.92$\pm$0.05 & 1.57 & ${+0.021}_{-0.017}^{+0.034}$ & 242$\pm$10 & yes \\
SDSSJ1306+0600 & 0.173 & 0.472 & 1.298$\pm$0.013 & 1.250$\pm$0.044 & 2.18$\pm$0.05 & 1.79 & ${+0.011}_{-0.012}^{+0.022}$ & 248$\pm$14 & no \\
SDSSJ1402+6321 & 0.205 & 0.481 & 1.355$\pm$0.003 & 2.290$\pm$0.080 & 2.23$\pm$0.07 & 1.73 & ${+0.008}_{-0.008}^{+0.013}$ & 274$\pm$11 & no \\
SDSSJ1403+0006 & 0.189 & 0.473 & 0.830$\pm$0.020 & 1.140$\pm$0.040 & - & 1.51 & ${+0.004}_{-0.006}^{+0.010}$ & 202$\pm$12 & no \\
SDSSJ1432+6317 & 0.123 & 0.664 & 1.258$\pm$0.020 & 3.040$\pm$0.106 & - & 1.77 & ${+0.021}_{-0.011}^{+0.016}$ & 210$\pm$6 & no \\
SDSSJ1451-0239 & 0.1254 & 0.5203 & 1.040$\pm$0.020 & 2.640$\pm$0.200 & - & 1.08 & ${-0.001}_{-0.005}^{+0.006}$ & 204$\pm$10 & yes \\
SDSSJ1531-0105 & 0.16 & 0.744 & 1.704$\pm$0.008 & 1.970$\pm$0.069 & 1.92$\pm$0.11 & 1.36 & ${+0.010}_{-0.013}^{+0.023}$ & 261$\pm$10 & no \\
SDSSJ1621+3931 & 0.245 & 0.602 & 1.263$\pm$0.004 & 1.510$\pm$0.053 & 2.02$\pm$0.06 & 0.97 & ${-0.005}_{-0.004}^{+0.008}$ & 234$\pm$15 & no \\
SDSSJ1627-0053 & 0.208 & 0.524 & 1.227$\pm$0.002 & 1.980$\pm$0.069 & 1.85$\pm$0.14 & 1.47 & ${+0.004}_{-0.007}^{+0.014}$ & 274$\pm$11 & yes \\
SDSSJ1630+4520 & 0.248 & 0.793 & 1.786$\pm$0.029 & 1.650$\pm$0.058 & 2.00$\pm$0.03 & 1.29 & ${+0.004}_{-0.010}^{+0.019}$ & 283$\pm$13 & no \\
SDSSJ1644+2625 & 0.137 & 0.61 & 1.267$\pm$0.020 & 1.550$\pm$0.054 & - & 1.86 & ${+0.023}_{-0.014}^{+0.027}$ & 208$\pm$9 & no \\
SDSSJ2303+1422 & 0.155 & 0.517 & 1.613$\pm$0.007 & 2.940$\pm$0.103 & 2.00$\pm$0.04 & 1.56 & ${+0.006}_{-0.008}^{+0.020}$ & 251$\pm$13 & yes \\
SDSSJ2321-0939 & 0.082 & 0.532 & 1.599$\pm$0.020 & 4.110$\pm$0.144 & - & 1.23 & ${+0.000}_{-0.005}^{+0.008}$ & 240$\pm$6 & yes \\
SDSSJ2347-0005 & 0.417 & 0.714 & 1.107$\pm$0.020 & 1.140$\pm$0.040 & - & 1.39 & ${+0.006}_{-0.008}^{+0.015}$ & 404$\pm$59 & no \\

    \hline
\end{tabular}
\begin{tablenotes}
\end{tablenotes}
\end{threeparttable}
\end{center}
\label{table:slacs_data_summary}
\end{table*}

\end{document}